\documentclass[11pt]{cernrep}
\usepackage{graphicx}
\usepackage{here}
\usepackage{epsf}
\usepackage{epsfig}
\usepackage{psfrag}
\usepackage{axodraw}
\usepackage{rotate}

\renewcommand{\thesection}{\arabic{section}}
\renewcommand{\thesubsection}{\arabic{section}.\arabic{subsection}}

\newcommand{\lsim}{
\mathrel{\hbox{\rlap{\hbox{\lower4pt\hbox{$\sim$}}}\hbox{$<$}}}}

\newcommand{\gsim}{
\mathrel{\hbox{\rlap{\hbox{\lower4pt\hbox{$\sim$}}}\hbox{$>$}}}}

\begin{document}
 
\pagestyle{empty}


\begin{titlepage}
\begin{flushright}
\begin{tabular}{r}
CERN-PH-TH/2004-085\\
hep-ph/0405091\\
\end{tabular}
\end{flushright}

\vspace*{1.9truecm}

\begin{center}
\boldmath
{\LARGE \bf Flavour Physics and CP Violation}
\unboldmath

\vspace*{2.6cm}

\smallskip
\begin{center}
{\sc {\large Robert Fleischer}}\\
\vspace*{2mm}
{\sl {\large Theory Division, Department of Physics, CERN, 
CH-1211 Geneva 23, Switzerland}}
\end{center}

\vspace{2.6truecm}

{\large\bf Abstract\\[10pt]} \parbox[t]{\textwidth}{
The starting point of these lectures is an introduction to the weak 
interactions of quarks and the Standard-Model description of CP violation, 
where the key element is the Cabibbo--Kobayashi--Maskawa matrix and the
corresponding unitarity triangles. Since the $B$-meson system will govern 
the stage of (quark) flavour physics and CP violation in this decade, 
it will be -- after a brief look at the kaon system -- our main focus. We 
shall classify $B$-meson decays, introduce the theoretical tools to deal 
with them, explore the requirements for non-vanishing CP-violating 
asymmetries, and discuss $B^0_q$--$\bar B^0_q$ mixing ($q\in\{d,s\}$). We 
will then turn to $B$-factory benchmark modes, discuss the physics 
potential of $B^0_s$ mesons, which is particularly promising for $B$-decay 
experiments at hadron colliders, and emphasize the importance of studies of 
rare decays, which are absent at the tree level in the Standard Model, 
complement nicely the studies of CP violation, and provide interesting 
probes for new physics.   
}

\vspace{2.6cm}
 
{\sl 
Lectures given at the {\it 2003 European School of High-Energy Physics,}\\
Tsakhkadzor, Armenia, 24 August -- 6 September 2003\\
To appear in the Proceedings (CERN Report)
}
\end{center}

\end{titlepage}
 
\thispagestyle{empty}
\vbox{}
\newpage


\pagestyle{plain}

\setcounter{page}{1}
\pagenumbering{roman}

\tableofcontents

\newpage

\setcounter{page}{1}
\pagenumbering{arabic}

\title{FLAVOUR PHYSICS AND CP VIOLATION}
\author{Robert Fleischer}
\institute{CERN, Geneva, Switzerland}
\maketitle
\begin{abstract}
The starting point of these lectures is an introduction to the weak 
interactions of quarks and the Standard-Model description of CP violation, 
where the key element is the Cabibbo--Kobayashi--Maskawa matrix and the
corresponding unitarity triangles. Since the $B$-meson system will govern 
the stage of (quark) flavour physics and CP violation in this decade, 
it will be -- after a brief look at the kaon system -- our main focus. We 
shall classify $B$-meson decays, introduce the theoretical tools to deal 
with them, explore the requirements for non-vanishing CP-violating 
asymmetries, and discuss $B^0_q$--$\bar B^0_q$ mixing ($q\in\{d,s\}$). We 
will then turn to $B$-factory benchmark modes, discuss the physics 
potential of $B^0_s$ mesons, which is particularly promising for $B$-decay 
experiments at hadron colliders, and emphasize the importance of studies of 
rare decays, which are absent at the tree level in the Standard Model, 
complement nicely the studies of CP violation, and provide interesting 
probes for new physics.   
\end{abstract}

\pagestyle{plain}

\section{INTRODUCTION}\label{sec:intro}
\setcounter{equation}{0}
The violation of the CP symmetry, where C and P are the 
charge-conjugation and parity-transformation operators, respectively,
is one of the fundamental and most exciting phenomena in particle 
physics. Although weak interactions are not invariant under P (and C) 
transformations, as discovered in 1957, it was believed for several 
years that the product CP was preserved. Consider, for instance, the 
process
\begin{equation}
\pi^+\to e^+\nu_e~\stackrel{{\cal C}}{\longrightarrow}~\pi^-\to 
e^-\nu_e^{\rm C}~\stackrel{{\cal P}}{\longrightarrow}~\pi^-\to e^-\bar \nu_e,
\end{equation}
where the left-handed $\nu_e^{\rm C}$ state is not observed in nature; only 
after performing an additional parity transformation we obtain the usual 
right-handed electron antineutrino. Consequently, it appears as if
CP was conserved in weak interactions. However, in 1964, it was 
discovered through the observation of $K_{\rm L}\to \pi^+\pi^-$ decays 
that weak interactions are {\it not} invariant under CP transformations 
\cite{CP-discovery}. 

After its discovery, CP violation was, for a very long time, only 
accessible in the neutral kaon system, where it is described by two 
complex parameters, $\varepsilon$ and $\varepsilon'$; a non-zero
value of the latter could only be established -- after tremendous 
efforts -- in 1999 \cite{NA48-obs,KTeV-obs}. In 2001, CP violation 
could then also be observed in decays of neutral $B$ mesons 
\cite{babar-CP-obs,belle-CP-obs}, 
which represents the beginning of a new era in the exploration of this 
phenomenon. Despite this impressive progress, we still have few experimental 
insights into CP violation, which originates, within the Standard Model (SM) 
of electroweak interactions, from the flavour structure of the
charged-current interactions \cite{aitchison}. One of the main motivations 
for the exploration of CP violation is that ``new'' physics 
(NP), i.e.\ physics lying beyond the SM, is typically also associated 
with new sources of CP violation and new flavour structures 
\cite{gabadadze}--\cite{GNR97}. This is actually the case in many specific 
NP scenarios, for instance in supersymmetry (SUSY), 
left--right-symmetric models, and in models with extended Higgs sectors. 
In this context, it is also interesting to note that the evidence for 
non-vanishing neutrino masses that we obtained over the last years points 
towards on origin beyond the SM \cite{petkov,AF}, raising -- among other 
issues -- also the question of having CP violation in the neutrino sector, 
which could be studied, in the more distant future, at dedicated neutrino 
factories \cite{nu-fact}.

Interestingly, we may also obtain indirect information on CP violation from
cosmology. One of the characteristic features of our Universe is the 
cosmological baryon asymmetry of ${\cal O}(10^{-10})$ \cite{tkachev,buchm}. 
As was pointed out by Sakharov \cite{sakharov}, one of the necessary 
conditions to generate such an asymmetry of the Universe is -- in 
addition to baryon-number violation and deviations from thermal 
equilibrium -- that the elementary interactions violate CP (and C). 
Model calculations indicate, however, that the CP violation present 
in the SM is too small to generate the observed 
matter--antimatter asymmetry of ${\cal O}(10^{-10})$ 
\cite{shapos}. It is conceivable that the particular kind of NP
underlying the baryon asymmetry is associated with very short-distance
scales. In this case, it could not be seen in CP-violating effects in 
weak meson decays. However, as we have noted above, there are also
various scenarios for physics beyond the SM that would affect these 
processes. Moreover, we do not understand the observed patterns of quark 
and lepton masses, their mixings and the origin of flavour dynamics in 
general. It is likely that the NP required to understand these 
features is also related to new sources of CP violation.

The field of (quark) flavour physics and CP violation is very broad.
In this decade, it will be governed by studies of decays of $B$ mesons. 
The asymmetric $e^+e^-$ $B$ factories operating at the $\Upsilon(4S)$ 
resonance \cite{BaBar-Book}, with their detectors BaBar (SLAC) and 
Belle (KEK), have already been taking data for a couple of years and 
have produced plenty of exciting results. Moreover, 
also hadron colliders have a very 
promising potential for the exploration of $B$-meson decays. We may expect 
first interesting results on several processes from run II of the Tevatron 
soon \cite{TEV-Book}. The corresponding channels can then be fully 
exploited in the era of the LHC, in particular by LHCb (CERN) and 
BTeV (FNAL) \cite{LHC-Book}. The great interest in $B$ physics -- our
main topic -- originates from the fact that it provides a very fertile 
testing ground for the SM picture of flavour physics and CP violation, 
as we will see in these lectures. The outline is as follows: 
in Section~\ref{sec:CP-SM}, we have a closer look at the weak interactions 
of quarks, discuss the quark-mixing matrix, and introduce the unitarity 
triangle(s). After giving a brief introduction to the CP violation in the 
kaon system and making first contact with ``rare'' $K$ decays in 
Section~\ref{sec:kaon}, we enter the world of the $B$ mesons in 
Section~\ref{sec:Bdecays}, where we shall classify their decays, 
discuss the theoretical tools to deal with them, and
investigate the requirements for non-vanishing CP asymmetries. In 
Section~\ref{sec:mix}, we discuss features of neutral $B_q$
mesons ($q\in\{d,s\}$), including the very important 
phenomenon of $B^0_q$--$\bar B^0_q$ mixing, and 
introduce the corresponding CP-violating observables. These considerations
then allow us to have a closer look at important benchmark modes for the 
$B$ factories in Section~\ref{sec:b-fact-bench}, where we will also
address the current experimental status. In Section~\ref{sec:A-REL}, we 
discuss the exploration of CP violation with the help of amplitude relations,
whereas we shall focus on the $B_s$-meson system, which is particularly
interesting for $B$-decay studies at hadron colliders, in 
Section~\ref{sec:Bs}. In Section~\ref{sec:rare}, we emphasize the 
importance of studies of ``rare'' $B$- and $K$-meson decays, which are 
absent at the tree level in the SM, and offer important probes for the search 
of NP. Finally, we summarize our conclusions in Section~\ref{sec:concl}. 

For a collection of detailed textbooks and reviews on CP violation
and flavour physics, the reader is referred to 
\cite{CP-Books}--\cite{RF-Phys-Rep}. Since this field is evolving 
quickly, I will also address recent developments that took place after 
the school in Tsakhkadzor in order to complement the material that I
presented there. The data refer to the experimental situation in early 2004.

\section{CP VIOLATION IN THE STANDARD MODEL}\label{sec:CP-SM}
\setcounter{equation}{0}
\subsection{Weak Interactions of Quarks and the Quark-Mixing Matrix}
In the framework of the Standard Model of electroweak interactions 
\cite{aitchison,SM}, which is based on the spontaneously broken gauge group
\begin{equation}
SU(2)_{\rm L}\times U(1)_{\rm Y}
\stackrel{{\rm SSB}}
{\longrightarrow}U(1)_{\rm em},
\end{equation}
CP-violating effects may originate from the charged-current 
interactions of quarks, having the structure
\begin{equation}\label{cc-int}
D\to U W^-.
\end{equation}
Here $D\in\{d,s,b\}$ and $U\in\{u,c,t\}$ denote down- and up-type quark 
flavours, respectively, whereas the $W^-$ is the usual $SU(2)_{\rm L}$ 
gauge boson. From a phenomenological point of view, it is convenient to 
collect the generic ``coupling strengths'' $V_{UD}$ of the charged-current 
processes in (\ref{cc-int}) in the form of the following matrix:
\begin{equation}\label{ckm0}
\hat V_{\rm CKM}=
\left(\begin{array}{ccc}
V_{ud}&V_{us}&V_{ub}\\
V_{cd}&V_{cs}&V_{cb}\\
V_{td}&V_{ts}&V_{tb}
\end{array}\right),
\end{equation}
which is referred to as the Cabibbo--Kobayashi--Maskawa (CKM) matrix
\cite{cab,km}. 

From a theoretical point of view, this matrix connects the electroweak 
states $(d',s',b')$ of the down, strange and bottom quarks with their 
mass eigenstates $(d,s,b)$ through the following unitary transformation 
\cite{aitchison}:
\begin{equation}\label{ckm}
\left(\begin{array}{c}
d'\\
s'\\
b'
\end{array}\right)=\left(\begin{array}{ccc}
V_{ud}&V_{us}&V_{ub}\\
V_{cd}&V_{cs}&V_{cb}\\
V_{td}&V_{ts}&V_{tb}
\end{array}\right)\cdot
\left(\begin{array}{c}
d\\
s\\
b
\end{array}\right).
\end{equation}
Consequently, $\hat V_{\rm CKM}$ is actually a {\it unitary} matrix.
This feature ensures the absence of flavour-changing neutral-current 
(FCNC) processes at the tree level in the SM, and is hence at the basis 
of the famous Glashow--Iliopoulos--Maiani (GIM) mechanism \cite{GIM}. 
We shall return to the unitarity of the CKM matrix in
Subsection~\ref{ssec:UT}, discussing the ``unitarity triangles''.
If we express the non-leptonic charged-current interaction Lagrangian 
in terms of the mass eigenstates appearing in (\ref{ckm}), we arrive at 
\begin{equation}\label{cc-lag2}
{\cal L}_{\mbox{{\scriptsize int}}}^{\mbox{{\scriptsize CC}}}=
-\frac{g_2}{\sqrt{2}}\left(\begin{array}{ccc}
\bar u_{\mbox{{\scriptsize L}}},& \bar c_{\mbox{{\scriptsize L}}},
&\bar t_{\mbox{{\scriptsize L}}}\end{array}\right)\gamma^\mu\,\hat
V_{\mbox{{\scriptsize CKM}}}
\left(\begin{array}{c}
d_{\mbox{{\scriptsize L}}}\\
s_{\mbox{{\scriptsize L}}}\\
b_{\mbox{{\scriptsize L}}}
\end{array}\right)W_\mu^\dagger\,\,+\,\,\mbox{h.c.,}
\end{equation}
where the gauge coupling $g_2$ is related to the gauge group 
$SU(2)_{\mbox{{\scriptsize L}}}$, and the $W_\mu^{(\dagger)}$ field 
corresponds to the charged $W$ bosons. Looking at the
interaction vertices following from (\ref{cc-lag2}), we observe 
that the elements of the CKM matrix describe in fact the generic
strengths of the associated charged-current processes, as we have 
noted above.

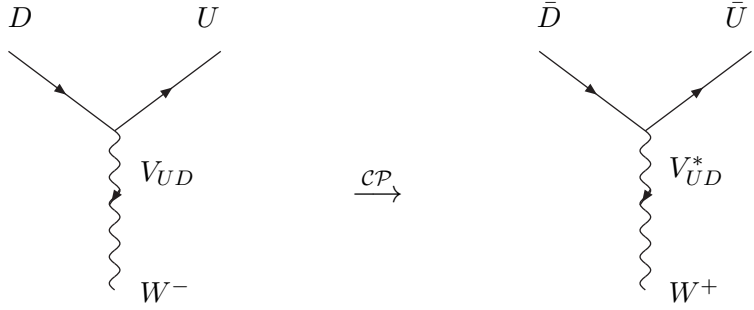
\begin{figure}
\begin{center}
\hspace*{-1truecm}\begin{picture}(360,100)(0,60)
\ArrowLine(60,150)(100,120)
\ArrowLine(100,120)(140,150)
\ArrowLine(260,150)(300,120)
\ArrowLine(300,120)(340,150)
\Photon(100,120)(100,60){2}{6}
\Photon(300,120)(300,60){2}{6}
\ArrowLine(102,98)(98,92)
\ArrowLine(302,98)(298,92)
\Text(60,160)[bl]{$D$}
\Text(260,160)[bl]{$\bar D$}
\Text(140,160)[br]{$U$}
\Text(340,160)[br]{$\bar U$}
\Text(110,100)[bl]{$V_{UD}$}
\Text(310,100)[bl]{$V_{UD}^\ast$}
\Text(110,60)[l]{$W^-$}
\Text(190,100)[l]{\large$\stackrel{{\cal 
CP}}{\longrightarrow}$}
\Text(310,60)[l]{$W^+$}
\end{picture}
\end{center}
\caption{CP-conjugate charged-current quark-level interaction processes
in the SM.}\label{fig:CC} 
\end{figure}

In Fig.~\ref{fig:CC}, we show the $D\to U W^-$ vertex and its CP 
conjugate. Since the corresponding CP transformation involves the 
replacement
\begin{equation}\label{CKM-CP}
V_{UD}\stackrel{{\cal CP}}{\longrightarrow}V_{UD}^\ast,
\end{equation}
CP violation could -- in principle -- be accommodated in the SM through 
complex phases in the CKM matrix. The crucial question in this context
is, of course, whether we may actually have physical complex phases in 
that matrix.

\subsection{Phase Structure of the CKM Matrix}
We have the freedom to redefine the up- and down-type quark fields
in the following manner:
\begin{equation}
U\to \exp(i\xi_U)U,\quad D\to \exp(i\xi_D)D. 
\end{equation}
If we perform such transformations in (\ref{cc-lag2}), the invariance 
of the charged-current interaction Lagrangian implies the 
following phase transformations of the CKM matrix elements:
\begin{equation}\label{CKM-trafo}
V_{UD}\to\exp(i\xi_U)V_{UD}\exp(-i\xi_D).
\end{equation}
Using these transformations to eliminate unphysical phases, it can be shown 
that the parametrization of the general $N\times N$ quark-mixing matrix, 
where $N$ denotes the number of fermion generations, involves the following 
parameters:
\begin{equation}
\underbrace{\frac{1}{2}N(N-1)}_{\mbox{Euler angles}} \, + \,
\underbrace{\frac{1}{2}(N-1)(N-2)}_{\mbox{complex phases}}=
(N-1)^2.
\end{equation}

If we apply this expression to the case of $N=2$ generations, we observe
that only one rotation angle -- the Cabibbo angle $\theta_{\rm C}$
\cite{cab} -- is required for the parametrization of the $2\times2$
quark-mixing matrix, which can be written in the following form:
\begin{equation}\label{Cmatrix}
\hat V_{\rm C}=\left(\begin{array}{cc}
\cos\theta_{\rm C}&\sin\theta_{\rm C}\\
-\sin\theta_{\rm C}&\cos\theta_{\rm C}
\end{array}\right),
\end{equation}
where $\sin\theta_{\rm C}=0.22$ can be determined from $K\to\pi\ell\bar\nu$ 
decays. On the other hand, in the case of $N=3$ generations, the 
parametrization of the corresponding $3\times3$ quark-mixing matrix involves 
three Euler-type angles and a single {\it complex} phase. This complex phase 
allows us to accommodate CP violation in the SM, as was pointed out by 
Kobayashi and Maskawa in 1973 \cite{km}. The corresponding picture
is referred to as the Kobayashi--Maskawa (KM) mechanism of CP violation.

In the ``standard parametrization'' advocated by the Particle Data Group
(PDG) \cite{PDG}, the three-generation CKM matrix takes the following 
form:
\begin{equation}\label{standard}
\hat V_{\rm CKM}=\left(\begin{array}{ccc}
c_{12}c_{13}&s_{12}c_{13}&s_{13}e^{-i\delta_{13}}\\ -s_{12}c_{23}
-c_{12}s_{23}s_{13}e^{i\delta_{13}}&c_{12}c_{23}-
s_{12}s_{23}s_{13}e^{i\delta_{13}}&
s_{23}c_{13}\\ s_{12}s_{23}-c_{12}c_{23}s_{13}e^{i\delta_{13}}&-c_{12}s_{23}
-s_{12}c_{23}s_{13}e^{i\delta_{13}}&c_{23}c_{13}
\end{array}\right),
\end{equation}
where $c_{ij}\equiv\cos\theta_{ij}$ and $s_{ij}\equiv\sin\theta_{ij}$. 
Performing appropriate redefinitions of the quark-field phases, the real 
angles $\theta_{12}$, $\theta_{23}$ and $\theta_{13}$ can all be made to
lie in the first quadrant. The advantage of this parametrization is that
the generation labels $i,j=1,2,3$ are introduced in such a manner that
the mixing between two chosen generations vanishes if the corresponding
mixing angle $\theta_{ij}$ is set to zero. In particular, for 
$\theta_{23}=\theta_{13}=0$, the third generation decouples, and the
$2\times2$ submatrix describing the mixing between the first and 
second generations takes the same form as (\ref{Cmatrix}).

Another interesting parametrization of the CKM matrix was proposed by 
Fritzsch and Xing \cite{FX}:
\begin{equation}
\hat V_{\rm CKM}=\left(\begin{array}{ccc}
s_{\rm u} s_{\rm d} c + c_{\rm u} c_{\rm d} e^{-i\varphi} & 
s_{\rm u} c_{\rm d} c - c_{\rm u} s_{\rm d} e^{-i\varphi} &s_{\rm u} s\\
c_{\rm u} s_{\rm d} c - s_{\rm u} c_{\rm d} e^{-i\varphi} & 
c_{\rm u} c_{\rm d} c + s_{\rm u} s_{\rm d} e^{-i\varphi} &c_{\rm u} s\\
-s_{\rm d}s & -c_{\rm d}s & c
\end{array}\right).
\end{equation}
It is inspired by the hierarchical structure of the quark-mass spectrum
and is particularly useful in the context of models for fermion masses and
mixings. The characteristic feature of this parametrization is that
the complex phase arises only in the $2\times2$ submatrix involving
the up, down, strange and charm quarks. 

Let us finally note that physical observables, for instance CP-violating
asymmetries, {\it cannot} depend on the chosen parametrization of the CKM 
matrix, i.e.\ have to be invariant under the phase transformations specified 
in (\ref{CKM-trafo}).

\subsection{Further Requirements for CP Violation}
As we have just seen, in order to be able to accommodate CP violation within 
the framework of the SM through a complex phase in the CKM matrix, at least 
three generations are required. However, this feature is not sufficient for 
observable CP-violating effects. To this end, further conditions have to 
be satisfied, which can be summarized as follows \cite{jarlskog,BBG}:
\begin{equation}\label{CP-req}
(m_t^2-m_c^2)(m_t^2-m_u^2)(m_c^2-m_u^2)
(m_b^2-m_s^2)(m_b^2-m_d^2)(m_s^2-m_d^2)\times J_{\rm CP}\,\not=\,0,
\end{equation}
where
\begin{equation}\label{JCP}
J_{\rm CP}=|\mbox{Im}(V_{i\alpha}V_{j\beta}V_{i\beta}^\ast 
V_{j\alpha}^\ast)|\quad(i\not=j,\,\alpha\not=\beta)\,.
\end{equation}

The mass factors in (\ref{CP-req}) are related to the fact that the 
CP-violating phase of the CKM matrix could be eliminated through an 
appropriate unitary transformation of the quark fields if any two quarks 
with the same charge had the same mass. Consequently, the origin 
of CP violation is closely related to the ``flavour problem'' in
elementary particle physics, and cannot be understood in a deeper 
way, unless we have fundamental insights into the hierarchy of quark 
masses and the number of fermion generations.

The second element of (\ref{CP-req}), the ``Jarlskog parameter'' 
$J_{\rm CP}$ \cite{jarlskog}, can be interpreted as a measure of the 
strength of CP violation in the SM. It does not depend on the chosen 
quark-field parametrization, i.e.\ it is invariant under (\ref{CKM-trafo}), 
and the unitarity of the CKM matrix implies that all combinations 
$|\mbox{Im}(V_{i\alpha}V_{j\beta}V_{i\beta}^\ast V_{j\alpha}^\ast)|$ 
are equal to one another. Using the standard parametrization of the
CKM matrix introduced in (\ref{standard}), we obtain
\begin{equation}\label{JCP-PDG}
J_{\rm CP}=s_{12}s_{13}s_{23}c_{12}c_{23}c_{13}^2\sin\delta_{13}.
\end{equation}
Since the current experimental information on the CKM parameters implies 
a value of $J_{\rm CP}$ at the $10^{-5}$ level, CP violation is a small 
effect in the SM. However, new complex couplings are typically present 
in scenarios for NP \cite{NP-revs,GNR97}, thereby yielding additional 
sources of CP violation.

\boldmath\subsection{Experimental Information on $|V_{\rm CKM}|$}
\begin{figure}
\vspace{0.10in}
\centerline{
\epsfysize=4.3truecm
\epsffile{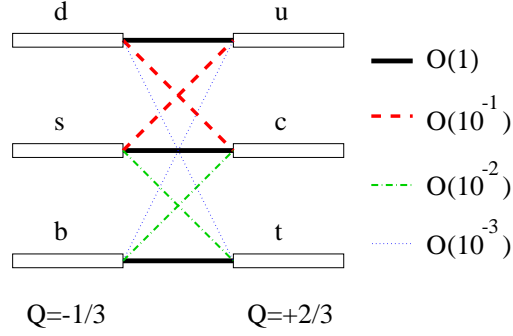}}
\caption{Hierarchy of the quark transitions mediated through 
charged-current processes.}\label{fig:term}
\end{figure}

\unboldmath
In order to determine the magnitudes $|V_{ij}|$ of the elements of the
CKM matrix, we may use the following tree-level processes:
\begin{itemize}
\item Nuclear beta decays, neutron decays $\Rightarrow$ $|V_{ud}|$.
\item $K\to\pi\ell\bar\nu$ decays $\Rightarrow$ $|V_{us}|$.
\item $\nu$ production of charm off valence $d$ quarks
$\Rightarrow$ $|V_{cd}|$.
\item Charm-tagged $W$ decays (as well as $\nu$ production and 
semileptonic $D$ decays)  $\Rightarrow$ $|V_{cs}|$.
\item Exclusive and inclusive $b\to c \ell \bar\nu$ decays 
$\Rightarrow$ $|V_{cb}|$.
\item Exclusive and inclusive 
$b\to u \ell \bar \nu$ decays $\Rightarrow$ $|V_{ub}|$.
\item $\bar t\to \bar b \ell \bar\nu$ processes $\Rightarrow$ (crude direct 
determination of) $|V_{tb}|$.
\end{itemize}
If we use the corresponding experimental information, together with the 
CKM unitarity condition, and assume that there are only three generations, 
we arrive at the following 90\% C.L. limits for the $|V_{ij}|$ \cite{PDG}:
\begin{equation}\label{CKM-mag}
|\hat V_{\rm CKM}|=\left(\begin{array}{ccc}
$0.9741\mbox{--}0.9756$ & $0.219\mbox{--}0.226$ & $0.0025\mbox{--}0.0048$\\ 
$0.219\mbox{--}0.226$ & $0.9732\mbox{--}0.9748$ & $0.038\mbox{--}0.044$\\ 
$0.004\mbox{--}0.014$ & $0.037\mbox{--}0.044$ & $0.9990\mbox{--}0.9993$\\ 
\end{array}\right).
\end{equation}
In Fig.~\ref{fig:term}, we have illustrated the resulting hierarchy 
of the strengths of the charged-current quark-level processes:
transitions within the same generation are governed by 
CKM matrix elements of ${\cal O}(1)$, those between the first and the second 
generation are suppressed by CKM factors of ${\cal O}(10^{-1})$, those 
between the second and the third generation are suppressed by 
${\cal O}(10^{-2})$, and the transitions between the first and the third 
generation are even suppressed by CKM factors of ${\cal O}(10^{-3})$. 
In the standard parametrization (\ref{standard}), this hierarchy is 
reflected by 
\begin{equation}
s_{12}=0.22 \,\gg\, s_{23}={\cal O}(10^{-2}) \,\gg\, 
s_{13}={\cal O}(10^{-3}). 
\end{equation}

\subsection{Wolfenstein Parametrization of the CKM Matrix}
For phenomenological applications, it would be useful to have a 
parametrization of the CKM matrix available that makes the 
hierarchy arising in (\ref{CKM-mag}) -- and illustrated in
Fig.~\ref{fig:term} -- explicit \cite{wolf}. In order to derive such 
a parametrization, we introduce a set of new parameters, 
$\lambda$, $A$, $\rho$ and $\eta$, by imposing the following 
relations \cite{blo}:
\begin{equation}\label{set-rel}
s_{12}\equiv\lambda=0.22,\quad s_{23}\equiv A\lambda^2,\quad 
s_{13}e^{-i\delta_{13}}\equiv A\lambda^3(\rho-i\eta).
\end{equation}
If we now go back to the standard parametrization (\ref{standard}), we 
obtain an {\it exact} parametrization of the CKM matrix as a function of 
$\lambda$ (and $A$, $\rho$, $\eta$), allowing us to expand each CKM 
element in powers of the small parameter $\lambda$. If we neglect terms of 
${\cal O}(\lambda^4)$, we arrive at the famous ``Wolfenstein 
parametrization'' \cite{wolf}:
\begin{equation}\label{W-par}
\hat V_{\mbox{{\scriptsize CKM}}} =\left(\begin{array}{ccc}
1-\frac{1}{2}\lambda^2 & \lambda & A\lambda^3(\rho-i\eta) \\
-\lambda & 1-\frac{1}{2}\lambda^2 & A\lambda^2\\
A\lambda^3(1-\rho-i\eta) & -A\lambda^2 & 1
\end{array}\right)+{\cal O}(\lambda^4),
\end{equation}
which makes the hierarchical structure of the CKM matrix very transparent 
and is an important tool for phenomenological considerations, as we 
will see throughout these lectures. 

For several applications, next-to-leading order corrections in $\lambda$ 
play an important r\^ole. Using the exact parametrization following from 
(\ref{standard}) and (\ref{set-rel}), they can be calculated straightforwardly 
by expanding each CKM element to the desired accuracy in 
$\lambda$ \cite{blo,Brev01}:
\begin{displaymath}
V_{ud}=1-\frac{1}{2}\lambda^2-\frac{1}{8}\lambda^4+{\cal O}(\lambda^6),\quad
V_{us}=\lambda+{\cal O}(\lambda^7),\quad
V_{ub}=A\lambda^3(\rho-i\,\eta),
\end{displaymath}
\begin{displaymath}
V_{cd}=-\lambda+\frac{1}{2}A^2\lambda^5\left[1-2(\rho+i\eta)\right]+
{\cal O}(\lambda^7),
\end{displaymath}
\begin{equation}\label{NLO-wolf}
V_{cs}=1-\frac{1}{2}\lambda^2-\frac{1}{8}\lambda^4(1+4A^2)+
{\cal O}(\lambda^6),
\end{equation}
\begin{displaymath}
V_{cb}=A\lambda^2+{\cal O}(\lambda^8),\quad
V_{td}=A\lambda^3\left[1-(\rho+i\eta)\left(1-\frac{1}{2}\lambda^2\right)
\right]+{\cal O}(\lambda^7),
\end{displaymath}
\begin{displaymath}
V_{ts}=-A\lambda^2+\frac{1}{2}A(1-2\rho)\lambda^4-i\eta A\lambda^4
+{\cal O}(\lambda^6),\quad
V_{tb}=1-\frac{1}{2}A^2\lambda^4+{\cal O}(\lambda^6).
\end{displaymath}
It should be noted that 
\begin{equation}
V_{ub}\equiv A\lambda^3(\rho-i\eta)
\end{equation}
receives {\it by definition} no power corrections in $\lambda$ within
this prescription. If we follow \cite{blo} and introduce the generalized
Wolfenstein parameters
\begin{equation}\label{rho-eta-bar}
\bar\rho\equiv\rho\left(1-\frac{1}{2}\lambda^2\right),\quad
\bar\eta\equiv\eta\left(1-\frac{1}{2}\lambda^2\right),
\end{equation}
we may simply write, up to corrections of ${\cal O}(\lambda^7)$,
\begin{equation}\label{Vtd-expr}
V_{td}=A\lambda^3(1-\bar\rho-i\,\bar\eta).
\end{equation}
Moreover, we have to an excellent accuracy
\begin{equation}\label{Def-A}
V_{us}=\lambda\quad \mbox{and}\quad 
V_{cb}=A\lambda^2,
\end{equation}
as these quantities receive only corrections at the $\lambda^7$ and
$\lambda^8$ levels, respectively. In comparison with other generalizations
of the Wolfenstein parametrization found in the literature, the advantage
of (\ref{NLO-wolf}) is the absence of relevant corrections to $V_{us}$
and $V_{cb}$, and that $V_{ub}$ and $V_{td}$ take forms similar to those 
in (\ref{W-par}). As far as the Jarlskog parameter introduced in
(\ref{JCP}) is concerned, we obtain the simple expression
\begin{equation}
J_{\rm CP}=\lambda^6A^2\eta,
\end{equation}
which should be compared with (\ref{JCP-PDG}).

\subsection{Unitarity Triangles of the CKM Matrix}\label{ssec:UT}
The unitarity of the CKM matrix, which is described by
\begin{equation}
\hat V_{\mbox{{\scriptsize CKM}}}^{\,\,\dagger}\cdot\hat 
V_{\mbox{{\scriptsize CKM}}}=
\hat 1=\hat V_{\mbox{{\scriptsize CKM}}}\cdot\hat V_{\mbox{{\scriptsize 
CKM}}}^{\,\,\dagger},
\end{equation}
leads to a set of 12 equations, consisting of 6 normalization 
and 6 orthogonality relations. The latter can be represented as 6 
triangles in the complex plane \cite{AKL}, all having the same area, 
$2 A_{\Delta}=J_{\rm CP}$ \cite{JS}. Let us now have a closer look at 
these relations: those describing the orthogonality of different columns 
of the CKM matrix are given by
\begin{eqnarray}
\underbrace{V_{ud}V_{us}^\ast}_{{\cal O}(\lambda)}+
\underbrace{V_{cd}V_{cs}^\ast}_{{\cal O}(\lambda)}+
\underbrace{V_{td}V_{ts}^\ast}_{{\cal O}(\lambda^5)} & = &
0\\
\underbrace{V_{us}V_{ub}^\ast}_{{\cal O}(\lambda^4)}+
\underbrace{V_{cs}V_{cb}^\ast}_{{\cal O}(\lambda^2)}+
\underbrace{V_{ts}V_{tb}^\ast}_{{\cal O}(\lambda^2)} & = &
0\\
\underbrace{V_{ud}V_{ub}^\ast}_{(\rho+i\eta)A\lambda^3}+
\underbrace{V_{cd}V_{cb}^\ast}_{-A\lambda^3}+
\underbrace{V_{td}V_{tb}^\ast}_{(1-\rho-i\eta)A\lambda^3} & = &
0,\label{UT1}
\end{eqnarray}
whereas those associated with the orthogonality of different rows 
take the following form:
\begin{eqnarray}
\underbrace{V_{ud}^\ast V_{cd}}_{{\cal O}(\lambda)}+
\underbrace{V_{us}^\ast V_{cs}}_{{\cal O}(\lambda)}+
\underbrace{V_{ub}^\ast V_{cb}}_{{\cal O}(\lambda^5)} & = &
0\\
\underbrace{V_{cd}^\ast V_{td}}_{{\cal O}(\lambda^4)}+
\underbrace{V_{cs}^\ast V_{ts}}_{{\cal O}(\lambda^2)}+
\underbrace{V_{cb}^\ast V_{tb}}_{{\cal O}(\lambda^2)} & = & 0\\
\underbrace{V_{ud}^\ast V_{td}}_{(1-\rho-i\eta)A\lambda^3}+
\underbrace{V_{us}^\ast V_{ts}}_{-A\lambda^3}+
\underbrace{V_{ub}^\ast V_{tb}}_{(\rho+i\eta)A\lambda^3}
& = & 0.\label{UT2}
\end{eqnarray}
Here we have also indicated the structures that arise if we apply the 
Wolfenstein parametrization by keeping just the leading, non-vanishing 
terms. We observe that only in (\ref{UT1}) and (\ref{UT2}), which 
describe the orthogonality of the first and third columns 
and of the first and third rows, respectively, all three sides are 
of comparable magnitude, ${\cal O}(\lambda^3)$, while in the 
remaining relations, one side is suppressed with respect to the others 
by factors of ${\cal O}(\lambda^2)$ or ${\cal O}(\lambda^4)$. Consequently,
we have to deal with only {\it two} non-squashed unitarity triangles 
in the complex plane. However, as we have already indicated in (\ref{UT1}) 
and (\ref{UT2}), the corresponding orthogonality relations agree with each 
other at the $\lambda^3$ level, yielding
\begin{equation}\label{UTLO}
\left[(\rho+i\eta)+(-1)+(1-\rho-i\eta)\right]A\lambda^3=0.
\end{equation}
Consequently, they describe the same triangle, which is usually referred 
to as {\it the} unitarity triangle of the CKM matrix 
\cite{JS,ut}.

\begin{figure}
\centerline{
\begin{tabular}{lr}
   \epsfysize=4.5cm
   \epsffile{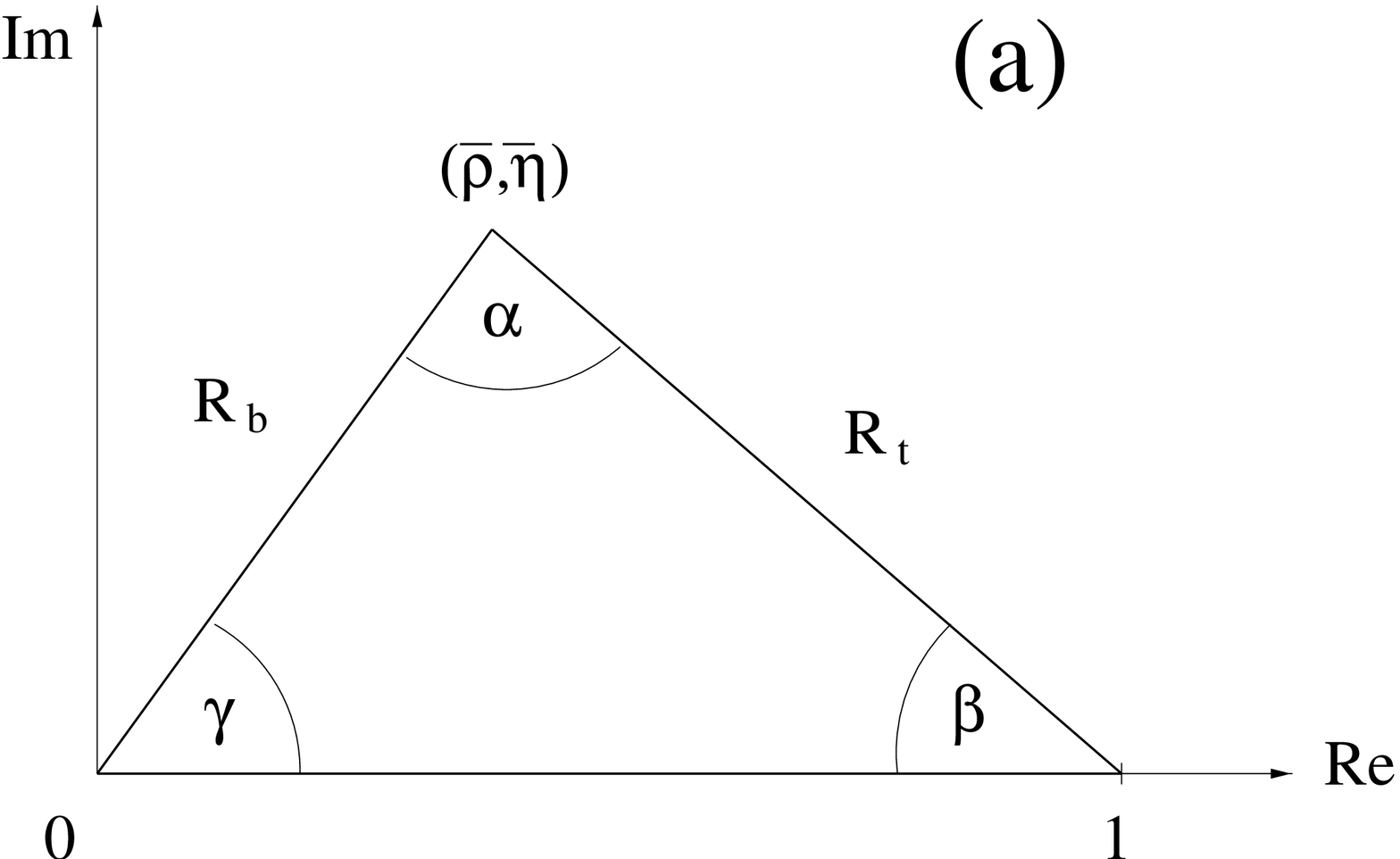}
&
   \epsfysize=4.5cm
   \epsffile{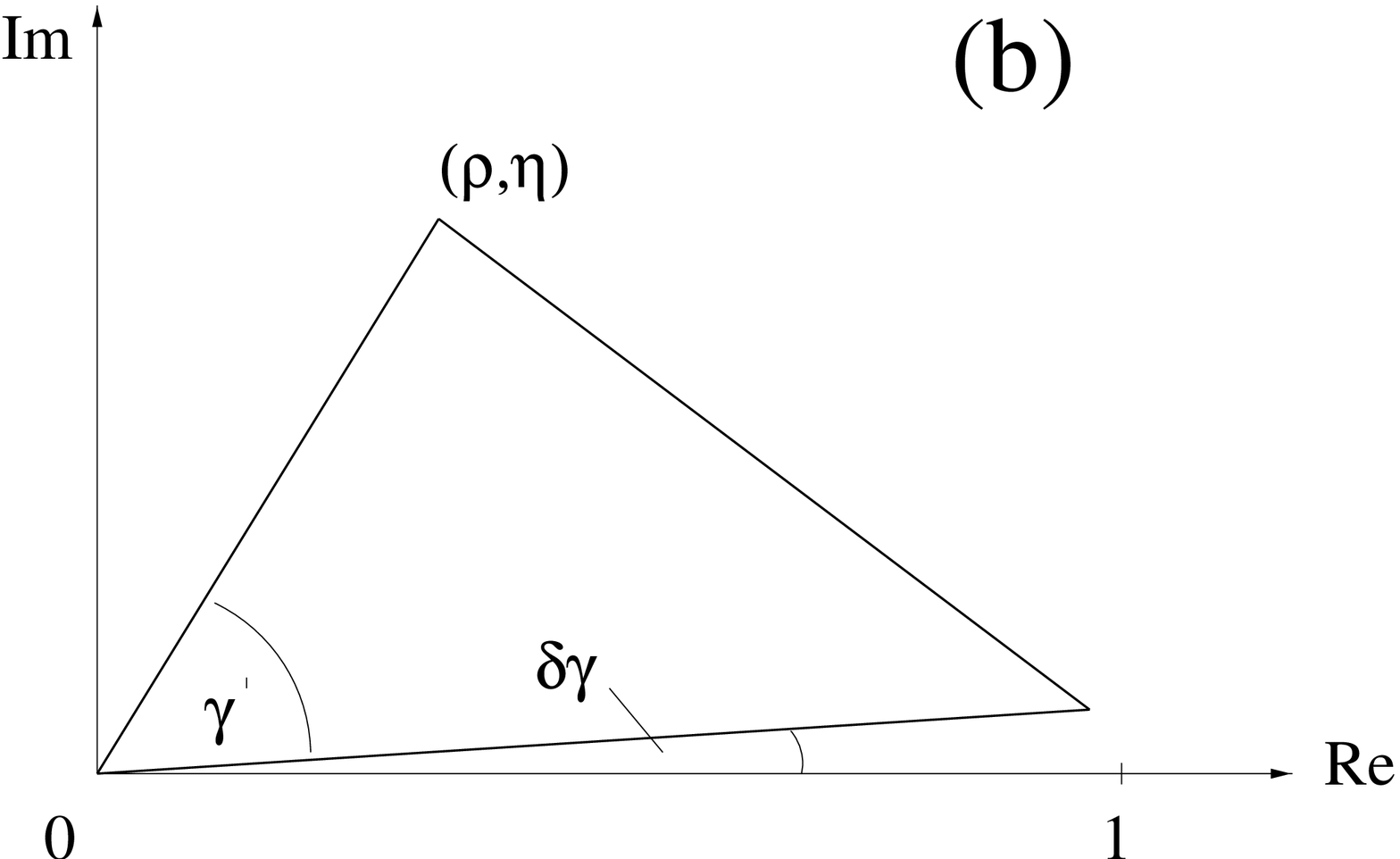}
\end{tabular}}
\caption{The two non-squashed unitarity triangles of the CKM matrix, as
explained in the text: (a) and (b) correspond to the orthogonality 
relations (\ref{UT1}) and (\ref{UT2}), respectively.}
\label{fig:UT}
\end{figure}

In the era of second-generation $B$-decay experiments of the LHC era, 
the experimental accuracy will be so tremendous that we will also have 
to take the next-to-leading order terms of the Wolfenstein expansion
into account, and will have to distinguish between the unitarity triangles 
following from (\ref{UT1}) and (\ref{UT2}). Let us first have a closer
look at the former relation. Including terms of ${\cal O}(\lambda^5)$, 
we obtain the following generalization of (\ref{UTLO}):
\begin{equation}\label{UT1-NLO}
\left[(\bar\rho+i\bar\eta)+(-1)+(1-\bar\rho-
i\bar\eta)\right]A\lambda^3 +{\cal O}(\lambda^7)=0, 
\end{equation}
where $\bar\rho$ and $\bar\eta$ are as defined in (\ref{rho-eta-bar}). 
If we divide this relation by the overall normalization factor $A\lambda^3$, 
and introduce
\begin{equation}\label{Rb-def}
R_b\equiv\sqrt{\overline{\rho}^2+\overline{\eta}^2}=\left(1-\frac{\lambda^2}{2}
\right)\frac{1}{\lambda}\left|\frac{V_{ub}}{V_{cb}}\right|
\end{equation}
\begin{equation}\label{Rt-def}
R_t\equiv\sqrt{(1-\overline{\rho})^2+\overline{\eta}^2}=
\frac{1}{\lambda}\left|\frac{V_{td}}{V_{cb}}\right|,
\end{equation}
we arrive at the unitarity triangle illustrated in Fig.\ \ref{fig:UT} (a). 
It is a straightforward generalization of the leading-order
case described by (\ref{UTLO}): instead of $(\rho,\eta)$, the apex
is now simply given by $(\bar\rho,\bar\eta)$ \cite{blo}. The two sides 
$R_b$ and $R_t$, as well as the three angles $\alpha$, $\beta$ and $\gamma$, 
will show up at several places throughout these lectures. Moreover, the 
relations
\begin{equation}
V_{ub}=A\lambda^3\left(\frac{R_b}{1-\lambda^2/2}\right)e^{-i\gamma},\quad 
V_{td}=A\lambda^3 R_t e^{-i\beta}
\end{equation}
are also useful for phenomenological applications, since they make the
dependences of $\gamma$ and $\beta$ explicit; they correspond to the
phase convention chosen both in the standard parametrization 
(\ref{standard}) and in the generalized Wolfenstein parametrization 
(\ref{NLO-wolf}). Finally, if we take also (\ref{set-rel}) into account, 
we obtain
\begin{equation}
\delta_{13}=\gamma.
\end{equation}

Let us now turn to (\ref{UT2}). Here we arrive at an expression 
that is more complicated than (\ref{UT1-NLO}): 
\begin{equation}
\left[\left\{1-\frac{\lambda^2}{2}-
(1-\lambda^2)\rho-i(1-\lambda^2)\eta
\right\}\!+\!\left\{-1+\left(\frac{1}{2}-\rho\right)
\lambda^2-i\eta\lambda^2\right\}
\!+\!\left\{\rho+i\eta\right\}\right]A\lambda^3+{\cal O}(\lambda^7)=0.
\end{equation}
If we divide again by $A\lambda^3$, we obtain the unitarity triangle
sketched in Fig.\ \ref{fig:UT} (b), where the apex is given by $(\rho,\eta)$ 
and {\it not} by $(\bar\rho,\bar\eta)$. On the other hand, we encounter a 
tiny angle 
\begin{equation}
\delta\gamma\equiv\lambda^2\eta={\cal O}(1^\circ)
\end{equation}
between real axis and basis of the triangle, which satisfies 
\begin{equation}
\gamma=\gamma'+\delta\gamma, 
\end{equation}
where $\gamma$ coincides with the corresponding angle in  
Fig.\ \ref{fig:UT} (a).

Whenever we will refer to a ``unitarity triangle'' (UT) in the following 
discussion, we mean the one illustrated in Fig.\ \ref{fig:UT} (a), which
is the generic generalization of the leading-order case described
by (\ref{UTLO}). As we will see below, the UT is the central target 
of the experimental tests of the SM description of CP violation. 
Interestingly, also the tiny angle $\delta\gamma$ can be probed directly 
through certain CP-violating effects that can be explored at 
hadron colliders, in particular at the LHC.

\begin{figure}
\vspace{0.10in}
\centerline{
\epsfysize=5.3truecm
\epsffile{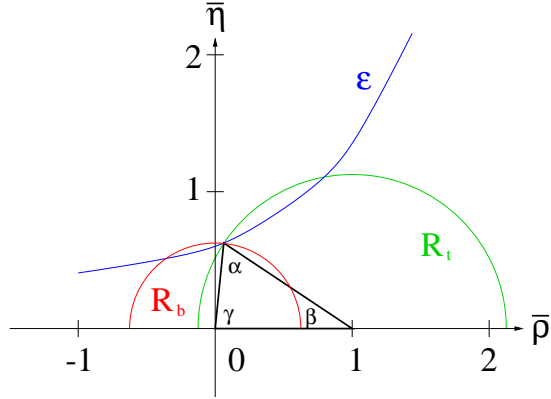}
}
\caption{Contours in the $\bar\rho$--$\bar\eta$ plane, allowing us to
determine the apex of the UT.}\label{fig:cont-scheme}
\end{figure}

\boldmath\subsection{Towards an Allowed Region in the 
$\bar\rho$--$\bar\eta$ Plane}\unboldmath\label{subsec:CKM-fits}
It is possible to constrain -- and even determine -- the apex of the UT 
in the $\bar\rho$--$\bar\eta$ plane with the help of experimental data. 
Unfortunately, we do not yet have the theoretical framework available to 
discuss in detail how this can actually be done (but this will become 
obvious in the course of these lectures). However, it is nevertheless 
useful to sketch the corresponding procedure -- the ``CKM fits'' -- 
already now, consisting of the following elements:
\begin{itemize}
\item The parameter $R_b$ introduced in (\ref{Rb-def}), which involves
the ratio $|V_{ub}/V_{cb}|$. It can be determined experimentally through 
$b\to u \ell\bar \nu$ and $b\to c \ell \bar \nu$  decay processes. 
Following these lines, we may fix a circle in the $\bar\rho$--$\bar\eta$ 
plane that is centred at the origin $(0,0)$ and has the radius $R_b$.
\item The parameter $R_t$ introduced in (\ref{Rt-def}), 
which involves the ratio $|V_{td}/V_{cb}|$. It can be determined with 
the help of the mass differences $\Delta M_{d,s}$ of the mass eigenstates 
of the neutral $B_d$- and $B_s$-meson systems. Experimental information on 
these quantities then allows us to fix another circle in the 
$\bar\rho$--$\bar\eta$ plane, which is centred at $(1,0)$ and has 
the radius $R_t$. 
\item  Finally, we may convert the measurement of the observable 
$\varepsilon$, which describes the CP violation in the neutral kaon 
system that was discovered in 1964, into a hyperbola in the 
$\bar\rho$--$\bar\eta$ plane.
\end{itemize}
In Fig.~\ref{fig:cont-scheme}, we have illustrated these contours;
their intersection allows us to determine the apex of the UT within
the SM. The curves that are implied by $\Delta M_{d}$ and $\varepsilon$ 
depend on the CKM parameter $A$ and the top-quark mass $m_t$, as well as
on certain perturbatively calculable QCD corrections and non-perturbative 
parameters. Consequently, strong correlations between the theoretical 
and experimental uncertainties arise in the CKM fits. As discussed 
in detail in \cite{CKM-Book}, several different approaches can be found in 
the literature to deal with the corresponding error propagation. The 
typical (conservative) ranges for the UT angles that follow from the 
CKM fits read as follows:
\begin{equation}\label{UT-range}
70^\circ\lsim\alpha\lsim130^\circ, \quad
20^\circ\lsim\beta\lsim30^\circ, \quad
50^\circ\lsim\gamma\lsim70^\circ.
\end{equation}

On the other hand, CP violation in the $B$-meson system provides various 
strategies to determine these angles {\it directly}, thereby offering 
different ways to fix the apex of the UT in the $\bar\rho$--$\bar\eta$ 
plane. Following these lines, a powerful test of the KM mechanism can 
be performed. This very interesting feature is also reflected by the 
tremendous efforts to explore CP violation in $B$ decays experimentally 
in this decade. Before having a closer look at $B$ mesons, their decays, 
the theoretical tools to deal with them and the general requirements 
for having non-vanishing CP asymmetries, let us first turn to the kaon 
system.

\section{A FIRST LOOK AT CP VIOLATION AND RARE DECAYS IN THE 
KAON SYSTEM}\label{sec:kaon}
\setcounter{equation}{0}
\boldmath
\subsection{CP Violation: $\varepsilon$ and 
$\varepsilon'$}\label{subsec:kaon-CP}
\unboldmath
As we have already noted, in 1964, CP violation was discovered -- as a 
big surprise -- in the famous experiment by Christenson et 
al.~\cite{CP-discovery}, who observed $K_{\rm L}\to\pi^+\pi^-$ decays. 
If the weak interactions {\it were} invariant under CP transformations,
the mass eigenstates $K_{\rm S}$ and $K_{\rm L}$ of the Hamilton operator
describing $K^0$--$\bar K^0$ mixing {\it were} eigenstates of the CP 
operator, with eigenvalues $+1$ and $-1$, respectively. Since the 
$\pi^+\pi^-$ final state of $K_{\rm L}\to\pi^+\pi^-$ is CP-even, the 
detection of this transition signals indeed the violation of the CP symmetry 
in weak interaction processes. The discussion in this subsection serves 
mainly to make a first contact with this phenomenon; for detailed 
presentations of CP violation in kaon decays, we refer the reader to 
\cite{BF-rev,B-LH98,Brev01}. 

\begin{figure}
\begin{center}
\begin{picture}(360,100)(0,60)
\Text(70,140)[bc]{$({\cal CP})$}
\Text(109,140)[bc]{$(-)$}
\Text(145,140)[bc]{$(+)$}
\Text(276,140)[bc]{$(+)$}
\Text(160,130)[tc]{direct: $\varepsilon'$}
\Text(190,94)[tc]{indirect: $\varepsilon$}
\Text(60,81.5)[bl]{$K_{\rm L}\,\,=\,\,K_2\,\,+\,\,\overline{\varepsilon}K_1
\hspace*{3.5truecm}\left\{\begin{array}{c}\pi^+\pi^-\\
\pi^0\pi^0\end{array}\right.$}
\Line(105,105)(105,115)
\Line(105,115)(225,115)
\ArrowLine(225,115)(239,96)
\Line(145,90)(145,80)
\Line(145,80)(225,80)
\ArrowLine(225,80)(239,96)
\Line(239,96)(255,96)
\end{picture}
\end{center}
\vspace*{-1truecm}
\caption{Illustration of indirect and direct CP violation in 
$K_{\rm L}\to\pi\pi$ decays.}\label{fig:CP-kaon}
\end{figure}
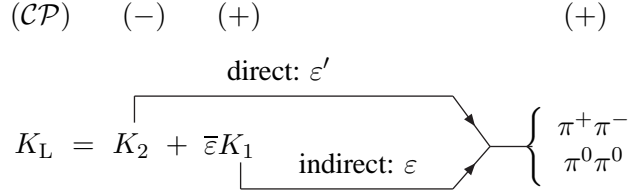

In the neutral $K$-meson system, CP violation is described by two complex 
quantities, called $\varepsilon$ and $\varepsilon'$, which are defined 
by the following ratios of decay amplitudes:
\begin{equation}\label{defs-eps}
\frac{A(K_{\rm L}\to\pi^+\pi^-)}{A(K_{\rm S}
\to\pi^+\pi^-)}\approx\varepsilon+\varepsilon',\quad
\frac{A(K_{\rm L}\to\pi^0\pi^0)}{A(K_{\rm S}
\to\pi^0\pi^0)}\approx\varepsilon-2\,\varepsilon'.
\end{equation}
These parameters are associated with ``indirect'' and ``direct'' CP 
violation, as we have illustrated in Fig.~\ref{fig:CP-kaon}, where $K_1$ 
and $K_2$ denote the CP eigenstates of the neutral kaon system with CP 
eigenvalues $+1$ and $-1$, respectively. The terminology of ``indirect 
CP violation'' originates from the fact that the mass eigenstate 
$K_{\rm L}$ of the neutral kaon system is {\it not} an eigenstate of the 
CP operator because of the small admixture of the CP-even $K_1$ state, 
which may decay -- through a CP-conserving transition -- into a $\pi\pi$ 
final state. On the other hand, direct CP violation originates from 
{\it direct} transitions of the CP-odd $K_2$ state into the CP-even 
$\pi\pi$ final state.

After the discovery of indirect CP violation through 
$K_{\rm L}\to\pi^+\pi^-$ decays, this phenomenon could also
be observed in $K_{\rm L}\to\pi^0\pi^0$, $\pi \ell\bar\nu$, 
$\pi^+\pi^-\gamma$ modes, and recently in $K_{\rm L}\to\pi^+\pi^-e^+e^-$ 
transitions. All these effects can be described by 
\begin{equation}
\varepsilon=(2.280\pm0.013)\times e^{i\frac{\pi}{4}}\times 10^{-3}.
\end{equation}
As we noted in Subsection~\ref{subsec:CKM-fits}, the knowledge of
the CKM parameter $A$ and the top-quark mass $m_t$ allows us --
in combination with the calculation of perturbative QCD corrections
and estimates of non-perturbative parameters  -- to convert 
the observable $\varepsilon$ into a hyperbola in the $\bar\rho$--$\bar\eta$ 
plane, as is explicitly shown in \cite{BF-rev,B-LH98,Brev01}. 
This analysis implies in particular $\bar\eta>0$, i.e.\ that 
the appex of the UT lies in the 
{\it upper} half of the $\bar\rho$--$\bar\eta$ plane.

Direct CP violation in neutral $K\to\pi\pi$ decays can be described through
the quantity Re$(\varepsilon'/\varepsilon)$. In 1999, measurements 
at CERN (NA48) \cite{NA48-obs} and FNAL (KTeV) \cite{KTeV-obs} have 
demonstrated -- after tremendous efforts over many years -- that 
this observable is actually {\it different} from zero, thereby establishing 
the phenomenon of {\it direct} CP violation. The experimental status is 
now given as follows:
\begin{equation}\label{epsprime-res}
\mbox{Re}(\varepsilon'/\varepsilon)=\left\{\begin{array}{ll}
(14.7\pm2.2)\times10^{-4}&\mbox{(NA48 \cite{NA48}),}\\
(20.7\pm2.8)\times10^{-4}&\mbox{(KTeV \cite{KTeV}).}
\end{array}\right.
\end{equation}
If we take also the previous results of the NA31 and E731 collaborations 
into account, we obtain the world average 
\begin{equation}\label{epspeps-WA}
\mbox{Re}(\varepsilon'/\varepsilon)=(16.6\pm1.6)\times 10^{-4}.
\end{equation}
Within the SM, calculations of Re$(\varepsilon'/\varepsilon)$
give the same order of magnitude (for an overview of the current 
status, see \cite{BuJa}). However, these analyses are affected
by large hadronic uncertainties; the situation is 
particularly unfavourable, since Re$(\varepsilon'/\varepsilon)$ is 
governed by the competition between two different decay topologies 
and suffers from a strong cancellation between them. Consequently, 
although the measurement of Re$(\varepsilon'/\varepsilon)$ led to the
discovery of a new kind of CP violation, this observable does 
unfortunately not allow us to perform stringent tests of the KM mechanism 
of CP violation, unless better techniques to deal with the hadronic 
uncertainties are available.

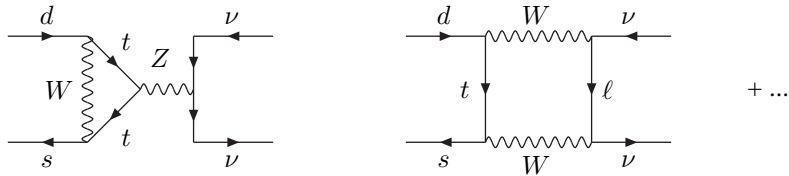
\begin{figure}
\begin{center}
{\small
\hspace*{-6.0truecm}\begin{picture}(250,70)(0,45)
\ArrowLine(60,100)(90,100)
\ArrowLine(160,100)(130,100)
\ArrowLine(90,60)(60,60)
\ArrowLine(130,60)(160,60)
\ArrowLine(90,100)(110,80)
\ArrowLine(110,80)(90,60)
\ArrowLine(130,80)(130,60)
\ArrowLine(130,100)(130,80)
\Photon(90,100)(90,60){2}{8}
\Photon(110,80)(130,80){2}{4}
\Text(75,105)[b]{$d$}\Text(105,95)[b]{$t$}\Text(145,105)[b]{$\nu$}
\Text(75,55)[t]{$s$}\Text(105,65)[t]{$t$}\Text(145,55)[t]{$\nu$}
\Text(85,80)[r]{$W$}\Text(114,92)[l]{$Z$}
\ArrowLine(210,100)(240,100)\Photon(240,100)(280,100){2}{8}
\ArrowLine(310,100)(280,100)
\ArrowLine(240,60)(210,60)\Photon(240,60)(280,60){2}{8}
\ArrowLine(280,60)(310,60)
\ArrowLine(240,100)(240,60)\ArrowLine(280,100)(280,60)
\Text(225,105)[b]{$d$}\Text(260,105)[b]{$W$}\Text(295,105)[b]{$\nu$}
\Text(225,55)[t]{$s$}\Text(260,55)[t]{$W$}\Text(295,55)[t]{$\nu$}
\Text(235,80)[r]{$t$}\Text(285,80)[l]{$\ell$}
\Text(340,80)[l]{+ ...}
\end{picture}}
\end{center}
\vspace*{-0.3truecm}
\caption{Decay processes contributing to $K_{\rm L}\to\pi^0\nu\bar\nu$
in the SM.}\label{fig:KLpi0nunu-diag}
\end{figure}

\boldmath
\subsection{Rare Decays: $K\to\pi\nu\bar\nu$}\label{subsec:rare-kaon-brief}
\unboldmath
From a theoretical point of view, the decays $K_{\rm L}\to\pi^0\nu\bar\nu$ 
and $K^+\to\pi^+\nu\bar\nu$ are very interesting. Since we will have a 
detailed look at them in Subsection~\ref{ssec:Kpinunu-detail}, let us here 
just sketch their most interesting features. As can easily be seen, these 
transitions originate from FCNC processes. Consequently, because of the GIM 
mechanism, they receive no contributions at the tree level in the SM.
However, they may be induced through loop processes of the kind shown 
in Fig.~\ref{fig:KLpi0nunu-diag}, and are therefore strongly suppressed 
transitions, which are referred to as ``rare'' decays. One of the most
exciting features of the $K\to\pi\nu\bar\nu$ modes is that they are 
theoretically very clean. Moreover, it can be shown that the measurement 
of the $K_{\rm L}\to\pi^0\nu\bar\nu$ branching ratio allows us to determine 
$|\bar\eta|$, whereas the one of $K^+\to\pi^+\nu\bar\nu$ can be converted 
into an ellipse in the $\bar\rho$--$\bar\eta$ plane. The intersection of 
these contours provides an interesting determination of the UT, where in 
particular $\sin2\beta$ can be extracted with respectable accuracy 
\cite{BBSIN}. We may hence perform a stringent test of the SM description 
of CP violation by comparing the UT thus determined with the ones 
following from the construction illustrated in Fig.~\ref{fig:cont-scheme} 
and the studies of CP violation in the $B$-meson system. In particular, 
as we will see in Subsection~\ref{subsec:BpsiK}, 
$B_d\to J/\psi K_{\rm S}$ decays allow also a clean determination 
of $\sin2\beta$, so that a violation of the SM relation
\begin{equation}\label{s2b-K-B}
(\sin2\beta)_{\pi\nu\bar\nu} = (\sin2\beta)_{\psi K_{\rm S}}
\end{equation}
would indicate sources of CP violation lying beyond the SM. Moreover,
also the determination of the angle $\gamma$ of the UT is interesting for
the search of NP with $K\to\pi\nu\bar\nu$ decays \cite{AI,FIM}.

Unfortunately, the $K\to\pi\nu\bar\nu$ branching ratios are extremely
small. A recent update of the corresponding calculations within the
SM yields the following results \cite{BFRS3}:
\begin{equation}\label{Kpinunu-SM}
{\rm BR}(K^+\to\pi^+\nu\bar\nu)= (8.0 \pm 1.1)\times 10^{-11}, \quad
{\rm BR}(K_{\rm L}\to\pi^0\nu\bar\nu)= (3.2 \pm 0.6)\times 10^{-11},
\end{equation}
which are in the ballpark of other recent analyses \cite{Gino03,KLN}.
Interestingly, a third event for the former channel was very recently
observed by the E949 experiment at BNL \cite{E949}, thereby complementing 
the previous observation of the two events by the E787 collaboration 
\cite{Adler01}. The three observed $K^+\to\pi^+\nu\bar\nu$ events 
can be converted into the following branching ratio:
\begin{equation}\label{Brook}
{\rm BR}(K^+\to\pi^+\nu\bar\nu)= (14.7^{+13.0}_{-8.9})\times 10^{-11}.
\end{equation}
On the other hand, for the $K_{\rm L}\to\pi^0\nu\bar\nu$ channel, only the
experimental upper bound
\begin{equation}\label{KTeV-Bound}
{\rm BR}(K_{\rm L}\to\pi^0\nu\bar\nu)< 5.9 \times 10^{-7}
\end{equation}
is available from the KTeV collaboration \cite{KTEV-KL-BOUND}. 

In the presence of NP, the $K\to\pi \nu\bar\nu$ branching ratios may
differ strongly from the SM expectations given in (\ref{Kpinunu-SM}). 
For instance, in a recent NP analysis \cite{BFRS3,BFRS2}, which is
motivated by certain puzzling patterns in the $B$-factory data and
will be discussed in Subsection~\ref{ssec:NP-rare}, a spectacular 
enhancement of the $K_{\rm L}\to\pi^0\nu\bar\nu$ branching ratio, by 
one order of magnitude, is found, and the relation in (\ref{s2b-K-B}) 
would in fact be dramatically violated. 

Concerning the experimental aspects of the $K\to\pi\nu\bar\nu$
modes, we refer the reader to the recent overview given in
\cite{Kpinunu-EXP}. Let us now move on to the central topic of 
these lectures, the $B$-meson system.

\section{DECAYS OF {\boldmath$B$\unboldmath} MESONS}\label{sec:Bdecays}
\setcounter{equation}{0}
The $B$-meson system consists of charged and neutral $B$ mesons, which are 
characterized by the
\begin{displaymath}
\begin{array}{ccc}
B^+\sim u\,\bar b, && B^-\sim \bar u\,b\\
B^+_c\sim c\,\bar b, && B^+_c\sim\bar c\,b
\end{array}
\end{displaymath}
and 
\begin{displaymath}
\begin{array}{ccc}
B^0_d\sim d\,\bar b, && \bar B^0_d\sim\bar d\,b\\
B^0_s\sim s\,\bar b, && \bar B^0_s\sim\bar s\,b\\
\end{array}
\end{displaymath}
valence-quark contents, respectively. The characteristic feature
of the neutral $B_q$ ($q\in \{d,s\}$) mesons is the phenomenon
of $B_q^0$--$\bar B_q^0$ mixing (the counterpart of
$K^0$--$\bar K^0$ mixing), which will be discussed in 
Subsection~\ref{ssec:BBbar-mix}. As far as the weak decays of 
$B$ mesons are concerned, we distinguish between leptonic, 
semileptonic and non-leptonic transitions.

\subsection{Leptonic Decays}
\begin{figure}
\begin{center}
\leavevmode
\epsfysize=4.0truecm 
\epsffile{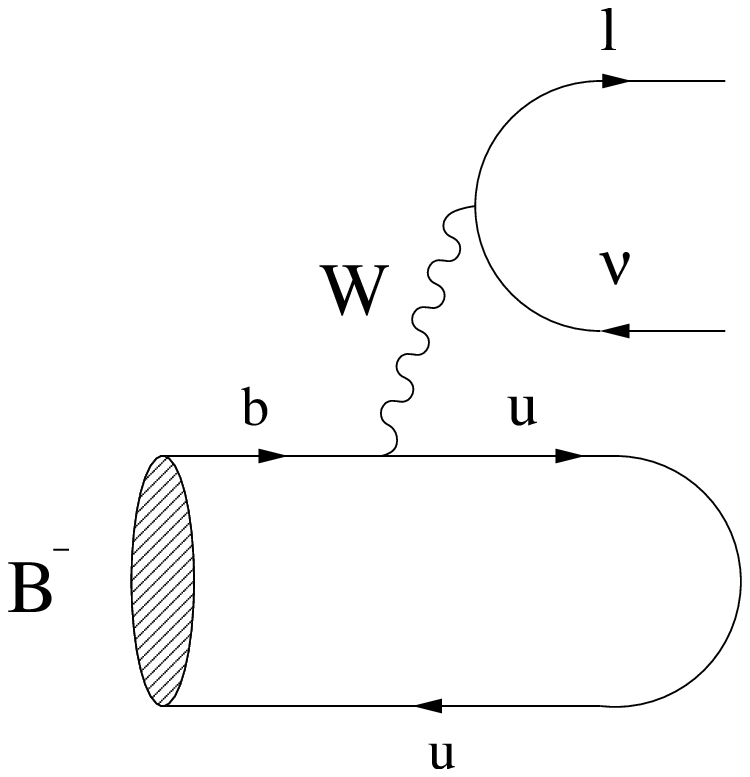} \hspace*{2truecm}
\epsfysize=3.0truecm 
\epsffile{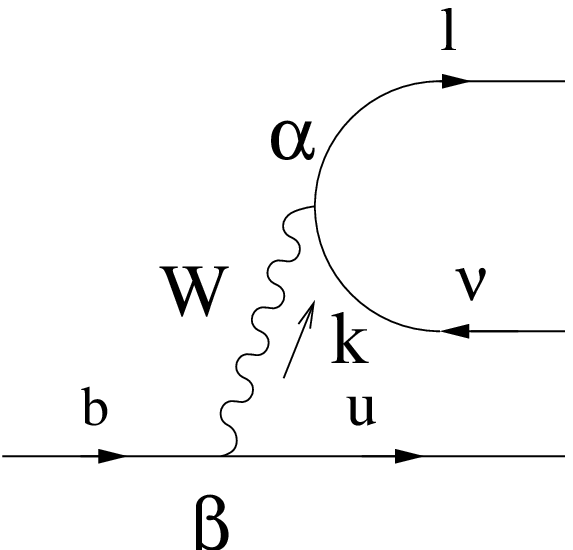}
\end{center}
\caption{Feynman diagram contributing to the leptonic decay
$B^-\to \ell\bar\nu$.}\label{fig:lep}
\end{figure}

The simplest $B$-meson decay class is given by leptonic decays 
of the kind $B^-\to \ell\bar\nu$, as illustrated in Fig.~\ref{fig:lep}.
If we evaluate the corresponding Feynman diagram, we arrive at the 
following transition amplitude:
\begin{equation}\label{Tfi-lept}
T_{fi}=-\,\frac{g_2^2}{8} V_{ub}
\underbrace{\left[\bar u_\ell\gamma^\alpha(1-\gamma_5)v_\nu
\right]}_{\mbox{Dirac spinors}}
\left[\frac{g_{\alpha\beta}}{k^2-M_W^2}\right]
\underbrace{\langle 0|\bar u\gamma^\beta
(1-\gamma_5)b|B^-\rangle}_{\mbox{hadronic ME}},
\end{equation}
where $g_2$ is the $SU(2)_{\rm L}$ gauge coupling, $V_{ub}$ the corresponding
element of the CKM matrix, $\alpha$ and $\beta$ are Lorentz indices,
and $M_W$ denotes the mass of the $W$ gauge boson. Since the four-momentum
$k$ that is carried by the $W$ satisfies $k^2=M_B^2\ll M_W^2$, we may
write
\begin{equation}\label{W-int-out}
\frac{g_{\alpha\beta}}{k^2-M_W^2}\quad\longrightarrow\quad
-\,\frac{g_{\alpha\beta}}{M_W^2}\equiv-\left(\frac{8G_{\rm F}}{\sqrt{2}g_2^2}
\right)g_{\alpha\beta},
\end{equation}
where $G_{\rm F}$ is Fermi's constant. Consequently, we may ``integrate out'' 
the $W$ boson in (\ref{Tfi-lept}), which yields
\begin{equation}\label{Tfi-lept-2}
T_{fi}=\frac{G_{\rm F}}{\sqrt{2}}V_{ub}\left[\bar u_\ell\gamma^\alpha
(1-\gamma_5)v_\nu\right]\langle 0|\bar u\gamma_\alpha(1-\gamma_5)b
|B^-\rangle.
\end{equation}
In this simple expression, {\it all} the hadronic physics is encoded 
in the {\it hadronic matrix element} 
\begin{displaymath}
\langle 0|\bar u\gamma_\alpha(1-\gamma_5)b
|B^-\rangle,
\end{displaymath}
i.e.\ there are no other strong-interaction (QCD) effects. Since the 
$B^-$ meson is a pseudoscalar particle, we have
\begin{equation}\label{ME-rel1}
\langle 0|\overline{u}\gamma_\alpha b|B^-\rangle=0,
\end{equation}
and may write
\begin{equation}\label{ME-rel2}
\langle 0|\bar u\gamma_\alpha\gamma_5 b|B^-(q)\rangle =
i f_B q_\alpha,
\end{equation}
where $f_B$ is the $B$-meson {\it decay constant}, which is an important
input for phenomenological studies. In order to determine this
quantity, which is a very challenging task, non-perturbative 
techniques, such as lattice \cite{luscher} or QCD sum-rule analyses 
\cite{khodjamirian},
are required. If we use (\ref{Tfi-lept-2}) with (\ref{ME-rel1}) and 
(\ref{ME-rel2}), and perform the corresponding phase-space integrations, 
we obtain the following decay rate:
\begin{equation}
\Gamma(B^-\to\ell \bar \nu)=\frac{G_{\rm F}^2}{8\pi}
|V_{ub}|^2 M_Bm_\ell^2\left(1-\frac{m_\ell^2}{M_B^2}\right)^2f_B^2,
\end{equation}
where $M_B$ and $m_\ell$ denote the masses of the $B^-$ and $\ell$,
respectively. Because of the tiny value of $|V_{ub}|\propto\lambda^3$ 
and a helicity-suppression mechanism, we obtain unfortunately very small 
branching ratios of ${\cal O}(10^{-10})$ and ${\cal O}(10^{-7})$ for 
$\ell=e$ and $\ell=\mu$, respectively \cite{fulvia}. The helicity 
suppression is not effective for $\ell=\tau$, but -- because of the 
required $\tau$ reconstruction -- these modes are also very challenging from 
an experimental point of view. A measurement of leptonic 
$B$-meson decays would nevertheless be very interesting, as it would allow 
an experimental determination of $f_B$, thereby providing tests 
of non-perturbative calculations of this important 
parameter.\footnote{Leptonic decays of $D_{(s)}$ mesons allow the
extraction of the corresponding decay constants $f_{D_{(s)}}$, which
are defined in analogy to (\ref{ME-rel2}). These measurements are
an important element of the CLEO-c research programme \cite{CLEO-c}.}
The CKM element $|V_{ub}|$ can be extracted from semileptonic $B$
decays, our next topic.

\subsection{Semileptonic Decays}\label{subsec:semi-lept}
\subsubsection{General Structure}
\begin{figure}
\begin{center}
\leavevmode
\epsfysize=4.0truecm 
\epsffile{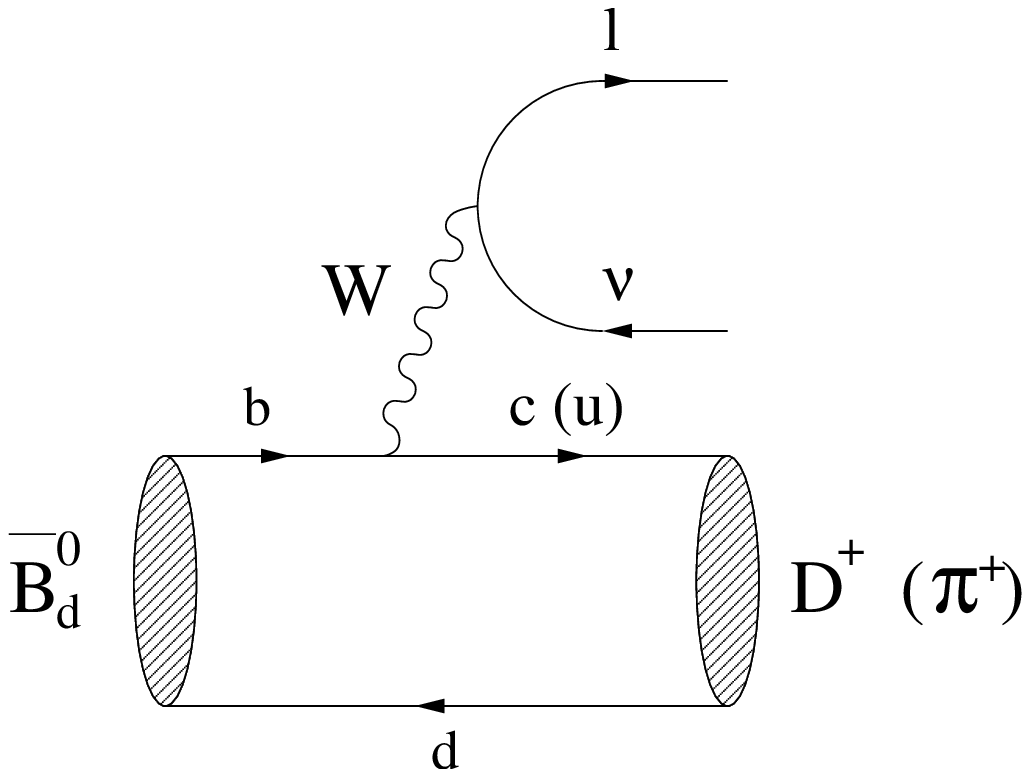} \hspace*{2truecm}
\epsfysize=3.0truecm 
\epsffile{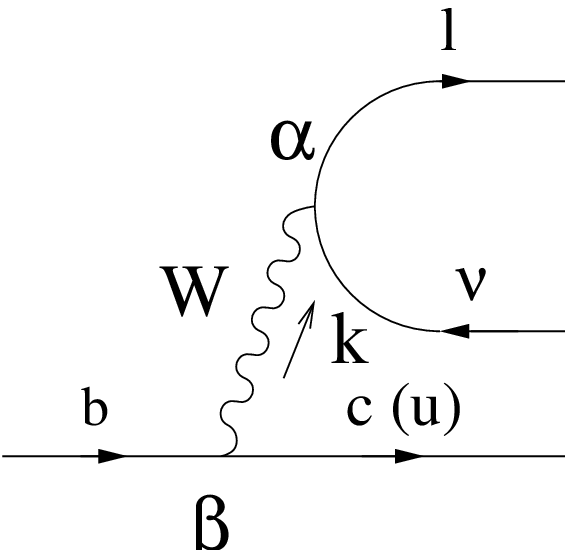}
\end{center}
\caption{Feynman diagram contributing to semileptonic 
$\bar B^0_d\to D^+ (\pi^+) \ell \bar \nu$ decays.}\label{fig:semi}
\end{figure}

Semileptonic $B$-meson decays of the kind shown in Fig.~\ref{fig:semi}
have a structure that is more complicated than the one of the 
leptonic transitions. If we evaluate the corresponding Feynman diagram
for the $b\to c$ case, we obtain
\begin{equation}\label{Tfi-semi-full}
T_{fi}=-\,\frac{g_2^2}{8} V_{cb}
\underbrace{\left[\bar u_\ell\gamma^\alpha(1-\gamma_5)v_\nu
\right]}_{\mbox{Dirac spinors}}
\left[\frac{g_{\alpha\beta}}{k^2-M_W^2}\right]
\underbrace{\langle D^+|\bar c\gamma^\beta
(1-\gamma_5)b|\bar B^0_d\rangle}_{\mbox{hadronic ME}}.
\end{equation}
Because of $k^2\sim M_B^2\ll M_W^2$, we may again -- as in (\ref{Tfi-lept}) --
integrate out the $W$ boson with the help of (\ref{W-int-out}), which
yields
\begin{equation}\label{Tfi-semi}
T_{fi}=\frac{G_{\rm F}}{\sqrt{2}}V_{cb}\left[\bar u_\ell\gamma^\alpha
(1-\gamma_5)v_\nu\right]\langle  D^+|\bar c\gamma_\alpha(1-\gamma_5)b
|\bar B^0_d\rangle,
\end{equation}
where {\it all} the hadronic physics is encoded in the hadronic
matrix element
\begin{displaymath}
\langle D^+|\bar c\gamma_\alpha
(1-\gamma_5)b|\bar B^0_d\rangle,
\end{displaymath}
i.e.\ there are {\it no} other strong-interaction (QCD) effects.
Since the $\bar B^0_d$ and $D^+$ are pseudoscalar mesons, we have
\begin{equation}
\langle D^+|\bar c\gamma_\alpha\gamma_5b|
\bar B^0_d\rangle=0,
\end{equation}
and may write
\begin{equation}\label{BD-ME}
\langle D^+(k)|\bar c\gamma_\alpha b|\bar B^0_d(p)
\rangle=F_1(q^2)\left[(p+k)_\alpha -
\left(\frac{M_B^2-M_D^2}{q^2}\right)q_\alpha\right]
+F_0(q^2)\left(\frac{M_B^2-M_D^2}{q^2}\right)q_\alpha, 
\end{equation}
where $q\equiv p-k$, and the $F_{1,0}(q^2)$ denote the {\it form factors}
of the $\bar B\to D$ transitions. Consequently, in contrast to the simple 
case of the leptonic transitions, semileptonic decays involve {\it two} 
hadronic form factors instead of the decay constant $f_B$. In order to 
calculate these parameters, which depend on the momentum transfer $q$, 
again non-perturbative techniques (lattice, QCD sum rules, etc.) are 
required.

\subsubsection{Aspects of the Heavy-Quark Effective Theory}
If the mass $m_Q$ of a quark $Q$ is much larger than the QCD scale parameter
$\Lambda_{\rm QCD}={\cal O}(100\,\mbox{MeV})$, it is referred to as 
a ``heavy'' quark. Since the bottom and charm quarks have masses at the 
level of $5\,\mbox{GeV}$ and $1\,\mbox{GeV}$, respectively, they belong 
to this important category. As far as the extremely heavy top quark, 
with $m_t\sim 170\,\mbox{GeV}$ is concerned, it decays unfortunately 
through weak interactions before a hadron can be formed. Let us now 
consider a heavy quark that is bound inside a hadron, i.e.\ a bottom 
or a charm quark. The heavy quark then moves almost with the 
hadron's four velocity $v$ and is almost on-shell, so that
\begin{equation}
p_Q^\mu=m_Qv^\mu + k^\mu,
\end{equation}
where $v^2=1$ and $k\ll m_Q$ is the ``residual'' momentum. Owing to 
the interactions of the heavy quark with the light degrees of freedom of 
the hadron, the residual momentum may only change by 
$\Delta k\sim\Lambda_{\rm QCD}$, and $\Delta v \to 0$ for 
$\Lambda_{\rm QCD}/m_Q\to 0$. 

It is now instructive to have a look at the elastic scattering process 
$\bar B(v)\to \bar B(v')$ in the limit of $\Lambda_{\rm QCD}/m_b \to 0$, 
which is characterized by the following matrix element:
\begin{equation}\label{BB-ME}
\frac{1}{M_B}\langle\bar B(v')|\bar b_{v'}\gamma_\alpha b_v
|\bar B(v)\rangle=\xi(v'\cdot v)(v+v')_\alpha.
\end{equation}
Since the contraction of this matrix element with $(v-v')^\alpha$ has to 
vanish because of $\not \hspace*{-0.1truecm}v b_v= b_v$ and 
$\overline{b}_{v'} \hspace*{-0.2truecm}\not \hspace*{-0.1truecm}v' = 
\overline{b}_{v'}$, no $(v-v')_\alpha$ term arises in the parametrization
in (\ref{BB-ME}). On the other hand, the $1/M_B$ factor is related
to the normalization of states, i.e.\ the right-hand side of
\begin{equation}
\left(\frac{1}{\sqrt{M_B}}\langle\bar B(p')|\right)
\left(|\bar B(p)\rangle\frac{1}{\sqrt{M_B}}\right)=2v^0(2\pi)^3
\delta^3(\vec p-\vec p')
\end{equation}
does not depend on $M_B$. Finally, current conservation implies 
the following normalization condition:
\begin{equation}
\xi(v'\cdot v=1)=1,
\end{equation}
where  the ``Isgur--Wise'' function $\xi(v'\cdot v)$ does 
{\it not} depend on the flavour of the heavy quark (heavy-quark symmetry)
\cite{IW}. Consequently, for $\Lambda_{\rm QCD}/m_{b,c}\to 0$, we may write
\begin{equation}\label{BD-ME-HQ}
\frac{1}{\sqrt{M_D M_B}}\langle D(v')|\bar c_{v'}\gamma_\alpha b_v
|\bar B(v)\rangle=\xi(v'\cdot v)(v+v')_\alpha,
\end{equation}
and observe that this transition amplitude is governed -- in the 
heavy-quark limit -- by {\it one} hadronic form factor $\xi(v'\cdot v)$, 
which satisfies $\xi(1)=1$. If we now compare (\ref{BD-ME-HQ})
with (\ref{BD-ME}), we obtain 
\begin{equation}
F_1(q^2)=\frac{M_D+M_B}{2\sqrt{M_DM_B}}\xi(w)
\end{equation}
\begin{equation}
F_0(q^2)=\frac{2\sqrt{M_DM_B}}{M_D+M_B}\left[\frac{1+w}{2}\right]\xi(w),
\end{equation}
with
\begin{equation}
w\equiv v_D\cdot v_B=\frac{M_D^2+M_B^2-q^2}{2M_DM_B}.
\end{equation}
Similar relations hold also for the $\bar B\to D^\ast$ form factors
because of the heavy-quark spin symmetry, since the $D^\ast$ is 
related to the $D$ by a rotation of the heavy-quark spin. A detailed 
discussion of these interesting features and the associated ``heavy-quark 
effective theory'' (HQET) is beyond the scope of these lectures. For
a detailed overview, we refer the reader to \cite{neubert-rev}, where
also a comprehensive list of the original references can be found. 
For a more phenomenological discussion, also \cite{BaBar-Book} is very
useful.

\subsubsection{Applications}
An important application of the formalism sketched above is the
extraction of the CKM element $|V_{cb}|$. To this end, 
$\bar B\to D^*\ell\bar \nu$ decays are particularly promising. 
The corresponding rate can be written as 
\begin{equation}\label{BD-rate}
\frac{{\rm d}\Gamma}{{\rm d}w}=G_{\rm F}^2 K(M_B,M_{D^\ast},w) 
F(w)^2 |V_{cb}|^2,
\end{equation}
where $K(M_B,M_{D^\ast},w)$ is a known kinematic function, and 
$F(w)$ agrees with the Isgur--Wise function, up to perturbative
QCD corrections and $\Lambda_{\rm QCD}/m_{b,c}$ terms. The form 
factor $F(w)$ is a non-perturbative quantity. However, it satisfies 
the following normalization condition:
\begin{equation}\label{F1-norm}
F(1)=\eta_A(\alpha_s)\left[1+\frac{0}{m_c}+
\frac{0}{m_b}+
{\cal O}(\Lambda_{\rm QCD}^2/m_{b,c}^2)\right],
\end{equation} 
where $\eta_A(\alpha_s)$ is a perturbatively calculable short-distance
QCD factor, and the $\Lambda_{\rm QCD}/m_{b,c}$ corrections {\it vanish}
\cite{neubert-rev,neu-BDast}. The important latter feature is an 
implication of Luke's theorem \cite{luke}. Consequently, 
if we extract $F(w)|V_{cb}|$ from a measurement of (\ref{BD-rate}) 
as a function of $w$ and extrapolate to the ``zero-recoil point'' $w=1$
(where the rate vanishes), we may determine $|V_{cb}|$. In the case of 
$\bar B\to D\ell\bar \nu$ decays, we have 
${\cal O}(\Lambda_{\rm QCD}/m_{b,c})$ corrections to the 
corresponding rate ${\rm d}\Gamma/{\rm d}w$ at $w=1$. 
In order to determine $|V_{cb}|$, inclusive $B\to X_c\ell\bar \nu$ decays 
offer also very attractive avenues. As becomes obvious from (\ref{Def-A})
and the considerations in Subsection~\ref{ssec:UT}, $|V_{cb}|$ fixes 
the normalization of the UT. Moreover, this quantity is an important 
input parameter for various theoretical calculations. Its current 
experimental status can be summarized as follows:
\begin{equation}\label{Vcb-det}
|V_{cb}|=0.04\times[1\pm0.05] \quad\Rightarrow\quad
A=0.83\times [1\pm0.05].
\end{equation}

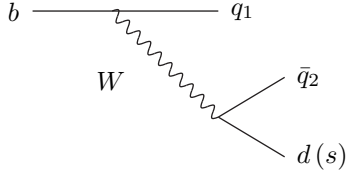
\begin{figure}
\begin{center}
{\small
\hspace*{4truecm}\begin{picture}(80,50)(80,20)
\Line(10,45)(80,45)\Photon(40,45)(80,5){2}{10}
\Line(80,5)(105,20)\Line(80,5)(105,-10)
\Text(5,45)[r]{$b$}\Text(85,45)[l]{$q_1$}
\Text(110,20)[l]{$\bar q_2$}
\Text(110,-10)[l]{$d\,(s)$}
\Text(45,22)[tr]{$W$}
\end{picture}}
\end{center}
\vspace*{1.0truecm}
\caption{Tree diagrams ($q_1,q_2\in\{u,c\}$).}\label{fig:tree-top}
\end{figure}

Let us now turn to $\bar B\to \pi\ell\bar\nu, \rho\ell\bar\nu$ decays, 
which originate from $b\to u\ell \bar\nu$ quark-level processes, as
can be seen in Fig.~\ref{fig:semi}, and provide access to $|V_{ub}|$. 
If we complement this CKM matrix element with $|V_{cb}|$, we may determine 
the side $R_b$ of the UT with the help of (\ref{Rb-def}). The 
determination of $|V_{ub}|$ is hence a very important aspect of
flavour physics. Since the $\pi$ and $\rho$ are ``light'' mesons, 
the HQET symmetry relations 
cannot be applied to the $\bar B\to \pi\ell\bar\nu, \rho\ell\bar\nu$ modes. 
Consequently, in order to determine $|V_{ub}|$ from these exclusive 
channels, the corresponding heavy-to-light form factors have to be
described by models. An important alternative is provided by inclusive 
decays. The corresponding decay rate takes the following form:
\begin{equation}\label{inclusive-rate}
\Gamma(\bar B\to X_u \ell \bar \nu)=
\frac{G_{\rm F}^2|V_{ub}|^2}{192\pi^3}m_b^5
\left[1-2.41\frac{\alpha_s}{\pi}+\frac{\lambda_1-9\lambda_2}{2m_b^2}
+\ldots\right],
\end{equation}
where $\lambda_1$ and $\lambda_2$ are non-perturbative parameters, 
which describe the hadronic matrix elements of certain ``kinetic'' 
and ``chromomagnetic'' operators appearing within the framework of
the HQET. Using the heavy-quark expansions
\begin{equation}\label{mass-exp}
M_B=m_b+\bar\Lambda-\frac{\lambda_1+3\lambda_2}{2m_b}+\ldots, \quad
M_{B^\ast}=m_b+\bar\Lambda-\frac{\lambda_1-\lambda_2}{2m_b}+\ldots
\end{equation}
for the $B^{(\ast)}$-meson masses, where $\bar\Lambda\sim\Lambda_{\rm QCD}$ 
is another non-perturbative parameter that is related to the light degrees
of freedom, the parameter $\lambda_2$ can be determined from the measured
values of the $M_{B^{(\ast)}}$. The strong dependence of 
(\ref{inclusive-rate}) on $m_b$ is a significant
source of uncertainty. On the other hand, the $1/m_b^2$ corrections
can be better controlled than in the exclusive case (\ref{F1-norm}), 
where we have, moreover, to deal with $1/m_c^2$ corrections. From an 
experimental point of view, we have to struggle with large backgrounds, 
which originate from $b\to c \ell \bar\nu$ processes and require also 
a model-dependent treatment. The determination of $|V_{ub}|$ from 
exclusive and inclusive $B$-meson decays caused by $b\to u\ell \bar\nu$ 
quark-level processes is therefore a very challenging issue; a summary 
of the current status is given by 
\begin{equation}\label{Vub-det}
|V_{ub}|=0.0037\times[1\pm0.15].
\end{equation}
If we now insert (\ref{Vub-det}) and (\ref{Vcb-det}) into (\ref{Rb-def})
and use $\lambda=0.22$, we obtain
\begin{equation}
R_b=0.41\pm0.07.
\end{equation}

For a much more detailed discussion of the determinations of $|V_{cb}|$ 
and $|V_{ub}|$, addressing also the various interesting recent developments 
and the future prospects, we refer the reader to \cite{CKM-Book}, where 
also the references to the vast original literature can be found. Another 
excellent presentation is given in \cite{BaBar-Book}.

\begin{figure}
\vspace*{0.7truecm}
\begin{center}
{\small
\begin{picture}(140,60)(0,20)
\Line(10,50)(130,50)\Text(5,50)[r]{$b$}\Text(140,50)[l]{$d\,(s)$}
\PhotonArc(70,50)(30,0,180){3}{15}
\Text(69,56)[b]{$u,c,t$}\Text(109,75)[b]{$W$}
\Gluon(70,50)(120,10){2}{10}
\Line(120,10)(135,23)\Line(120,10)(135,-3)
\Text(85,22)[tr]{$G$}\Text(140,-3)[l]{$q_1$}
\Text(140,23)[l]{$\bar q_2=\bar q_1$}
\end{picture}}
\end{center}
\vspace*{0.7truecm}
\caption{QCD penguin diagrams ($q_1=q_2\in\{u,d,c,s\}$).}\label{fig:QCD-top}
\end{figure}
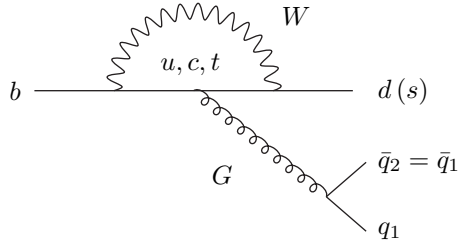

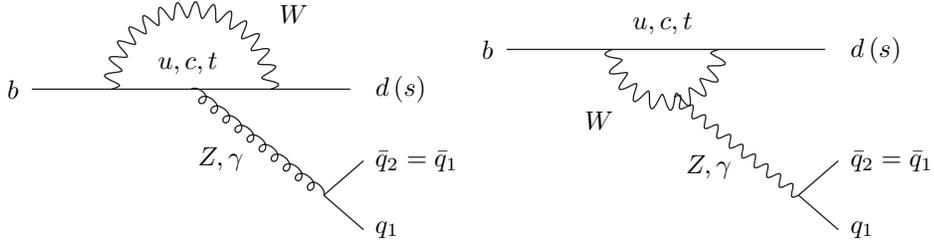
\begin{figure}
\vspace*{0.7truecm}
\begin{center}
{\small
\begin{picture}(140,60)(0,20)
\Line(10,50)(130,50)\Text(5,50)[r]{$b$}\Text(140,50)[l]{$d\,(s)$}
\PhotonArc(70,50)(30,0,180){3}{15}
\Text(69,56)[b]{$u,c,t$}\Text(109,75)[b]{$W$}
\Gluon(70,50)(120,10){2}{10}
\Line(120,10)(135,23)\Line(120,10)(135,-3)
\Text(90,28)[tr]{$Z,\gamma$}\Text(140,-3)[l]{$q_1$}
\Text(140,23)[l]{$\bar q_2=\bar q_1$}
\end{picture}}
\hspace*{1.2truecm}
{\small
\begin{picture}(140,60)(0,20)
\Line(10,65)(130,65)\Text(5,65)[r]{$b$}\Text(140,65)[l]{$d\,(s)$}
\PhotonArc(70,65)(20,180,360){3}{10}
\Text(69,71)[b]{$u,c,t$}\Text(45,35)[b]{$W$}
\Photon(73,47)(120,10){2}{10}
\Line(120,10)(135,23)\Line(120,10)(135,-3)
\Text(95,25)[tr]{$Z,\gamma$}\Text(140,-3)[l]{$q_1$}
\Text(140,23)[l]{$\bar q_2=\bar q_1$}
\end{picture}}
\end{center}
\vspace*{1.0truecm}
\caption{Electroweak penguin diagrams 
($q_1=q_2\in\{u,d,c,s\}$).}\label{fig:EWP-top}
\end{figure}

\subsection{Non-Leptonic Decays}\label{subsec:non-lept}
\subsubsection{Classification}\label{sec:class}
The most complicated $B$ decays are the non-leptonic transitions, 
which are mediated by 
$b\to q_1\,\bar q_2\,d\,(s)$ quark-level processes, with 
$q_1,q_2\in\{u,d,c,s\}$. There are two kinds of 
topologies contributing to such decays: tree-diagram-like and ``penguin'' 
topologies. The latter consist of gluonic (QCD) and electroweak (EW) 
penguins. In Figs.~\ref{fig:tree-top}--\ref{fig:EWP-top}, the corresponding 
leading-order Feynman diagrams are shown. Depending
on the flavour content of their final states, we may classify 
$b\to q_1\,\bar q_2\,d\,(s)$ decays as follows:
\begin{itemize}
\item $q_1\not=q_2\in\{u,c\}$: {\it only} tree diagrams contribute.
\item $q_1=q_2\in\{u,c\}$: tree {\it and} penguin diagrams contribute.
\item $q_1=q_2\in\{d,s\}$: {\it only} penguin diagrams contribute.
\end{itemize}

\subsubsection{Low-Energy Effective Hamiltonians}\label{subsec:ham}
In order to analyse non-leptonic $B$ decays theoretically, one uses 
low-energy effective Hamiltonians, which are calculated by making use 
of the ``operator product expansion'', yielding transition 
matrix elements of the following structure:
\begin{equation}\label{ee2}
\langle f|{\cal H}_{\rm eff}|i\rangle=\frac{G_{\rm F}}{\sqrt{2}}
\lambda_{\rm CKM}\sum_k C_k(\mu)\langle f|Q_k(\mu)|i\rangle\,.
\end{equation}
The technique of the operator product expansion allows us to separate 
the short-distance contributions to this transition amplitude from the 
long-distance ones, which are described by perturbative quantities 
$C_k(\mu)$ (``Wilson coefficient functions'') and non-perturbative 
quantities $\langle f|Q_k(\mu)|i\rangle$ (``hadronic matrix elements''), 
respectively. As before, $G_{\rm F}$ is the Fermi constant, whereas
$\lambda_{\rm CKM}$ is a CKM factor and $\mu$ denotes an appropriate 
renormalization scale. The $Q_k$ are local operators, which 
are generated by electroweak interactions and QCD, and govern ``effectively'' 
the decay in question. The Wilson coefficients $C_k(\mu)$ can be 
considered as scale-dependent couplings related to the vertices described
by the $Q_k$.

In order to illustrate this rather abstract formalism, let us consider
the decay $\bar B^0_d\to D^+K^-$, which allows a transparent discussion 
of the evaluation of the corresponding low-energy effective Hamiltonian.
Since this transition originates from a $b\to c \bar u s$ quark-level 
process, it is -- as we have seen in our classification in 
Subsection~\ref{sec:class} -- a pure ``tree'' decay, i.e.\ we do not have
to deal with penguin topologies, which simplifies the analysis
considerably. The leading-order Feynman diagram contributing to
$\bar B^0_d\to D^+K^-$ can straightforwardly be obtained from 
Fig.~\ref{fig:semi} by substituting $\ell$ and $\nu$ by $s$ and $u$, 
respectively. Consequently, the lepton current is simply replaced by a 
quark current, which will have important implications shown below. 
Evaluating the corresponding Feynman diagram yields
\begin{equation}\label{trans-ampl}
-\,\frac{g_2^2}{8}V_{us}^\ast V_{cb}
\left[\bar s\gamma^\nu(1-\gamma_5)u\right]
\left[\frac{g_{\nu\mu}}{k^2-M_W^2}\right]
\left[\bar c\gamma^\mu(1-\gamma_5)b\right].
\end{equation}
Because of $k^2\sim m_b^2\ll M_W^2$, we may -- as in (\ref{Tfi-semi-full}) -- 
``integrate out'' the $W$ boson with the help of (\ref{W-int-out}),
and arrive at
\begin{eqnarray}
\lefteqn{{\cal H}_{\rm eff}=\frac{G_{\rm F}}{\sqrt{2}}V_{us}^\ast V_{cb}
\left[\bar s_\alpha\gamma_\mu(1-\gamma_5)u_\alpha\right]
\left[\bar c_\beta\gamma^\mu(1-\gamma_5)b_\beta\right]}\nonumber\\
&&=\frac{G_{\rm F}}{\sqrt{2}}V_{us}^\ast V_{cb}
(\bar s_\alpha u_\alpha)_{\mbox{{\scriptsize 
V--A}}}(\bar c_\beta b_\beta)_{\mbox{{\scriptsize V--A}}}
\equiv\frac{G_{\rm F}}{\sqrt{2}}V_{us}^\ast V_{cb}O_2\,,
\end{eqnarray}
where $\alpha$ and $\beta$ denote the colour indices of the $SU(3)_{\rm C}$
gauge group of QCD. Effectively, our $b\to c \bar u s$ decay process 
is now described by the ``current--current'' operator $O_2$. 

If we take QCD corrections into account, operator mixing induces a 
second ``current--current'' operator, which is given by
\begin{equation}
O_1\equiv\left[\bar s_\alpha\gamma_\mu(1-\gamma_5)u_\beta\right]
\left[\bar c_\beta\gamma^\mu(1-\gamma_5)b_\alpha\right].
\end{equation}
Consequently, we obtain a low-energy effective Hamiltonian of the following
structure:
\begin{equation}\label{Heff-example}
{\cal H}_{\rm eff}=\frac{G_{\rm F}}{\sqrt{2}}V_{us}^\ast V_{cb}
\left[C_1(\mu)O_1+C_2(\mu)O_2\right],
\end{equation}
where $C_1(\mu)\not=0$ and $C_2(\mu)\not=1$ are due to QCD renormalization
effects \cite{HEFF-TREE}. In order to evaluate these coefficients, 
we must first calculate the QCD corrections to the decay processes 
both in the full theory, i.e. with $W$ exchange, and in the effective 
theory, where the $W$ is integrated out, and have then to express 
the QCD-corrected transition amplitude in terms of QCD-corrected matrix 
elements and Wilson coefficients as in (\ref{ee2}). This procedure is 
called ``matching'' between the full and the effective theory. 
The results for the $C_k(\mu)$ thus obtained contain 
terms of $\mbox{log}(\mu/M_W)$, which become large for $\mu={\cal O}(m_b)$, 
the scale governing the hadronic matrix elements of the $O_k$. Making use of 
the renormalization group, which exploits the fact that the transition 
amplitude (\ref{ee2}) cannot depend on the chosen renormalization scale 
$\mu$, we may sum up the following terms of the Wilson coefficients:
\begin{equation}
\alpha_s^n\left[\log\left(\frac{\mu}{M_W}\right)\right]^n 
\,\,\mbox{(LO)},\quad\,\,\alpha_s^n\left[\log\left(\frac{\mu}{M_W}\right)
\right]^{n-1}\,\,\mbox{(NLO)},\quad ...\quad ;
\end{equation}
detailed discussions of these rather technical aspects can be found in
\cite{B-LH98,BBL-rev}.

For the exploration of CP violation, the class of non-leptonic $B$ decays 
that receives contributions both from tree and from penguin topologies plays 
a key r\^ole. In this important case, the operator basis is much larger 
than in our example (\ref{Heff-example}), where we considered a pure 
``tree'' decay. If we apply the relation
\begin{equation}\label{CKM-UT-Rel}
V_{ur}^\ast V_{ub}+V_{cr}^\ast V_{cb}+V_{tr}^\ast V_{tb}=0
\quad (r\in\{d,s\}),
\end{equation}
which follows from the unitarity of the CKM matrix, and ``integrate out''
the top quark (which enters through the penguin loop processes) and 
the $W$ boson, we may write 
\begin{equation}\label{e4}
{\cal H}_{\mbox{{\scriptsize eff}}}=\frac{G_{\mbox{{\scriptsize 
F}}}}{\sqrt{2}}\left[\sum\limits_{j=u,c}V_{jr}^\ast V_{jb}\left\{
\sum\limits_{k=1}^2C_k(\mu)\,Q_k^{jr}
+\sum\limits_{k=3}^{10}C_k(\mu)\,Q_k^{r}\right\}\right].
\end{equation}
Here we have introduced another quark-flavour label $j\in\{u,c\}$,
and the $Q_k^{jr}$ can be divided as follows:
\begin{itemize}
\item Current--current operators:
\begin{equation}
\begin{array}{rcl}
Q_{1}^{jr}&=&(\bar r_{\alpha}j_{\beta})_{\mbox{{\scriptsize V--A}}}
(\bar j_{\beta}b_{\alpha})_{\mbox{{\scriptsize V--A}}}\\
Q_{2}^{jr}&=&(\bar r_\alpha j_\alpha)_{\mbox{{\scriptsize 
V--A}}}(\bar j_\beta b_\beta)_{\mbox{{\scriptsize V--A}}}.
\end{array}
\end{equation}
\item QCD penguin operators:
\begin{equation}\label{qcd-penguins}
\begin{array}{rcl}
Q_{3}^r&=&(\bar r_\alpha b_\alpha)_{\mbox{{\scriptsize V--A}}}\sum_{q'}
(\bar q'_\beta q'_\beta)_{\mbox{{\scriptsize V--A}}}\\
Q_{4}^r&=&(\bar r_{\alpha}b_{\beta})_{\mbox{{\scriptsize V--A}}}
\sum_{q'}(\bar q'_{\beta}q'_{\alpha})_{\mbox{{\scriptsize V--A}}}\\
Q_{5}^r&=&(\bar r_\alpha b_\alpha)_{\mbox{{\scriptsize V--A}}}\sum_{q'}
(\bar q'_\beta q'_\beta)_{\mbox{{\scriptsize V+A}}}\\
Q_{6}^r&=&(\bar r_{\alpha}b_{\beta})_{\mbox{{\scriptsize V--A}}}
\sum_{q'}(\bar q'_{\beta}q'_{\alpha})_{\mbox{{\scriptsize V+A}}}.
\end{array}
\end{equation}
\item EW penguin operators (the $e_{q'}$ denote the
electrical quark charges):
\begin{equation}
\begin{array}{rcl}
Q_{7}^r&=&\frac{3}{2}(\bar r_\alpha b_\alpha)_{\mbox{{\scriptsize V--A}}}
\sum_{q'}e_{q'}(\bar q'_\beta q'_\beta)_{\mbox{{\scriptsize V+A}}}\\
Q_{8}^r&=&
\frac{3}{2}(\bar r_{\alpha}b_{\beta})_{\mbox{{\scriptsize V--A}}}
\sum_{q'}e_{q'}(\bar q_{\beta}'q'_{\alpha})_{\mbox{{\scriptsize V+A}}}\\
Q_{9}^r&=&\frac{3}{2}(\bar r_\alpha b_\alpha)_{\mbox{{\scriptsize V--A}}}
\sum_{q'}e_{q'}(\bar q'_\beta q'_\beta)_{\mbox{{\scriptsize V--A}}}\\
Q_{10}^r&=&
\frac{3}{2}(\bar r_{\alpha}b_{\beta})_{\mbox{{\scriptsize V--A}}}
\sum_{q'}e_{q'}(\bar q'_{\beta}q'_{\alpha})_{\mbox{{\scriptsize V--A}}}.
\end{array}
\end{equation}
\end{itemize}
The current--current, QCD and EW penguin operators are related to the tree, 
QCD and EW penguin processes shown in 
Figs.~\ref{fig:tree-top}--\ref{fig:EWP-top}. At a renormalization scale
$\mu={\cal O}(m_b)$, the Wilson coefficients of the current--current operators
are $C_1(\mu)={\cal O}(10^{-1})$ and $C_2(\mu)={\cal O}(1)$, whereas those
of the penguin operators are ${\cal O}(10^{-2})$ \cite{B-LH98,BBL-rev}. 
Note that penguin 
topologies with internal charm- and up-quark exchanges \cite{BSS}
are described in this framework by penguin-like matrix elements of 
the corresponding current--current operators \cite{RF-DIPL}, and 
may also have important phenomenological consequences \cite{BF-PEN,CHARM-PEN}.

Since the ratio $\alpha/\alpha_s={\cal O}(10^{-2})$ of the QED and QCD 
couplings is very small, we would expect na\"\i vely that EW penguins 
should play a minor r\^ole in comparison with QCD penguins. This would 
actually be the case if the top quark was not ``heavy''. However, since 
the Wilson coefficient $C_9$ increases strongly with $m_t$, we obtain 
interesting EW penguin effects in several $B$ decays: $B\to K\phi$ 
modes are affected significantly by EW penguins, whereas $B\to\pi\phi$ 
and $B_s\to\pi^0\phi$ transitions are even {\it dominated} by such 
topologies \cite{RF-EWP,RF-rev}. EW penguins also have an important 
impact on the $B\to\pi K$ system \cite{EWP-BpiK}, as we will see 
in Subsection~\ref{ssec:BpiK}.

The low-energy effective Hamiltonians discussed above apply to all 
$B$ decays that are caused by the same quark-level transition, i.e.\ 
they are ``universal''. Consequently, the differences between the 
various exclusive modes of a given decay class arise within this formalism 
only through the hadronic matrix elements of the relevant four-quark 
operators. Unfortunately, the evaluation of such matrix elements is 
associated with large uncertainties and is a very challenging task. In this 
context, ``factorization'' is a widely used concept, which is our next 
topic.

\subsubsection{Factorization of Hadronic Matrix Elements}\label{ssec:ME-fact}
In order to discuss ``factorization'', let us consider once more 
the decay $\bar B^0_d\to D^+K^-$. Evaluating the corresponding 
transition amplitude, we encounter the hadronic matrix elements of the 
$O_{1,2}$ operators between the $\langle K^-D^+|$ final and the 
$|\bar B^0_d\rangle$ initial states. If we use the well-known 
$SU(N_{\rm C})$ colour-algebra relation
\begin{equation}
T^a_{\alpha\beta}T^a_{\gamma\delta}=\frac{1}{2}\left(\delta_{\alpha\delta}
\delta_{\beta\gamma}-\frac{1}{N_{\rm C}}\delta_{\alpha\beta}
\delta_{\gamma\delta}\right)
\end{equation}
to rewrite the operator $O_1$, we obtain
\begin{displaymath}
\langle K^-D^+|{\cal H}_{\rm eff}|\bar B^0_d\rangle=
\frac{G_{\rm F}}{\sqrt{2}}V_{us}^\ast V_{cb}\Bigl[a_1\langle K^-D^+|
(\bar s_\alpha u_\alpha)_{\mbox{{\scriptsize V--A}}}
(\bar c_\beta b_\beta)_{\mbox{{\scriptsize V--A}}}
|\bar B^0_d\rangle
\end{displaymath}
\vspace*{-0.3truecm}
\begin{equation}\label{ME-rewritten}
+2\,C_1\langle K^-D^+|
(\bar s_\alpha\, T^a_{\alpha\beta}\,u_\beta)_{\mbox{{\scriptsize 
V--A}}}(\bar c_\gamma 
\,T^a_{\gamma\delta}\,b_\delta)_{\mbox{{\scriptsize V--A}}}
|\bar B^0_d\rangle\Bigr],\nonumber
\end{equation}
with
\begin{equation}\label{a1-def}
a_1=C_1/N_{\rm C}+C_2 \sim 1.
\end{equation}
It is now straightforward to ``factorize'' the hadronic matrix elements
in (\ref{ME-rewritten}):
\begin{eqnarray}
\lefteqn{\left.\langle K^-D^+|
(\bar s_\alpha u_\alpha)_{\mbox{{\scriptsize 
V--A}}}(\bar c_\beta b_\beta)_{\mbox{{\scriptsize V--A}}}
|\bar B^0_d\rangle\right|_{\rm fact}}\nonumber\\
&&=\langle K^-|\left[\bar s_\alpha\gamma_\mu(1-\gamma_5)u_\alpha\right]
|0\rangle\langle D^+|\left[\bar c_\beta\gamma^\mu
(1-\gamma_5)b_\beta\right]|\bar B^0_d\rangle\nonumber\\
&&=\underbrace{i f_K}_{\mbox{decay constant}} \, \times \, 
\underbrace{F^{(BD)}_0(M_K^2)}_{\mbox{$B\to D$ form factor}} 
\, \times \,\underbrace{(M_B^2-M_D^2),}_{\mbox{kinematical factor}}
\end{eqnarray}
\begin{equation}
\left.\langle K^-D^+|
(\bar s_\alpha\, T^a_{\alpha\beta}\,u_\beta)_{\mbox{{\scriptsize 
V--A}}}(\bar c_\gamma 
\,T^a_{\gamma\delta}\,b_\delta)_{\mbox{{\scriptsize V--A}}}
|\bar B^0_d\rangle\right|_{\rm fact}=0.
\end{equation}
The quantity $a_1$ is a phenomenological ``colour factor'', 
which governs ``colour-allowed'' decays; the
decay $\bar B^0_d\to D^+K^-$ belongs to this category, since the 
colour indices of the $K^-$ meson and the $\bar B^0_d$--$D^+$ system 
run independently from each other in the corresponding leading-order 
diagram. On the other hand, in the case of ``colour-suppressed'' modes, 
for instance $\bar B^0_d\to \pi^0D^0$, where only one colour
index runs through the whole diagram, we have to deal with the combination
\begin{equation}\label{a2-def}
a_2=C_1+C_2/N_{\rm C}\sim0.25.
\end{equation}

The concept of factorizing the hadronic matrix elements of four-quark 
operators into the product of hadronic matrix elements of quark currents 
has a long history \cite{Neu-Ste}, and can be justified, for example, 
in the large-$N_{\rm C}$ limit \cite{largeN}. Interesting recent 
developments are the following:
\begin{itemize}
\item ``QCD factorization'' \cite{BBNS}, which is in 
accordance with the old picture that factorization should 
hold for certain decays in the limit of $m_b\gg\Lambda_{\rm QCD}$ 
\cite{QCDF-old}, provides a formalism to calculate the 
relevant amplitudes at the leading order of a $\Lambda_{\rm QCD}/m_b$ 
expansion. The resulting expression for the transition amplitudes 
incorporates elements both of the na\"\i ve factorization approach 
sketched above and of the hard-scattering picture. Let us consider a 
decay $\bar B\to M_1M_2$, where $M_1$ picks up the spectator quark. 
If $M_1$ is either a heavy ($D$) or a light ($\pi$, $K$) meson, and 
$M_2$ a light ($\pi$, $K$) meson, QCD factorization gives a transition 
amplitude of the following structure:
\begin{equation}
A(\bar B\to M_1M_2)=\left[\mbox{``na\"\i ve factorization''}\right]
\times\left[1+{\cal O}(\alpha_s)+{\cal O}(\Lambda_{\rm QCD}/m_b)\right].
\end{equation}
While the ${\cal O}(\alpha_s)$ terms, i.e.\ the radiative
non-factorizable corrections, can be calculated systematically, 
the main limitation of the theoretical accuracy originates from 
the ${\cal O}(\Lambda_{\rm QCD}/m_b)$ terms. 

\item Another QCD approach to deal with non-leptonic $B$-meson decays -- 
the ``perturbative hard-scattering approach '' (PQCD) -- was developed 
independently in \cite{PQCD}, and differs from the QCD factorization 
formalism in some technical aspects.

\item A very useful technique for ``factorization proofs'' is provided 
by the framework of the ``soft collinear effective theory'' (SCET) 
\cite{SCET}.

\item Non-leptonic $B$ decays can also be studied within 
QCD light-cone sum-rule approaches \cite{sum-rules}.
\end{itemize}
A detailed presentation of these topics would be very technical and is
beyond the scope of these lectures. However, for the discussion of the
CP-violating effects in the $B$-meson system, we must only be familiar 
with the general structure of the non-leptonic $B$ decay amplitudes and 
not enter the details of the techniques to deal with the
corresponding hadronic matrix elements. Let us finally note that the 
$B$-factory data will eventually decide how well factorization and 
the new concepts sketched above are actually working. For example, 
recent data on the $B\to\pi\pi$ system point towards 
large non-factorizable corrections \cite{BFRS3,BFRS2}, to which we 
shall return in Subsection~\ref{ssec:Bpipi-puzzle}.

\subsection{Towards Studies of CP Violation}\label{To-CP}
As we have seen above, leptonic and semileptonic $B$-meson decays
involve only a single weak (CKM) amplitude. On the other hand, the 
structure of non-leptonic transitions is considerably more complicated. 
However, because of the unitarity of the CKM matrix, which implies the 
relation in (\ref{CKM-UT-Rel}), we may write the amplitude of {\it any} 
non-leptonic $B$-meson decay within the SM in such a manner that
we encounter at most two contributions with different CKM 
factors (we will encounter explicit examples below):
\begin{eqnarray}
A(\bar B\to\bar f)&=&e^{+i\varphi_1}
|A_1|e^{i\delta_1}+e^{+i\varphi_2}|A_2|e^{i\delta_2}\label{par-ampl}\\
A(B\to f)&=&e^{-i\varphi_1}|A_1|e^{i\delta_1}+
e^{-i\varphi_2}|A_2|e^{i\delta_2}.\label{par-ampl-CP}
\end{eqnarray}
Here the $\varphi_{1,2}$ denote CP-violating weak phases, which are 
introduced by the elements of the CKM matrix, whereas the 
$|A_{1,2}|e^{i\delta_{1,2}}$ are CP-conserving ``strong'' amplitudes, 
which contain the whole hadron dynamics of the decay at hand:
\begin{equation}\label{ampl-struc}
|A|e^{i\delta}\sim\sum\limits_k
\underbrace{C_{k}(\mu)}_{\mbox{pert.\ QCD}} 
\times\,\,\, \underbrace{\langle\bar f|Q_{k}(\mu)|\bar B
\rangle}_{\mbox{non-pert.\ QCD}}.
\end{equation}
If we use (\ref{par-ampl}) and (\ref{par-ampl-CP}), it is an easy
exercise to calculate the following CP-violating rate asymmetry:
\begin{eqnarray}
{\cal A}_{\rm CP}&\equiv&\frac{\Gamma(B\to f)-
\Gamma(\bar B\to\bar f)}{\Gamma(B\to f)+\Gamma(\bar B
\to \bar f)}=\frac{|A(B\to f)|^2-|A(\bar B\to \bar f)|^2}{|A(B\to f)|^2+
|A(\bar B\to \bar f)|^2}\nonumber\\
&=&\frac{2|A_1||A_2|\sin(\delta_1-\delta_2)
\sin(\varphi_1-\varphi_2)}{|A_1|^2+2|A_1||A_2|\cos(\delta_1-\delta_2)
\cos(\varphi_1-\varphi_2)+|A_2|^2}.\label{direct-CPV}
\end{eqnarray}
Consequently, a non-vanishing CP asymmetry ${\cal A}_{\rm CP}$ arises from 
the interference effects between the two weak amplitudes, and requires both
a non-trivial weak phase difference $\varphi_1-\varphi_2$ and a
non-trivial strong phase difference $\delta_1-\delta_2$. This kind of
CP violation is referred to as ``direct'' CP violation, as it originates 
directly at the amplitude level of the considered decay. It is the 
$B$-meson counterpart of the effects that are probed through 
$\mbox{Re}(\varepsilon'/\varepsilon)$ in the neutral kaon 
system.\footnote{In order to calculate this quantity, an approriate 
low-energy effective Hamiltonian having the same structure as (\ref{e4}) 
is used. The large theoretical uncertainties mentioned in 
Subsection~\ref{subsec:kaon-CP} originate from a strong cancellation 
between the contributions of the QCD and EW penguins (caused by the
large top-quark mass) and the associated
hadronic matrix elements.}
Since $\varphi_1-\varphi_2$ is in general given by one of the angles of 
the UT -- usually $\gamma$ -- the goal is to determine this 
quantity from the measured value of ${\cal A}_{\rm CP}$. Unfortunately, 
the extraction of $\varphi_1-\varphi_2$ from ${\cal A}_{\rm CP}$ is affected 
by hadronic uncertainties, which are related to the poorly known 
hadronic matrix elements entering the expression (\ref{ampl-struc}) 
for the strong amplitudes $|A_{1,2}|e^{i\delta_{1,2}}$. In order to
deal with this problem, we may, in principle, proceed along one of the 
following three main avenues:
\begin{itemize}
\item[i)] The most obvious one -- but also the most challenging -- 
is to try to {\it calculate} the relevant hadronic matrix elements
$\langle \bar f|Q_k(\mu)|\bar B\rangle$. As we have noted
above, interesting progress has recently been made in this direction
through the development of the QCD factorization, PQCD, SCET  
and QCD light-cone sum-rule approaches. 

\item[ii)] We may search for fortunate cases, where relations between 
various decay amplitudes allow us to {\it eliminate} the poorly
known hadronic matrix elements. As we shall see, this avenue offers 
in particular determinations of the UT angle $\gamma$: we distinguish 
between exact relations, which are provided by pure ``tree'' decays 
of the kind $B\to KD$ or $B_c\to D_sD$, and relations, which follow from the 
flavour symmetries of strong interactions, involving 
$B_{(s)}\to\pi\pi,\pi K,KK$ transitions. 

\item[iii)] Finally, we may exploit the fact that in decays of  
neutral $B_q$ mesons ($q\in\{d,s\}$) interference effects 
between $B^0_q$--$\bar B^0_q$ mixing and decay processes may yield
another kind of CP violation, ``mixing-induced CP violation''.
In certain cases, the hadronic matrix elements {\it cancel} in such
CP asymmeties. 
\end{itemize}
In the remainder of these lectures, we will not consider (i) further. 
For the exploration of CP violation and the testing of the KM mechanism, 
the theoretical input related to strong-interaction physics should 
obviously be reduced as much as possible. In contrast to (i), this feature 
is present in (ii) and (iii), which provide -- as a by-product -- also 
important insights into hadron dynamics. In particular, we may extract 
various hadronic parameters from the data that can be calculated with 
the help of the theoretical frameworks listed in (i), thereby allowing
us to test them through a confrontation with nature. Since neutral 
$B_q$ mesons are a key element in this programme, offering also attractive 
connections between (ii) and (iii), let us next have a closer look at their 
most important features.

\section{FEATURES OF NEUTRAL {\boldmath$B_{d,s}$\unboldmath} 
MESONS}\label{sec:mix}
\setcounter{equation}{0}
\boldmath\subsection{$B^0_{d,s}$--$\bar B^0_{d,s}$ 
Mixing}\unboldmath\label{ssec:BBbar-mix}
Within the SM, $B^0_q$--$\bar B^0_q$ mixing ($q\in\{d,s\}$) arises from
the box diagrams shown in Fig.~\ref{fig:boxes}. Because of
this phenomenon, an initially, i.e.\ at time $t=0$, present 
$B^0_q$-meson state evolves into a time-dependent linear combination of 
$B^0_q$ and $\bar B^0_q$ states:
\begin{equation}
|B_q(t)\rangle=a(t)|B^0_q\rangle + b(t)|\bar B^0_q\rangle,
\end{equation}
where $a(t)$ and $b(t)$ are governed by a Schr\"odinger equation of 
the following form:
\begin{equation}\label{SG-OSZ}
i\,\frac{{\rm d}}{{\rm d} t}\left(\begin{array}{c} a(t)\\ b(t)
\end{array}
\right)= H \cdot\left(\begin{array}{c}
a(t)\\ b(t)\nonumber
\end{array}
\right) \equiv
\Biggl[\underbrace{\left(\begin{array}{cc}
M_{0}^{(q)} & M_{12}^{(q)}\\ M_{12}^{(q)\ast} & M_{0}^{(q)}
\end{array}\right)}_{\mbox{mass matrix}}-
\frac{i}{2}\underbrace{\left(\begin{array}{cc}
\Gamma_{0}^{(q)} & \Gamma_{12}^{(q)}\\
\Gamma_{12}^{(q)\ast} & \Gamma_{0}^{(q)}
\end{array}\right)}_{\mbox{decay matrix}}\Biggr]
\cdot\left(\begin{array}{c}
a(t)\\ b(t)\nonumber
\end{array}
\right).
\end{equation}
The special form $H_{11}=H_{22}$ of the Hamiltonian $H$ is an implication 
of the CPT theorem, i.e.\ of the invariance under combined CP and 
time-reversal (T) transformations.

\subsubsection{Solution of the Schr\"odinger Equation}
It is straightforward to calculate the eigenstates 
$\vert B_{\pm}^{(q)}\rangle$ and eigenvalues $\lambda_{\pm}^{(q)}$ 
of (\ref{SG-OSZ}):
\begin{equation}
\vert B_{\pm}^{(q)} \rangle  =
\frac{1}{\sqrt{1+\vert \alpha_q\vert^{2}}}
\left(\vert B^{0}_q\rangle\pm\alpha_q\vert\bar B^{0}_q\rangle\right)
\end{equation}
\begin{equation}\label{lam-pm}
\lambda_{\pm}^{(q)}  =
\left(M_{0}^{(q)}-\frac{i}{2}\Gamma_{0}^{(q)}\right)\pm
\left(M_{12}^{(q)}-\frac{i}{2}\Gamma_{12}^{(q)}\right)\alpha_q,
\end{equation}
where
\begin{equation}\label{alpha-q-expr}
\alpha_q e^{+i\left(\Theta_{\Gamma_{12}}^{(q)}+n'
\pi\right)}=
\sqrt{\frac{4\vert M_{12}^{(q)}\vert^{2}
e^{-i2\delta\Theta_{M/\Gamma}^{(q)}}+\vert\Gamma_{12}^{(q)}\vert^{2}}{4\vert 
M_{12}^{(q)}\vert^{2}+\vert\Gamma_{12}^{(q)}\vert^{2}- 
4\vert M_{12}^{(q)}\vert\vert\Gamma_{12}^{(q)}\vert
\sin\delta\Theta_{M/\Gamma}^{(q)}}}.
\end{equation}
Here we have written
\begin{equation}
M_{12}^{(q)}\equiv e^{i\Theta_{M_{12}}^{(q)}}\vert
M_{12}^{(q)}\vert,\quad \Gamma_{12}^{(q)}\equiv
e^{i\Theta_{\Gamma_{12}}^{(q)}}\vert\Gamma_{12}^{(q)}\vert,\quad
\delta\Theta_{M/\Gamma}^{(q)}\equiv
\Theta_{M_{12}}^{(q)}-\Theta_{\Gamma_{12}}^{(q)},
\end{equation}
and have introduced the quantity $n'\in\{0,1\}$ to parametrize the 
sign of the square root in (\ref{alpha-q-expr}). 

\begin{figure}[t]
\begin{center}
{\small
\begin{picture}(250,70)(0,45)
\ArrowLine(10,100)(40,100)\Photon(40,100)(80,100){2}{8}
\ArrowLine(80,100)(110,100)
\ArrowLine(40,60)(10,60)\Photon(40,60)(80,60){2}{8}
\ArrowLine(110,60)(80,60)
\ArrowLine(40,100)(40,60)\ArrowLine(80,60)(80,100)
\Text(25,105)[b]{$q$}\Text(60,105)[b]{$W$}\Text(95,105)[b]{$b$}
\Text(25,55)[t]{$b$}\Text(60,55)[t]{$W$}\Text(95,55)[t]{$q$}
\Text(35,80)[r]{$u,c,t$}\Text(85,80)[l]{$u,c,t$}
\ArrowLine(160,100)(190,100)\ArrowLine(190,100)(230,100)
\ArrowLine(230,100)(260,100)
\ArrowLine(190,60)(160,60)\ArrowLine(230,60)(190,60)
\ArrowLine(260,60)(230,60)
\Photon(190,100)(190,60){2}{8}\Photon(230,60)(230,100){2}{8}
\Text(175,105)[b]{$q$}\Text(245,105)[b]{$b$}
\Text(175,55)[t]{$b$}\Text(245,55)[t]{$q$}
\Text(210,105)[b]{$u,c,t$}\Text(210,55)[t]{$u,c,t$}
\Text(180,80)[r]{$W$}\Text(240,80)[l]{$W$}
\end{picture}}
\end{center}
\vspace*{-0.4truecm}
\caption{Box diagrams contributing to $B^0_q$--$\bar B^0_q$ mixing 
in the SM ($q\in\{d,s\}$).}\label{fig:boxes}
\end{figure}
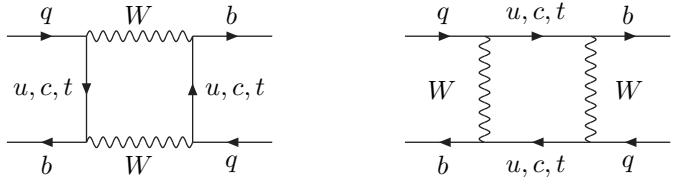

Evaluating the dispersive parts of the box diagrams shown in
Fig~\ref{fig:boxes}, which are dominated by internal top-quark 
exchanges, yields (for a more detailed discussion, see \cite{BF-rev}):
\begin{equation}\label{M12-calc}
M_{12}^{(q)}=
\frac{G_{\rm F}^2 M_{W}^{2}}{12\pi^{2}}
\eta_B M_{B_q}\hat B_{B_q}f_{B_q}^{2}\left(V_{tq}^\ast V_{tb}\right)^2 
S_0(x_{t}) e^{i(\pi-\phi_{\mbox{{\scriptsize CP}}}(B_q))},
\end{equation}
where $\eta_B=0.55\pm0.01$ is a perturbative QCD correction 
\cite{eta-B},\footnote{Note that the short-distance parameter 
$\eta_B$ does {\it not} depend on 
$q\in\{d,s\}$, i.e.\ is the same for $B_d$ and $B_s$ mesons.}
the non-perturbative
``bag'' parameter $\hat B_{B_q}$ is related to the hadronic
matrix element $\langle \bar B^0_q|(\bar bq)_{{\rm V}-{\rm A}}
(\bar bq)_{{\rm V}-{\rm A}}|B^0_q\rangle$, and $S_0(x_t\equiv m_t^2/M_W^2)$
is one of the ``Inami--Lim'' functions 
\cite{IL}, describing the dependence on the top-quark mass $m_t$. In the SM, 
we may write -- to a good approximation -- the following expression 
\cite{Buras-Schladming}:
\begin{equation}
S_0(x_t)=2.40\times\left[\frac{m_t}{167\,\mbox{GeV}}\right]^{1.52}.
\end{equation}
Finally, $\phi_{\mbox{{\scriptsize CP}}}(B_q)$ is a convention-dependent
phase, which is introduced through the CP transformation
\begin{equation}\label{CP-def}
({\cal CP})\vert B^{0}_q\rangle=
e^{i\phi_{\mbox{{\scriptsize CP}}}(B_q)}
\vert\bar B^{0}_q\rangle.
\end{equation}
If we calculate also the absorptive parts of the box diagrams in
Fig~\ref{fig:boxes}, we obtain
\begin{equation}
\frac{\Gamma_{12}^{(q)}}{M_{12}^{(q)}}\approx
-\frac{3\pi}{2S_0(x_{t})}\left(\frac{m_b^2}{M_W^2}\right)
={\cal O}(m_b^2/m_t^2)\ll 1.
\end{equation}
Consequently, we may expand (\ref{alpha-q-expr}) in 
$\Gamma_{12}^{(q)}/M_{12}^{(q)}$. Neglecting second-order 
terms, we arrive at
\begin{equation}
\alpha_q=\left[1+\frac{1}{2}\left|
\frac{\Gamma_{12}^{(q)}}{M_{12}^{(q)}}\right|\sin\delta
\Theta_{M/\Gamma}^{(q)}\right]e^{-i\left(\Theta_{M_{12}}^{(q)}+n'\pi\right)}.
\end{equation}

The deviation of $|\alpha_q|$ from 1 measures CP violation in 
$B^0_q$--$\bar B^0_q$ oscillations, and can be probed through
the following ``wrong-charge'' lepton asymmetries:
\begin{equation}
{\cal A}^{(q)}_{\mbox{{\scriptsize SL}}}\equiv
\frac{\Gamma(B^0_q(t)\to \ell^-\bar\nu X)-\Gamma(\bar B^0_q(t)\to
\ell^+\nu X)}{\Gamma(B^0_q(t)\to \ell^-\bar \nu X)+
\Gamma(\bar B^0_q(t)\to \ell^+\nu X)}
=\frac{|\alpha_q|^4-1}{|\alpha_q|^4+1}\approx\left|
\frac{\Gamma_{12}^{(q)}}{M_{12}^{(q)}}\right|
\sin\delta\Theta^{(q)}_{M/\Gamma}.
\end{equation}
Because of $|\Gamma_{12}^{(q)}|/|M_{12}^{(q)}|\propto
m_b^2/m_t^2$ and $\sin\delta\Theta^{(q)}_{M/\Gamma}\propto m_c^2/m_b^2$,
the asymmetry ${\cal A}^{(q)}_{\mbox{{\scriptsize SL}}}$ is suppressed by 
a factor of $m_c^2/m_t^2={\cal O}(10^{-4})$ and is hence tiny in the SM.
However, this observable may be enhanced through NP effects, thereby
representing an interesing probe for physics beyond the SM 
\cite{LLNP,BBLN-CFLMT}.  The current experimental constraints for 
${\cal A}^{(q)}_{\mbox{{\scriptsize SL}}}$
are at the $10^{-2}$ level.

\subsubsection{Mixing Parameters}\label{ssec:Mix-Par}
Let us denote the masses of the eigenstates of (\ref{SG-OSZ}) by 
$M^{(q)}_{\rm H}$ (``heavy'') and $M^{(q)}_{\rm L}$ (``light''). 
It is then useful to introduce 
\begin{equation}
M_q\equiv\frac{M^{(q)}_{\rm H}+M^{(q)}_{\rm L}}{2}=
M^{(q)}_0,
\end{equation}
as well as the mass difference
\begin{equation}\label{DeltaMq-def}
\Delta M_q\equiv M_{\rm H}^{(q)}-M_{\rm L}^{(q)}=2|M_{12}^{(q)}|>0,
\end{equation}
which is by definition {\it positive}. Using (\ref{Rt-def})
and (\ref{M12-calc}), we find that we may convert the mass difference
$\Delta M_d$ of the $B_d$-meson system into the side $R_t$ of the UT 
with the help of the following expression:
\begin{equation}\label{Rt-det}
R_t=\frac{1.10}{A\sqrt{|S_0(x_t)|}}
\sqrt{\frac{\Delta M_d}{0.50\,{\rm ps}^{-1}}}\left[\frac{230 \, 
{\rm MeV}}{\sqrt{\hat B_{B_d}}f_{B_d}}\right]\sqrt{\frac{0.55}{\eta_B}},
\end{equation}
where $A$ is the usual Wolfenstein parameter. We shall return to this
important issue in Subsection~\ref{ssec:DMs-UT}.

On the other hand, the decay widths $\Gamma_{\rm H}^{(q)}$ and 
$\Gamma_{\rm L}^{(q)}$ of the mass eigenstates, which correspond to 
$M^{(q)}_{\rm H}$ and $M^{(q)}_{\rm L}$, respectively, satisfy 
\begin{equation}
\Delta\Gamma_q\equiv\Gamma_{\rm H}^{(q)}-\Gamma_{\rm L}^{(q)}=
\frac{4\mbox{\,Re}\left[M_{12}^{(q)}\Gamma_{12}^{(q)\ast}\right]}{\Delta M_q},
\end{equation}
whereas 
\begin{equation}
\Gamma_q\equiv\frac{\Gamma^{(q)}_{\rm H}+\Gamma^{(q)}_{\rm L}}{2}=
\Gamma^{(q)}_0.
\end{equation}
There is the following interesting relation:
\begin{equation}\label{DGoG}
\frac{\Delta\Gamma_q}{\Gamma_q}\approx-\frac{3\pi}{2S_0(x_t)}
\left(\frac{m_b^2}{M_W^2}\right)x_q=-{\cal O}(10^{-2})\times x_q,
\end{equation}
where
\begin{equation}\label{mix-par}
x_q\equiv\frac{\Delta M_q}{\Gamma_q}=\left\{\begin{array}{cc}
0.771\pm0.012&(q=d)\\
{\cal O}(20)& (q=s)
\end{array}\right.
\end{equation}
denotes the $B^0_q$--$\bar B^0_q$ ``mixing parameter''.\footnote{Note that
$\Delta\Gamma_q/\Gamma_q$ is negative in the SM because of the
minus sign in (\ref{DGoG}).}
Consequently, we observe that $\Delta\Gamma_d/\Gamma_d\sim 10^{-2}$ is 
negligibly small, while $\Delta\Gamma_s/\Gamma_s\sim 10^{-1}$ may
be sizeable. For a discussion of the experimental status of the $B_q$ 
mixing parameters, the reader is referred to \cite{LEPBOSC,HFAG}.

\subsubsection{Time-Dependent Decay Rates}
The time evolution of initially, i.e.\ at $t=0$, pure $B^0_q$- and 
$\bar B^0_q$-meson states is given by
\begin{equation}
|B^0_q(t)\rangle=f_+^{(q)}(t)|B^{0}_q\rangle
+\alpha_qf_-^{(q)}(t)|\bar B^{0}_q\rangle
\end{equation}
and
\begin{equation}
|\bar B^0_q(t)\rangle=\frac{1}{\alpha_q}f_-^{(q)}(t)
|B^{0}_q\rangle+f_+^{(q)}(t)|\bar B^{0}_q\rangle,
\end{equation}
respectively, with
\begin{equation}\label{f-functions}
f_{\pm}^{(q)}(t)=\frac{1}{2}\left[e^{-i\lambda_+^{(q)}t}\pm
e^{-i\lambda_-^{(q)}t}\right].
\end{equation}
These time-dependent state vectors allow the calculation of the 
corresponding transition rates. To this end, it is useful to introduce
\begin{equation}\label{g-funct-1}
|g^{(q)}_{\pm}(t)|^2=\frac{1}{4}\left[e^{-\Gamma_{\rm L}^{(q)}t}+
e^{-\Gamma_{\rm H}^{(q)}t}\pm2\,e^{-\Gamma_q t}\cos(\Delta M_qt)\right]
\end{equation}
\begin{equation}\label{g-funct-2}
g_-^{(q)}(t)\,g_+^{(q)}(t)^\ast=\frac{1}{4}\left[e^{-\Gamma_{\rm L}^{(q)}t}-
e^{-\Gamma_{\rm H}^{(q)}t}+2\,i\,e^{-\Gamma_q t}\sin(\Delta M_qt)\right],
\end{equation}
as well as
\begin{equation}\label{xi-def}
\xi_f^{(q)}=e^{-i\Theta_{M_{12}}^{(q)}}
\frac{A(\bar B_q^0\to f)}{A(B_q^0\to f)},\quad
\xi_{\bar f}^{(q)}=e^{-i\Theta_{M_{12}}^{(q)}}
\frac{A(\bar B_q^0\to \bar f)}{A(B_q^0\to \bar f)}.
\end{equation}
Looking at (\ref{M12-calc}), we find
\begin{equation}\label{theta-def}
\Theta_{M_{12}}^{(q)}=\pi+2\mbox{arg}(V_{tq}^\ast V_{tb})-
\phi_{\mbox{{\scriptsize CP}}}(B_q),
\end{equation}
and observe that this phase depends on the chosen CKM and 
CP phase conventions specified in (\ref{CKM-trafo}) and (\ref{CP-def}), 
respectively. However, these dependences are cancelled through the 
amplitude ratios in (\ref{xi-def}), so that $\xi_f^{(q)}$ and 
$\xi_{\bar f}^{(q)}$ are {\it convention-independent} observables. 
Whereas $n'$ enters the functions in (\ref{f-functions}) through 
(\ref{lam-pm}), the dependence on this parameter is cancelled in 
(\ref{g-funct-1}) and (\ref{g-funct-2}) through the introduction of 
the {\it positive} mass difference $\Delta M_q$ (see (\ref{DeltaMq-def})). 
Combining the formulae listed above, we eventually arrive at the 
following transition rates for decays of initially, i.e.\ at $t=0$, 
present $B^0_q$ or $\bar B^0_q$ mesons:
\begin{equation}\label{rates}
\Gamma(\stackrel{{\mbox{\tiny (--)}}}{B^0_q}(t)\to f)
=\left[|g_\mp^{(q)}(t)|^2+|\xi_f^{(q)}|^2|g_\pm^{(q)}(t)|^2-
2\mbox{\,Re}\left\{\xi_f^{(q)}
g_\pm^{(q)}(t)g_\mp^{(q)}(t)^\ast\right\}
\right]\tilde\Gamma_f,
\end{equation}
where the time-independent rate $\tilde\Gamma_f$ corresponds to the 
``unevolved'' decay amplitude $A(B^0_q\to f)$, and can be calculated by 
performing the usual phase-space integrations. The rates into the 
CP-conjugate final state $\bar f$ can straightforwardly be obtained from 
(\ref{rates}) by making the substitutions
\begin{equation}
\tilde\Gamma_f  \,\,\,\to\,\,\, 
\tilde\Gamma_{\bar f},
\quad\,\,\xi_f^{(q)} \,\,\,\to\,\,\, 
\xi_{\bar f}^{(q)}.
\end{equation}

\subsection{CP Asymmetries}\label{subsec:CPasym}
A particularly simple -- but also very interesting  -- situation arises 
if we restrict ourselves to decays of neutral $B_q$ mesons 
into final states $f$ that are eigenstates of the CP operator, i.e.\
satisfy the relation 
\begin{equation}
({\cal CP})|f\rangle=\pm |f\rangle. 
\end{equation}
Consequently, we have $\xi_f^{(q)}=\xi_{\bar f}^{(q)}$ in this case, 
as can be seen in (\ref{xi-def}). Using the decay rates in (\ref{rates}), 
we find that the corresponding time-dependent CP asymmetry is given by
\begin{eqnarray}
{\cal A}_{\rm CP}(t)&\equiv&\frac{\Gamma(B^0_q(t)\to f)-
\Gamma(\bar B^0_q(t)\to f)}{\Gamma(B^0_q(t)\to f)+
\Gamma(\bar B^0_q(t)\to f)}\nonumber\\
&=&\left[\frac{{\cal A}_{\rm CP}^{\rm dir}(B_q\to f)\,\cos(\Delta M_q t)+
{\cal A}_{\rm CP}^{\rm mix}(B_q\to f)\,\sin(\Delta 
M_q t)}{\cosh(\Delta\Gamma_qt/2)-{\cal A}_{\rm 
\Delta\Gamma}(B_q\to f)\,\sinh(\Delta\Gamma_qt/2)}\right],\label{ee6}
\end{eqnarray}
with
\begin{equation}\label{CPV-OBS}
{\cal A}^{\mbox{{\scriptsize dir}}}_{\mbox{{\scriptsize CP}}}(B_q\to f)\equiv
\frac{1-\bigl|\xi_f^{(q)}\bigr|^2}{1+\bigl|\xi_f^{(q)}\bigr|^2},\qquad
{\cal A}^{\mbox{{\scriptsize mix}}}_{\mbox{{\scriptsize
CP}}}(B_q\to f)\equiv\frac{2\,\mbox{Im}\,\xi^{(q)}_f}{1+
\bigl|\xi^{(q)}_f\bigr|^2}.
\end{equation}
Because of the relation
\begin{equation}
{\cal A}^{\mbox{{\scriptsize dir}}}_{\mbox{{\scriptsize CP}}}(B_q\to f)=
\frac{|A(B^0_q\to f)|^2-|A(\bar B^0_q\to \bar f)|^2}{|A(B^0_q\to f)|^2+
|A(\bar B^0_q\to \bar f)|^2},
\end{equation}
this observable measures the direct CP violation in the decay
$B_q\to f$, which originates from the interference between different
weak amplitudes, as we have seen in (\ref{direct-CPV}). On the other
hand, the interesting {\it new} aspect of (\ref{ee6}) is due to 
${\cal A}^{\mbox{{\scriptsize mix}}}_{\mbox{{\scriptsize
CP}}}(B_q\to f)$, which originates from interference effects between 
$B_q^0$--$\bar B_q^0$ mixing and decay processes, and describes
``mixing-induced'' CP violation. Finally, the width difference 
$\Delta\Gamma_q$, which may be sizeable in the $B_s$-meson system, 
provides another observable,
\begin{equation}\label{ADGam}
{\cal A}_{\rm \Delta\Gamma}(B_q\to f)\equiv
\frac{2\,\mbox{Re}\,\xi^{(q)}_f}{1+\bigl|\xi^{(q)}_f
\bigr|^2},
\end{equation}
which is, however, not independent from ${\cal A}^{\mbox{{\scriptsize 
dir}}}_{\mbox{{\scriptsize CP}}}(B_q\to f)$ and 
${\cal A}^{\mbox{{\scriptsize mix}}}_{\mbox{{\scriptsize CP}}}(B_q\to f)$,
satisfying 
\begin{equation}\label{Obs-rel}
\Bigl[{\cal A}_{\rm CP}^{\rm dir}(B_q\to f)\Bigr]^2+
\Bigl[{\cal A}_{\rm CP}^{\rm mix}(B_q\to f)\Bigr]^2+
\Bigl[{\cal A}_{\Delta\Gamma}(B_q\to f)\Bigr]^2=1.
\end{equation}

In order to calculate the quantity $\xi_f^{(q)}$, which contains 
essentially all the information that is required for the evaluation
of the observables provided by the time-dependent CP asymmetry
introduced in (\ref{ee6}), we employ the low-energy effective 
Hamiltonian (\ref{e4}):
\begin{eqnarray}
\lefteqn{A(\bar B^0_q\to f)=\langle f\vert
{\cal H}_{\mbox{{\scriptsize eff}}}\vert\bar B^0_q\rangle}\nonumber\\
&&=\frac{G_{\mbox{{\scriptsize F}}}}{\sqrt{2}}\left[
\sum\limits_{j=u,c}V_{jr}^\ast V_{jb}\left\{\sum\limits_{k=1}^2
C_{k}(\mu)\langle f\vert Q_{k}^{jr}(\mu)\vert\bar B^0_q\rangle
+\sum\limits_{k=3}^{10}C_{k}(\mu)\langle f\vert Q_{k}^r(\mu)
\vert\bar B^0_q\rangle\right\}\right].~~~\mbox{}
\end{eqnarray}
On the other hand, we also have 
\begin{eqnarray}
\lefteqn{A(B^0_q\to f)=\langle f|
{\cal H}_{\mbox{{\scriptsize 
eff}}}^\dagger|B^0_q\rangle}\nonumber\\
&&\hspace*{-0.5truecm}=\frac{G_{\mbox{{\scriptsize F}}}}{\sqrt{2}}
\left[\sum\limits_{j=u,c}V_{jr}V_{jb}^\ast \left\{\sum\limits_{k=1}^2
C_{k}(\mu)\langle f\vert Q_{k}^{jr\dagger}(\mu)\vert B^0_q\rangle
+\sum\limits_{k=3}^{10}C_{k}(\mu)\langle f\vert Q_k^{r\dagger}(\mu)
\vert B^0_q\rangle\right\}\right].~~~\mbox{}
\end{eqnarray}
If we now insert the operator $({\cal CP})^\dagger({\cal CP})=\hat 1$ 
both after the $\langle f|$ and in front of the $|B^0_q\rangle$, we obtain
\begin{eqnarray}
\lefteqn{A(B^0_q\to f)=\pm e^{i\phi_{\mbox{{\scriptsize CP}}}
(B_q)}}\nonumber\\
&&\times\frac{G_{\mbox{{\scriptsize F}}}}{\sqrt{2}}\left[\sum\limits_{j=u,c}
V_{jr}V_{jb}^\ast\left\{\sum\limits_{k=1}^2
C_{k}(\mu)\langle f\vert Q_{k}^{jr}(\mu)\vert\bar B^0_q\rangle
+\sum\limits_{k=3}^{10}
C_{k}(\mu)\langle f\vert Q_{k}^r(\mu)\vert\bar B^0_q\rangle
\right\}\right],
\end{eqnarray}
where we have also applied the relation 
$({\cal CP})Q_k^{jr\dagger}({\cal CP})^\dagger=Q_k^{jr}$, and have 
furthermore taken (\ref{CP-def}) into account. Using then (\ref{xi-def})
and (\ref{theta-def}), we  observe that the phase-convention-dependent 
quantity $\phi_{\mbox{{\scriptsize CP}}}(B_q)$ cancels, and finally 
arrive at
\begin{equation}\label{xi-expr}
\xi_f^{(q)}=\mp\,e^{-i\phi_q}\,
\left[\frac{\sum\limits_{j=u,c}V_{jr}^\ast V_{jb}
\langle f|{\cal Q}^{jr}|\bar B^0_q\rangle}{\sum\limits_{j=u,c}
V_{jr}V_{jb}^\ast\langle f|{\cal Q}^{jr}|\bar B^0_q\rangle}\right].
\end{equation}
Here we have introduced the abbreviation
\begin{equation}
{\cal Q}^{jr}\equiv\sum\limits_{k=1}^2C_k(\mu)\,Q_k^{jr}+
\sum\limits_{k=3}^{10}C_k(\mu)\,Q_k^{r},
\end{equation}
and 
\begin{equation}\label{phiq-def}
\phi_q\equiv 2\,\mbox{arg} (V_{tq}^\ast V_{tb})=\left\{\begin{array}{cl}
+2\beta&\mbox{($q=d$)}\\
-2\delta\gamma&\mbox{($q=s$)}\end{array}\right.
\end{equation}
(where $\beta$ and $\delta\gamma$ are the angles in the
unitarity triangles illustrated in Fig.\ \ref{fig:UT}) is the
CP-violating weak phase introduced by $B_q^0$--$\bar B_q^0$ mixing
within the SM. 

Using the notation of (\ref{par-ampl}) and (\ref{par-ampl-CP}), we may
rewrite (\ref{xi-expr}) as follows:
\begin{equation}\label{xi-re}
\xi_f^{(q)}=\mp\, e^{-i\phi_q}\left[
\frac{e^{+i\varphi_1}|A_1|e^{i\delta_1}+
e^{+i\varphi_2}|A_2|e^{i\delta_2}}{
e^{-i\varphi_1}|A_1|e^{i\delta_1}+
e^{-i\varphi_2}|A_2|e^{i\delta_2}}\right].
\end{equation}
In analogy to the discussion of direct CP violation in 
Subsection~\ref{To-CP}, the calculation of $\xi_f^{(q)}$ suffers -- in
general -- from large hadronic uncertainties. However, if one CKM 
amplitude plays the {\it dominant} r\^ole in the transition 
$B_q\to f$, we obtain
\begin{equation}\label{xi-si}
\xi_f^{(q)}=\mp\, e^{-i\phi_q}\left[
\frac{e^{+i\phi_f/2}|M_f|e^{i\delta_f}}{e^{-i\phi_f/2}|M_f|e^{i\delta_f}}
\right]=\mp\, e^{-i(\phi_q-\phi_f)},
\end{equation}
and observe that the hadronic matrix element $|M_f|e^{i\delta_f}$ 
{\it cancels} in this expression. Since the requirements for 
direct CP violation discussed in the context of (\ref{direct-CPV}) 
are no longer satisfied, we have vanishing direct CP violation in 
this important special case, i.e.\ 
${\cal A}^{\mbox{{\scriptsize dir}}}_{\mbox{{\scriptsize CP}}}
(B_q\to f)=0$, which is also obvious from (\ref{CPV-OBS})
and (\ref{xi-si}). On the other hand, we still have mixing-induced
CP violation. In particular, 
\begin{equation}\label{Amix-simple}
{\cal A}^{\rm mix}_{\rm CP}(B_q\to f)=\pm\sin\phi
\end{equation}
is now governed by the CP-violating weak phase difference 
$\phi\equiv\phi_q-\phi_f$ and is {\it not} affected by hadronic 
uncertainties. The corresponding time-dependent CP asymmetry then
takes the simple form
\begin{equation}\label{Amix-t-simple}
\left.\frac{\Gamma(B^0_q(t)\to f)-
\Gamma(\bar B^0_q(t)\to \bar f)}{\Gamma(B^0_q(t)\to f)+
\Gamma(\bar B^0_q(t)\to \bar f)}\right|_{\Delta\Gamma_q=0}
=\pm\sin\phi\,\sin(\Delta M_q t),
\end{equation}
and allows an elegant determination of $\sin\phi$.

Let us next apply the formalism developed above to discuss decays of 
(neutral) $B$ mesons that are particularly important for the physics 
programme of the $B$ factories.

\begin{figure}
\begin{center}
\leavevmode
\epsfysize=4.0truecm 
\epsffile{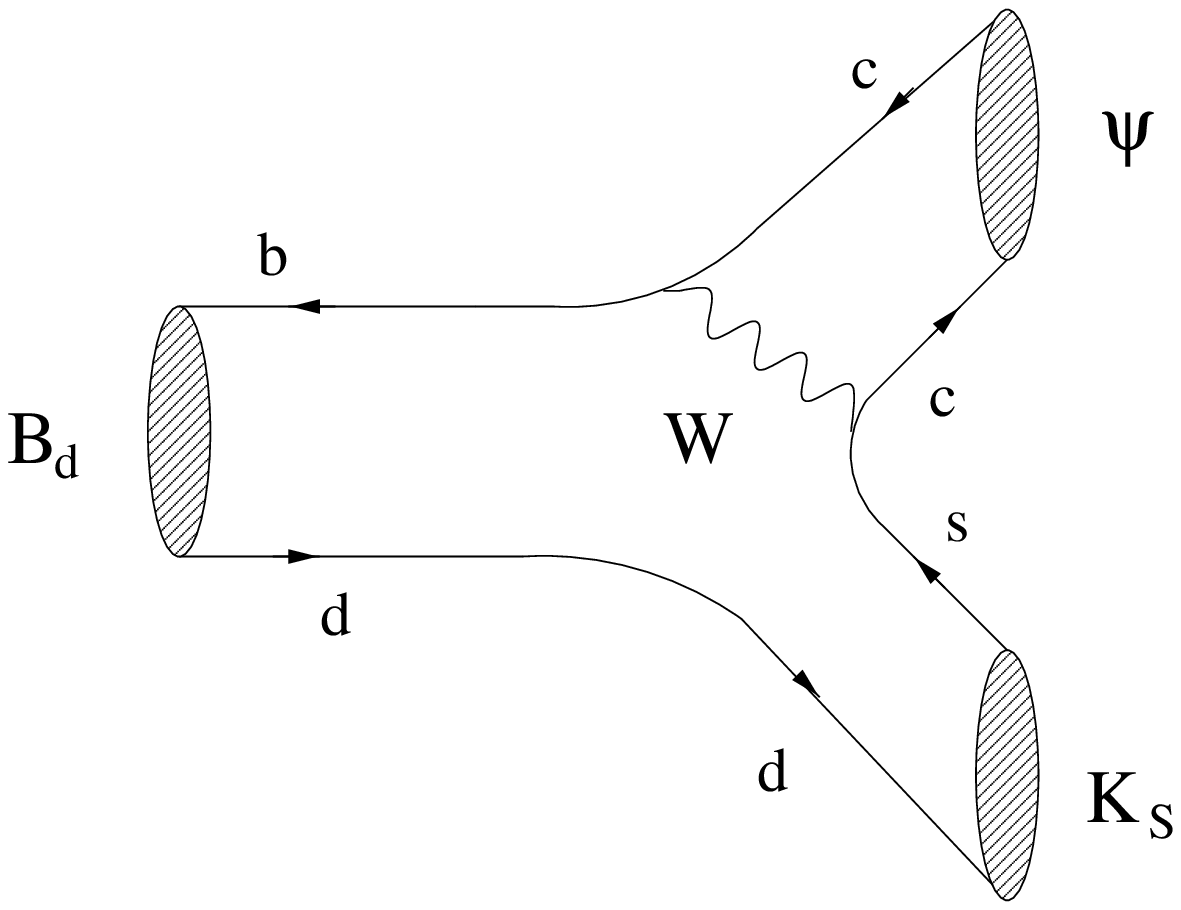} \hspace*{1truecm}
\epsfysize=4.0truecm 
\epsffile{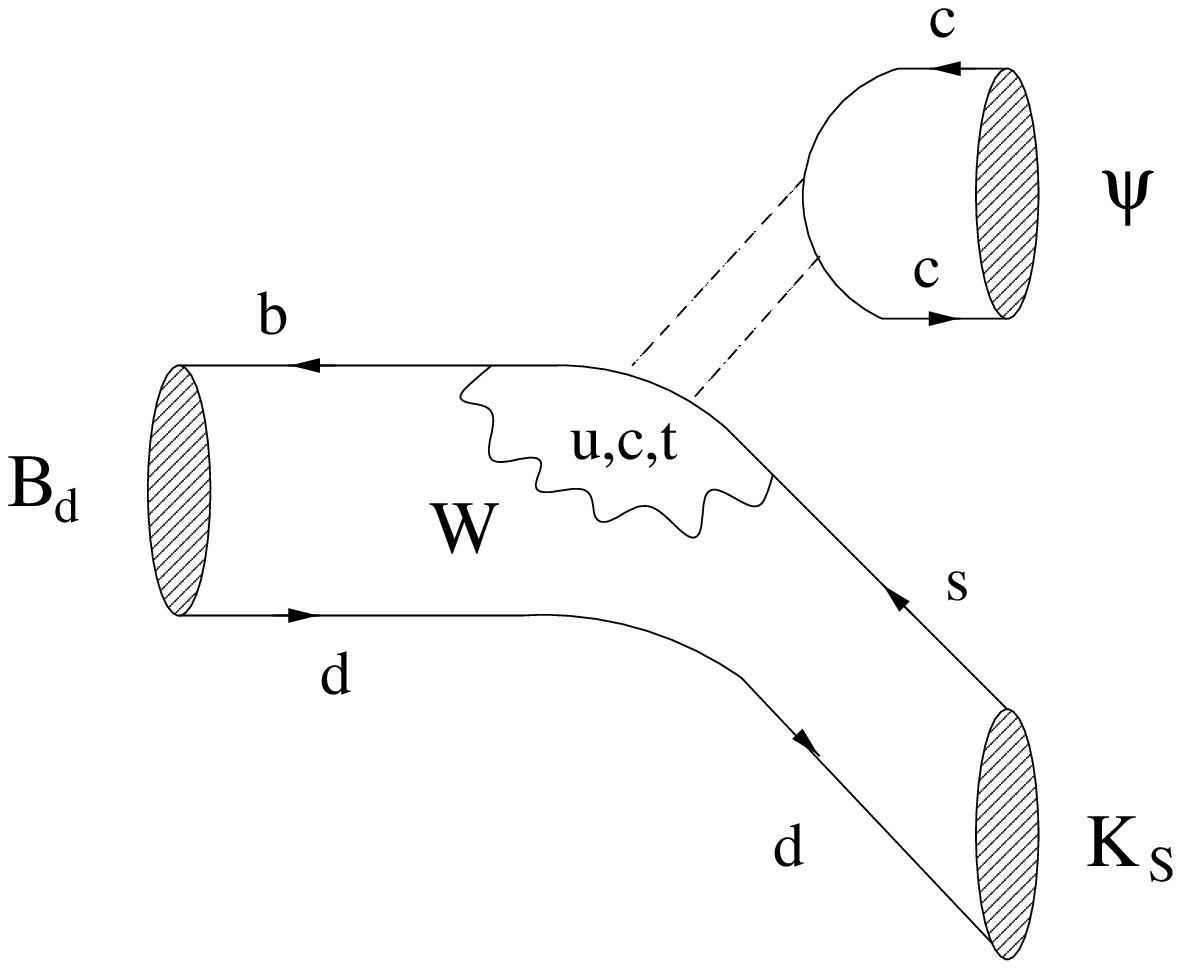}
\end{center}
\vspace*{-0.3truecm}
\caption{Feynman diagrams contributing to $B_d^0\to J/\psi K_{\rm S}$.
The dashed lines in the penguin topology represent a colour-singlet 
exchange.}\label{fig:BdPsiKS}
\end{figure}

\section{BENCHMARK MODES FOR THE {\boldmath$B$\unboldmath} 
FACTORIES}\label{sec:b-fact-bench}
\setcounter{equation}{0}
\boldmath\subsection{Exploring CP Violation 
through $B\to J/\psi K$}\unboldmath\label{subsec:BpsiK}
\subsubsection{Amplitude Structure and CP Asymmetries}
One of the most prominent $B$ decays is given by 
$B_d\to J/\psi K_{\rm S}$. If we take the CP parities 
of the $J/\psi$ and $K_{\rm S}$ into account,\footnote{Here we neglect
the tiny indirect CP violation in the neutral kaon system.} and note that
these mesons are produced in a $P$ wave with angular momentum $L=1$, 
we find that the final state of this transition is an eigenstate of the 
CP operator, with eigenvalue 
\begin{displaymath}
\underbrace{(+1)}_{J/\psi}\times\underbrace{(+1)}_{K_{\rm S}}
\times\underbrace{(-1)^1}_{L=1}=-1.
\end{displaymath}
As can be seen in Fig.~\ref{fig:BdPsiKS}, $B_d^0\to J/\psi K_{\rm S}$ 
originates from $\bar b\to\bar c c \bar s$ quark-level decays, and 
receives contributions from tree and penguin topologies (see the 
classification in Subsection~\ref{sec:class}). Consequently, we may 
write the decay amplitude as follows \cite{RF-BdsPsiK}:
\begin{equation}\label{Bd-ampl1}
A(B_d^0\to J/\psi K_{\rm S})=\lambda_c^{(s)}\left(A_{\rm T}^{c'}+
A_{\rm P}^{c'}\right)+\lambda_u^{(s)}A_{\rm P}^{u'}
+\lambda_t^{(s)}A_{\rm P}^{t'},
\end{equation}
where $A_{\rm T}^{c'}$ corresponds to the tree process in 
Fig.\ \ref{fig:BdPsiKS}, and the strong amplitudes $A_{\rm P}^{q'}$ 
describe the penguin topologies with internal $q$-quark exchanges 
($q\in\{u,c,t\})$, including QCD and EW penguins; the primes remind us 
that we are dealing with a $\bar b\to\bar s$ transition. Finally, the
\begin{equation}\label{lamqs-def}
\lambda_q^{(s)}\equiv V_{qs}V_{qb}^\ast
\end{equation}
are CKM factors. If we eliminate now $\lambda_t^{(s)}$ through 
(\ref{CKM-UT-Rel}) and apply the Wolfenstein parametrization, 
we straightforwardly arrive at
\begin{equation}\label{BdpsiK-ampl2}
A(B_d^0\to J/\psi K_{\rm S})\propto\left[1+\lambda^2 a e^{i\theta}
e^{i\gamma}\right],
\end{equation}
where
\begin{equation}
a e^{i\vartheta}\equiv\left(\frac{R_b}{1-\lambda^2}\right)
\left[\frac{A_{\rm P}^{u'}-A_{\rm P}^{t'}}{A_{\rm T}^{c'}+
A_{\rm P}^{c'}-A_{\rm P}^{t'}}\right]
\end{equation}
is a hadronic parameter that is a measure for the ratio of the
$B_d^0\to J/\psi K_{\rm S}$ penguin to tree contributions. Using 
the results derived in Subsection~\ref{subsec:CPasym}, we obtain
\begin{equation}\label{xi-BdpsiKS}
\xi_{\psi K_{\rm S}}^{(d)}=+e^{-i\phi_d}\left[\frac{1+
\lambda^2a e^{i\vartheta}e^{-i\gamma}}{1+\lambda^2a e^{i\vartheta}
e^{+i\gamma}}\right].
\end{equation}
Unfortunately, the parameter $a e^{i\vartheta}$ can only be estimated with 
large hadronic uncertainties. However, since it enters (\ref{xi-BdpsiKS}) 
in a doubly Cabibbo-suppressed way, its impact on the CP-violating
observables is practically negligible. We can put this statement 
on a more quantitative basis by making the plausible assumption that 
$a={\cal O}(\bar\lambda)={\cal O}(0.2)={\cal O}(\lambda)$,
where $\bar\lambda$ is a ``generic'' expansion parameter. Applying now
(\ref{CPV-OBS}) yields
\begin{eqnarray}
{\cal A}^{\mbox{{\scriptsize dir}}}_{\mbox{{\scriptsize
CP}}}(B_d\to J/\psi K_{\mbox{{\scriptsize S}}})&=&0+
{\cal O}(\overline{\lambda}^3)\label{Adir-BdpsiKS}\\
{\cal A}^{\mbox{{\scriptsize mix}}}_{\mbox{{\scriptsize
CP}}}(B_d\to J/\psi K_{\mbox{{\scriptsize S}}})&=&-\sin\phi_d +
{\cal O}(\overline{\lambda}^3) \, \stackrel{\rm SM}{=} \, -\sin2\beta+
{\cal O}(\overline{\lambda}^3).\label{Amix-BdpsiKS}
\end{eqnarray}
These expressions are one of the most important applications of the 
general features that we discussed in the context of
(\ref{xi-si})--(\ref{Amix-t-simple}).

\subsubsection{Experimental Status and Theoretical Uncertainties}
Looking at (\ref{Amix-BdpsiKS}), we observe that the mixing-induced 
CP violation in $B_d\to J/\psi K_{\mbox{{\scriptsize S}}}$ allows us to 
determine $\sin2\beta$ in an essentially {\it clean} manner \cite{bisa}.
Because of this feature, this transition is referred to as the
``golden'' mode to measure the angle $\beta$ of the UT. 
After important first steps by the OPAL, CDF and ALEPH collaborations, 
the $B_d\to J/\psi K_{\rm S}$ mode (and similar decays) eventually led, 
in 2001, to the observation of CP violation in the $B$ system 
\cite{babar-CP-obs,belle-CP-obs}. The current status of $\sin2\beta$ is 
given as follows:
\begin{equation}
\sin2\beta=\left\{\begin{array}{ll}
0.741\pm 0.067  \pm0.033  &
\mbox{(BaBar \cite{Babar-s2b-02})}\\
0.733\pm 0.057  \pm0.028  &
\mbox{(Belle \cite{Belle-s2b-03}),}
\end{array}\right.
\end{equation}
yielding the world average 
\begin{equation}\label{s2b-average}
\sin2\beta=0.736\pm0.049. 
\end{equation}
On the other hand, the CKM fits of the UT described in
Subsection~\ref{subsec:CKM-fits} imply the ranges in (\ref{UT-range}), where
the one for $\beta$ can be converted into
\begin{equation}
0.6\lsim\sin2\beta\lsim0.9, 
\end{equation}
which agrees well with the {\it direct} determination summarized in
(\ref{s2b-average}). 

As far as the theoretical accuracy of (\ref{Adir-BdpsiKS}) and 
(\ref{Amix-BdpsiKS}) is concerned, the corrections, which originate from 
the penguin contributions and are at most of 
${\cal O}(1\%)$,\footnote{In this case, the penguin topologies would 
{\it not} be suppressed with respect to the tree contributions, 
i.e.\ $a={\cal O}(1)$.} 
are not yet an issue. However, in the era of the 
LHC \cite{LHC-Book}, the experimental accuracy will be so tremendous 
that we have to start to deal with these terms. A possibility to 
control them is provided by the $B_s\to J/\psi K_{\rm S}$ channel, 
which can be combined with $B_d\to J/\psi K_{\rm S}$
through flavour-symmetry relations \cite{RF-BdsPsiK}.
Moreover, also the direct CP violation in the $B\to J/\psi K$ system 
allows us to probe such penguin effects \cite{RF-rev}, where a combined 
analysis of the neutral $B_d\to J/\psi K_{\rm S}$ and charged 
$B^\pm\to J/\psi K^\pm$ modes provides the whole picture \cite{FM-BpsiK}; 
the current $B$-factory data for the corresponding direct CP asymmetries 
are consistent with zero.
In a very recent analysis \cite{BMR}, this issue was also addressed from
a more theoretical point of view. The corresponding estimates lead to
tiny corrections at the $10^{-3}$ level, in accordance with the picture
developed in \cite{FM-BpsiK}.

Although the agreement between (\ref{s2b-average}) and the results of the
CKM fits is striking, it should not be forgotten that NP may -- in
principle -- nevertheless hide in 
${\cal A}_{\rm CP}^{\rm mix}(B_d\to J/\psi K_{\rm S})$. 
The point is that the key quantity is actually $\phi_d$, which is fixed 
through $\sin\phi_d=0.736\pm0.049$ up to a twofold ambiguity,
\begin{equation}\label{phid-det}
\phi_d=(47\pm 4)^\circ \, \lor \, (133 \pm 4)^\circ.
\end{equation}
Here the former solution would be in perfect agreement with CKM fits, 
implying $40^\circ\lsim2\beta\stackrel{\rm SM}{=}
\phi_d\lsim60^\circ$, whereas the latter would correspond to NP. 
The two solutions can be distinguished through a measurement of the 
sign of $\cos\phi_d$: 
in the case of $\cos\phi_d=+0.7>0$, we would conclude
$\phi_d=47^\circ$, whereas $\cos\phi_d=-0.7<0$ would point towards
$\phi_d=133^\circ$, i.e.\ to NP. There are several strategies on the 
market to resolve the twofold ambiguity in the extraction of $\phi_d$ 
\cite{ambig}. Unfortunately, they are rather challenging from a 
practical point of view. For instance, in the $B\to J/\psi K$ system, 
$\cos\phi_d$ can be extracted from the time-dependent angular distribution 
of the decay products of $B_d\to J/\psi[\to\ell^+\ell^-] 
K^\ast[\to\pi^0K_{\rm S}]$, if the sign of a hadronic parameter 
$\cos\delta$ involving a strong phase $\delta$ is fixed through 
factorization \cite{DDF2,DFN}.

\begin{figure}[t]
\begin{center}
\leavevmode
\epsfysize=3.8truecm 
\epsffile{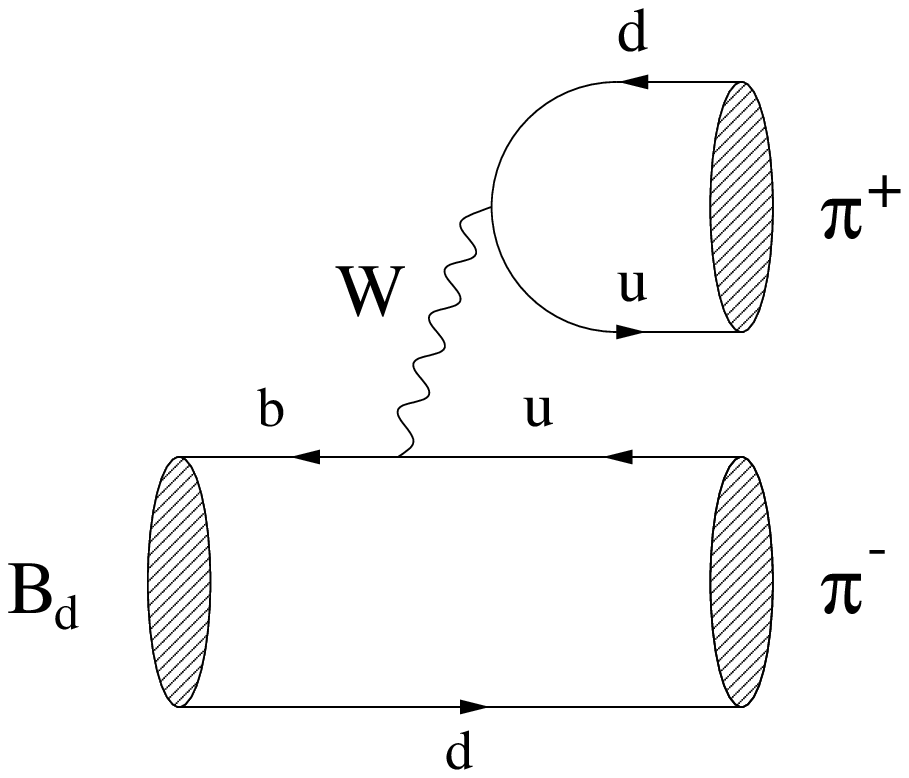} \hspace*{1truecm}
\epsfysize=4.0truecm 
\epsffile{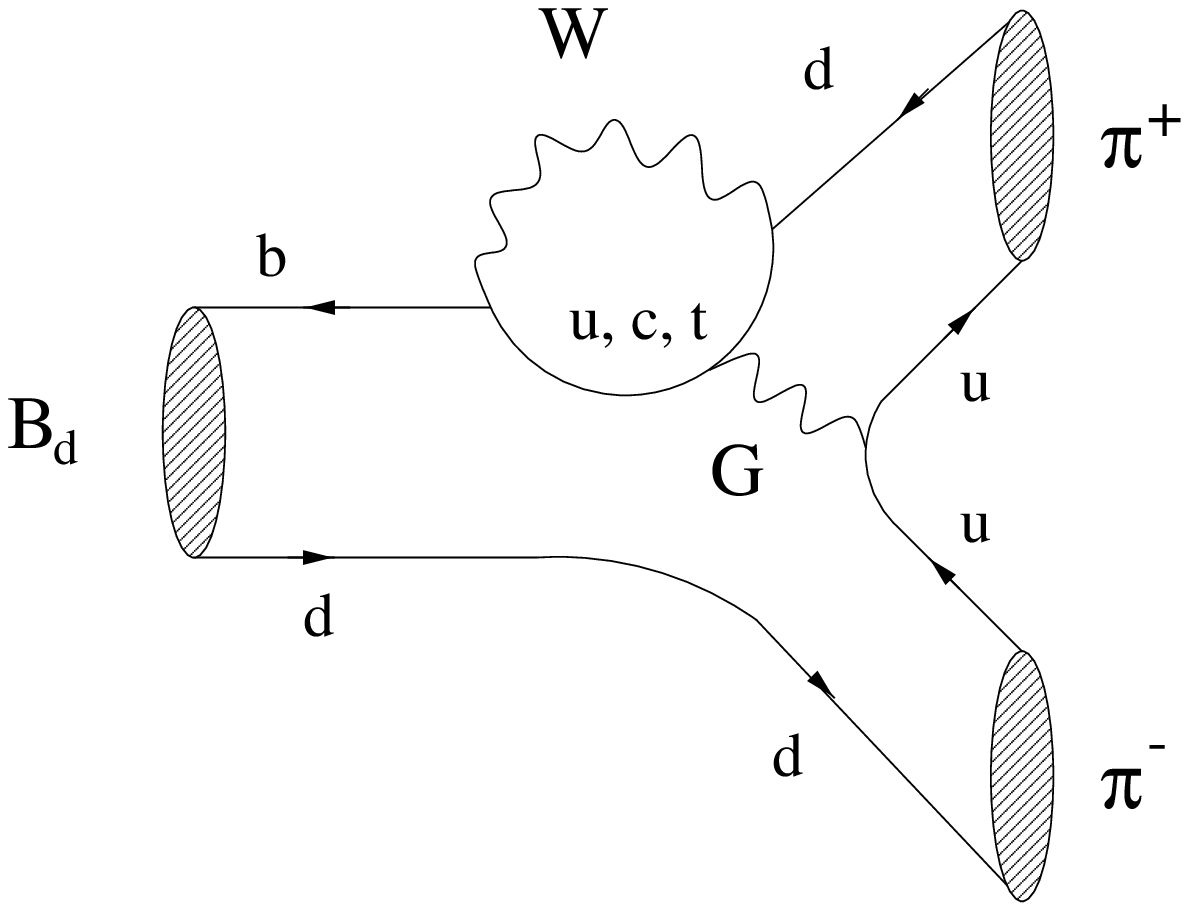}
\end{center}
\vspace*{-0.3truecm}
\caption{Feynman diagrams contributing to 
$B_d^0\to\pi^+\pi^-$.}\label{fig:bpipi}
\end{figure}

\boldmath\subsection{Exploring CP Violation through 
$B\to\pi\pi$}\unboldmath\label{ssec:Bpipi}
\subsubsection{Amplitude Structure and CP Asymmetries}
Another benchmark mode for the $B$ factories is the decay
$B_d^0\to\pi^+\pi^-$, which is a transition into a CP eigenstate 
with eigenvalue $+1$, and originates from $\bar b\to\bar u u \bar d$ 
quark-level processes, as can be seen in Fig.~\ref{fig:bpipi}. 
In analogy to (\ref{Bd-ampl1}), the decay amplitude 
can be written in the following form \cite{RF-BsKK}:
\begin{equation}
A(B_d^0\to\pi^+\pi^-)=
\lambda_u^{(d)}\left(A_{\rm T}^{u}+
A_{\rm P}^{u}\right)+\lambda_c^{(d)}A_{\rm P}^{c}+
\lambda_t^{(d)}A_{\rm P}^{t}.
\end{equation}
If we use again (\ref{CKM-UT-Rel}) to eliminate the CKM factor 
$\lambda_t^{(d)}=V_{td}V_{tb}^\ast$ and apply once more the Wolfenstein 
parametrization, we obtain 
\begin{equation}\label{Bpipi-ampl}
A(B_d^0\to\pi^+\pi^-)\propto\left[e^{i\gamma}-de^{i\theta}\right],
\end{equation}
where the hadronic parameter
\begin{equation}\label{D-DEF}
d e^{i\theta}\equiv\frac{1}{R_b}
\left[\frac{A_{\rm P}^{c}-A_{\rm P}^{t}}{A_{\rm T}^{u}+
A_{\rm P}^{u}-A_{\rm P}^{t}}\right]
\end{equation}
is a measure for the ratio of the $B_d\to\pi^+\pi^-$ penguin to tree 
amplitudes. The formalism discussed in Subsection~\ref{subsec:CPasym} 
then implies 
\begin{equation}\label{xi-Bdpipi}
\xi_{\pi^+\pi^-}^{(d)}=-e^{-i\phi_d}\left[\frac{e^{-i\gamma}-
d e^{i\theta}}{e^{+i\gamma}-d e^{i\theta}}\right].
\end{equation}
In contrast to the expression for the 
$B_d^0\to J/\psi K_{\rm S}$ counterpart given in (\ref{xi-BdpsiKS}), 
the hadronic parameter $d e^{i\theta}$, which suffers from large 
theoretical uncertainties, does {\it not} enter in (\ref{xi-Bdpipi}) 
in a doubly Cabibbo-suppressed way. This feature is at the basis of the
famous ``penguin problem'' in $B_d\to\pi^+\pi^-$, which was addressed
in many papers over the recent years (see, for instance, 
\cite{GL}--\cite{GLSS}). If we had negligible 
penguin contributions in this channel, i.e.\ $d=0$, the corresponding
CP-violating observables were simply given as follows:
\begin{eqnarray}
{\cal A}_{\rm CP}^{\rm dir}(B_d\to\pi^+\pi^-)&=&0 \\
{\cal A}_{\rm CP}^{\rm mix}(B_d\to\pi^+\pi^-)&=&\sin(\phi_d+2\gamma)
\stackrel{\rm SM}{=}\sin(\underbrace{2\beta+2\gamma}_{2\pi-2\alpha})
=-\sin 2\alpha.
\end{eqnarray}
Consequently, ${\cal A}_{\rm CP}^{\rm mix}(B_d\to\pi^+\pi^-)$ would
allow us to determine $\alpha$. However, in the general case of $d\not=0$, 
we obtain formulae with the help of (\ref{CPV-OBS}) and (\ref{xi-Bdpipi}), 
which are considerably more complicated:
\begin{equation}\label{Adir-Bpipi-gen}
{\cal A}_{\rm CP}^{\rm dir}(B_d\to\pi^+\pi^-)
=-\left[\frac{2d\sin\theta\sin\gamma}{1-
2d\cos\theta\cos\gamma+d^2}\right]
\end{equation}
\begin{equation}\label{Amix-Bpipi-gen}
{\cal A}_{\rm CP}^{\rm mix}(B_d\to\pi^+\pi^-)
=\frac{\sin(\phi_d+2\gamma)-2d\cos\theta
\sin(\phi_d+\gamma)+d^2\sin\phi_d}{1-2d\cos\theta\cos\gamma+d^2}.
\end{equation}
We observe that actually the phases $\phi_d$ and $\gamma$ 
enter directly in the $B_d\to\pi^+\pi^-$ observables, and not $\alpha$. 
Consequently, since $\phi_d$ can be fixed straightforwardly through
the mixing-induced CP violation in the ``golden'' mode 
$B_d\to J/\psi K_{\rm S}$, as we have seen in (\ref{Amix-BdpsiKS}), 
we may use $B_d\to\pi^+\pi^-$ to probe $\gamma$. This is advantageous 
to deal with penguins and possible NP effects.

\subsubsection{Experimental Status and the ``$B\to\pi\pi$ 
Puzzle''}\label{ssec:Bpipi-puzzle}
Measurements of the $B_d\to\pi^+\pi^-$ CP asymmetries are already 
available:
\begin{eqnarray}
{\cal A}_{\rm CP}^{\rm dir}(B_d\to\pi^+\pi^-)
&=&\left\{
\begin{array}{ll}
-0.19\pm0.19\pm0.05 & \mbox{(BaBar \cite{BaBar-Bpipi})}\\
-0.77\pm0.27\pm0.08 & \mbox{(Belle \cite{Belle-Bpipi})}
\end{array}
\right.\label{Adir-exp}\\
{\cal A}_{\rm CP}^{\rm mix}(B_d\to\pi^+\pi^-)
&=&\left\{
\begin{array}{ll}
+0.40\pm0.22\pm0.03& \mbox{(BaBar \cite{BaBar-Bpipi})}\\
+1.23\pm0.41 ^{+0.07}_{-0.08} & \mbox{(Belle \cite{Belle-Bpipi}).}
\end{array}
\right.\label{Amix-exp}
\end{eqnarray}
Unfortunately, the BaBar and Belle results are not fully consistent 
with each other, although both experiments point towards the same
signs, and the last BaBar update of 
${\cal A}_{\rm CP}^{\rm mix}(B_d\to\pi^+\pi^-)$ has moved 
towards Belle. In \cite{HFAG}, the Heavy Flavour Averaging Group (HFAG)
gave the following averages:
\begin{eqnarray}
{\cal A}_{\rm CP}^{\rm dir}(B_d\to\pi^+\pi^-)&=&-0.38\pm0.16
\label{Bpipi-CP-averages}\\
{\cal A}_{\rm CP}^{\rm mix}(B_d\to\pi^+\pi^-)&=&+0.58\pm0.20.
\label{Bpipi-CP-averages2}
\end{eqnarray}
Direct CP violation at this level would require large penguin 
contributions with large CP-conserving strong phases, as is evident from
(\ref{Adir-Bpipi-gen}). As we will see in Subsection~\ref{ssec:gam-Bpipi-piK}, 
the CP asymmetries in (\ref{Bpipi-CP-averages}) and 
(\ref{Bpipi-CP-averages2}) can be converted into the angle $\gamma$ of the 
UT, with a result around $65^\circ$, in remarkable accordance with 
the SM picture \cite{BFRS3,FlMa2}.

In addition to the decays $B_d\to\pi^+\pi^-$ and $B^\pm\to\pi^\pm\pi^0$, 
the $B$ factories have recently reported the observation of the 
$B_d\to\pi^0\pi^0$ channel, with the following CP-averaged branching ratios: 
\begin{equation}
\mbox{BR}(B_d\to\pi^0\pi^0)=\left\{
\begin{array}{ll}
(2.1\pm0.6\pm0.3)\times10^{-6} & \mbox{(BaBar \cite{Babar-Bpi0pi0})}\\
(1.7\pm0.6\pm0.2)\times10^{-6} & \mbox{(Belle \cite{Belle-Bpi0pi0});}
\end{array}
\right.
\end{equation}
CP-averaged branching ratios of this kind are generally defined through
\begin{equation}
\mbox{BR}\equiv\frac{1}{2}\left[\mbox{BR}(B\to f)+
\mbox{BR}(\bar B\to \bar f)\right].
\end{equation}
These measurements represent
quite a challenge for theory. For example, in a recent state-of-the-art 
calculation within QCD factorization \cite{Be-Ne}, a $B_d\to\pi^0\pi^0$
branching ratio that is about six times smaller is favoured, whereas the 
calculation of $B_d\to\pi^+\pi^-$ points towards a branching ratio about 
two times larger than the current experimental average. On the other hand, 
the calculation of $B^\pm\to\pi^\pm\pi^0$ reproduces the data rather well. 
This ``$B\to\pi\pi$ puzzle'' is reflected by the following quantities
\cite{BFRS3, BFRS2}:
\begin{eqnarray}
R_{+-}^{\pi\pi}&\equiv&2\left[\frac{\mbox{BR}(B^\pm\to\pi^\pm\pi^0)}{\mbox{BR}
(B_d\to\pi^+\pi^-)}\right]\frac{\tau_{B^0_d}}{\tau_{B^+}}=
2.12\pm0.37~~\mbox{}\label{Rpm-def}\\
R_{00}^{\pi\pi}&\equiv&2\left[\frac{\mbox{BR}(B_d\to\pi^0\pi^0)}{\mbox{BR}
(B_d\to\pi^+\pi^-)}\right]=0.83\pm0.23;\label{R00-def}
\end{eqnarray}
the central values calculated within QCD factorization give 
$R_{+-}^{\pi\pi}=1.24$ and $R_{00}^{\pi\pi}=0.07$ \cite{Be-Ne} . As was 
discussed in detail in \cite{BFRS3, BFRS2}, the $B\to\pi\pi$ puzzle 
can straightforwardly be accommodated within the SM through 
non-factorizable hadronic interference effects.\footnote{Similar conclusions
were also drawn very recently in \cite{ALP,CGRS}. In \cite{ALP}, also the
phenomenological implications of bounds on the UT that can be
derived from the CP-violating $B_d\to\pi^+\pi^-$ observables, as
pointed out in \cite{BS-bounds}, were discussed.} If we use
\begin{equation}\label{BFRS-gam}
\phi_d=(47\pm4)^\circ, \quad \gamma=(65\pm7)^\circ,
\end{equation}
as in the SM \cite{CKM-Book}, this analysis allows us to convert the 
$B\to\pi\pi$ data into certain hadronic parameters. In particular, we 
obtain 
\begin{equation}\label{d-det}
d=0.48^{+0.35}_{-0.22},\quad \theta=+(138^{+19}_{-23})^\circ,
\end{equation}
whereas QCD factorization favours $d\sim0.3$ and $\theta\sim180^\circ$. 
Moreover, the CP-violating observables of $B_d\to\pi^0\pi^0$ can be
predicted, with the result
\begin{equation}
{\cal A}_{\rm CP}^{\rm dir}(B_d\to\pi^0\pi^0)=-0.41^{+0.35}_{-0.17},\quad
{\cal A}_{\rm CP}^{\rm mix}(B_d\to\pi^0\pi^0)=-0.55^{+0.43}_{-0.45}.
\end{equation}
We shall return to $B_d\to\pi^+\pi^-$ in Subsection~\ref{ssec:BsKK},
in the context of $B_s\to K^+K^-$ \cite{RF-BsKK}.

\boldmath\subsection{Exploring CP Violation through 
$B\to \phi K$}\unboldmath
\subsubsection{Amplitude Structure and CP Asymmetries}
\begin{figure}[t]
\begin{center}
\leavevmode
\epsfysize=3.8truecm 
\epsffile{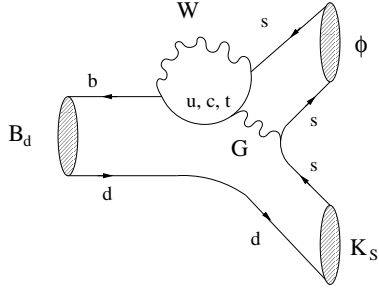} 
\end{center}
\vspace*{-0.6truecm}
\caption{Feynman diagrams contributing to 
$B_d\to \phi K_{\rm S}$.}\label{fig:BphiK}
\end{figure}

Another important mode for the testing of the KM mechanism of
CP violation is provided by $B_d\to \phi K_{\rm S}$, which is -- in
analogy to $B_d\to J/\psi K_{\rm S}$ -- a decay into a CP-odd final
state. As can be seen in Fig.~\ref{fig:BphiK}, $B_d^0\to \phi K_{\rm S}$
originates from $\bar b\to \bar s s \bar s$ quark-level processes,
i.e.\ is a pure penguin mode. Consequently, $B_d^0\to \phi K_{\rm S}$ 
and its charged counterpart $B^+\to \phi K^+$ are governed by QCD 
penguin topologies \cite{BphiK-old}, but also EW penguins have a 
sizeable impact because of the large top-quark mass \cite{RF-EWP,DH-PhiK}. 
Using the same notation as above, we may write the 
$B_d^0\to \phi K_{\rm S}$ decay amplitude within the SM as follows:
\begin{equation}
A(B_d^0\to \phi K_{\rm S})=\lambda_u^{(s)}\tilde A_{\rm P}^{u'}
+\lambda_c^{(s)}\tilde A_{\rm P}^{c'}+\lambda_t^{(s)}\tilde A_{\rm P}^{t'}.
\end{equation}
Applying now once more (\ref{CKM-UT-Rel}) to eliminate the CKM
factor $\lambda_t^{(s)}$, we obtain 
\begin{equation}
A(B_d^0\to \phi K_{\rm S})\propto
\left[1+\lambda^2 b e^{i\Theta}e^{i\gamma}\right],
\end{equation}
so that
\begin{equation}\label{xi-phiKS}
\xi_{\phi K_{\rm S}}^{(d)}=+e^{-i\phi_d}
\left[\frac{1+\lambda^2b e^{i\Theta}e^{-i\gamma}}{1+
\lambda^2b e^{i\Theta}e^{+i\gamma}}\right],
\end{equation}
with
\begin{equation}
b e^{i\Theta}=\left(\frac{R_b}{1-\lambda^2}\right)\left[
\frac{\tilde A_{\rm P}^{u'}-\tilde A_{\rm P}^{t'}}{\tilde A_{\rm P}^{c'}-
\tilde A_{\rm P}^{t'}}\right].
\end{equation}
The theoretical estimates of the hadronic parameter $b e^{i\Theta}$
suffer from large uncertainties. However, since this parameter enters 
(\ref{xi-phiKS}) in a doubly Cabibbo-suppressed way, we obtain the 
simple expressions
\begin{eqnarray}
{\cal A}_{\rm CP}^{\rm dir}(B_d\to \phi K_{\rm S})&=&0+
{\cal O}(\lambda^2)\\
{\cal A}_{\rm CP}^{\rm mix}(B_d\to \phi K_{\rm S})&=&-\sin\phi_d
+{\cal O}(\lambda^2),
\end{eqnarray}
where we made the plausible assumption that $b={\cal O}(1)$. On the other 
hand, the mixing-induced CP asymmetry of the ``golden'' mode $B_d\to J/\psi 
K_{\rm S}$ measures also $-\sin\phi_d$ (see (\ref{Amix-BdpsiKS})).
Consequently, we arrive at the following relation 
\cite{RF-rev,growo,loso,FM-BphiK}:
\begin{equation}\label{Bd-phiKS-SM-rel}
{\cal A}_{\rm CP}^{\rm mix}(B_d\to \phi K_{\rm S}) 
={\cal A}_{\rm CP}^{\rm mix}(B_d\to J/\psi K_{\rm S}) + 
{\cal O}(\lambda^2),
\end{equation}
which offers a very interesting test of the SM description of CP
violation. In order to obtain the whole picture and to search for 
NP systematically, it is useful to perform a combined analysis of the
neutral $B_d\to \phi K_{\rm S}$ and the charged $B^\pm \to \phi K^\pm$ 
modes \cite{FM-BphiK} (for a recent update, see \cite{BFRS3}).

\subsubsection{Experimental Status}
The experimental status of the CP-violating $B_d\to\phi K_{\rm S}$ 
observables is given as follows \cite{browder-talk}:\footnote{Note
that the very recent BaBar update in \cite{Babar-BphiK} uses also
$B_d\to \phi K_{\rm L}$ to extract the CP asymmetries of
$B^0_d\to \phi K^0$.}
\begin{equation}\label{aCP-Bd-phiK-dir}
{\cal A}_{\rm CP}^{\rm dir}(B_d\to \phi K_{\rm S})=
\left\{\begin{array}{ll}
+0.01\pm0.33\pm0.10 &\mbox{(BaBar \cite{Babar-BphiK})}\\
+0.15\pm0.29\pm0.07 &\mbox{(Belle \cite{Belle-BphiK})}
\end{array}\right.
\end{equation}
\begin{equation}\label{aCP-Bd-phiK-mix}
{\cal A}_{\rm CP}^{\rm mix}(B_d\to \phi K_{\rm S})=
\left\{\begin{array}{ll}
-0.47\pm0.34^{+0.06}_{-0.08} &\mbox{(BaBar \cite{Babar-BphiK})}\\
+0.96\pm0.50^{+0.11}_{-0.09} &\mbox{(Belle \cite{Belle-BphiK}),}
\end{array}\right.
\end{equation}
Since we have, on the other hand, 
${\cal A}_{\rm CP}^{\rm mix}(B_d\to J/\psi K_{\rm S})=-0.736 \pm 0.049$, 
we arrive at a puzzling situation, which has already stimulated many
speculations about NP effects in the decay $B_d\to\phi K_{\rm S}$ (see, for 
instance, \cite{BPhiK-NP}). However, because of the very unsatisfactory 
current experimental picture, it seems too early to get too excited by 
the possibility of having a violation of the SM relation 
(\ref{Bd-phiKS-SM-rel}). It will be very interesting to observe how the 
$B$-factory data will evolve, and to keep also an eye on 
$B_d\to \eta' K_{\rm S}$ and other related modes.

\subsection{Manifestations of New Physics}\label{ssec:NP}
\subsubsection{New-Physics Effects in $B^0_d$--$\bar B^0_d$ Mixing}
As we have seen in Subsection~\ref{ssec:BBbar-mix}, $B^0_d$--$\bar B^0_d$ 
mixing originates in the SM from box diagrams, which are characterized by 
the Inami--Lim function $S_0(x_t)$. Concerning the impact of NP, it may 
enter $B^0_d$--$\bar B^0_d$ mixing through new-particle exchanges in the 
loop diagrams shown in Fig.~\ref{fig:boxes}, or through new FCNC processes 
arising at the tree level. The impact on the mixing parameters is twofold:
\begin{itemize}
\item The mass difference of the mass eigenstates is generalized as
\begin{equation}\label{DMd-NP}
\Delta M_d = \Delta M_d^{\rm SM} + \Delta M_d^{\rm NP},
\end{equation}
so that the NP contribution would affect the determination of the 
UT side $R_t$ through (\ref{Rt-det}).
\item The CP-violating weak mixing phase is generalized as
\begin{equation}\label{phid-NP}
\phi_d = \phi_d^{\rm SM} + \phi_d^{\rm NP}=2\beta+\phi_d^{\rm NP},
\end{equation}
so that NP may enter the mixing-induced CP asymmetries through 
$\phi_d^{\rm NP}$.
\end{itemize}
On the basis of dimensional arguments borrowed from effective field
theory (see, for instance, \cite{FIM,FM-BpsiK}), and in specific NP
scenarios, the following pattern may -- in principle -- be
possible:
\begin{equation}
\Delta M_d^{\rm NP}/\Delta M_d^{\rm SM}\sim1,\quad 
\phi_d^{\rm NP}/\phi_d^{\rm SM}\sim1.
\end{equation}
The same is true for the case of $B^0_s$--$\bar B^0_s$ mixing, which may be 
significantly affected by NP as well.\footnote{Let us note that also 
$D^0$--$\bar D^0$ mixing offers an interesting probe to search for NP. 
Within the SM, this phenomenon is tiny, but it may be enhanced by 
the presence of NP. A similar comment applies to the CP-violating effects 
in $D$-meson decays. For a recent overview, we refer the reader to
\cite{D-rev}, and the references therein.}

\subsubsection{New-Physics Effects in Decay Amplitudes}
Another way for NP to manifest itself is through contributions to
decay amplitudes. If the decay does {\it not} arise at the tree level
in the SM, we may have potentially large NP effects. In particular,
NP may enter through new particles running in the loops, or through
new FCNC processes arising at the tree level. An important example 
for such decays is given by the $B\to\phi K$ system, which is governed 
by $\bar b\to \bar s s \bar s$ penguin processes, as we have seen above. 
On the basis of general dimensional arguments \cite{FM-BphiK}, and in 
specific NP scenarios \cite{BPhiK-NP}, significant effects may in fact 
arise in the $B\to\phi K$ amplitudes. The $B$-factory data may already
indicate the presence of such a kind of NP, although it is too early 
to draw definite conclusions on this exciting possibility. 

On the other hand, if a transition is dominated by a SM tree contribution, 
the impact of NP on the decay amplitude is generally small. An important 
example of this feature is given by the decay $B_d^0\to J/\psi K_{\rm S}$, 
which is governed by the $\bar b\to \bar c c \bar s$ process, arising
at the tree level in the SM. Generic dimensional arguments then indicate  
that we may have NP effects at the $B\to J/\psi K$ amplitude level of at 
most ${\cal O}(10\%)$ for a NP scale in the TeV regime. In order to 
search systematically for such effects, it is useful to perform a combined 
analysis of the neutral and charged $B\to J/\psi K$ modes, and to introduce 
appropriate observable combinations \cite{FM-BpsiK}; the current $B$-factory 
data do not indicate any anomaly (for a recent update, see \cite{BFRS3}). 
Since the determination of $\phi_d$ from the mixing-induced CP violation 
in $B_d\to J/\psi K_{\rm S}$ is very robust under NP, we may use the 
corresponding experimental result as an input for other studies of CP 
violation, as we have noted above.

\subsubsection{Back to the Status of the $B^0_d$--$\bar B^0_d$ Mixing
Phase $\phi_d$}
Let us now briefly come back to the two solutions for $\phi_d$
in (\ref{phid-det}). In this context, it is interesting to note that 
an upper bound on $\phi_d$ is implied by an upper bound on 
$R_b\propto|V_{ub}/V_{cb}|$, as can straightforwardly be seen in 
Fig.~\ref{fig:cont-scheme}. To be specific, we have 
\begin{equation}
\sin\beta_{\rm max}=R_b^{\rm max},
\end{equation}
which yields $(\phi_d)_{\rm max}^{\rm SM}\sim57^\circ$ for 
$R_b^{\rm max}\sim0.48$. Since the determination of $R_b$ from 
the semileptonic (tree-level) decays discussed in 
Subsection~\ref{subsec:semi-lept} is not expected to be sensitive to 
NP, $\phi_d\sim 133^\circ$ would require CP-violating NP contributions 
to $B^0_d$--$\bar B^0_d$ mixing. An interesting connection between 
the two solutions for $\phi_d$ and the UT angle $\gamma$ is 
provided by the CP asymmetries of $B_d\to\pi^+\pi^-$ \cite{FIM,FlMa2}. 
We shall return to this feature in Section~\ref{sec:Bs}.

\subsubsection{Models with Minimal Flavour Violation}\label{ssec:MFV}
An interesting scenario for NP is provided by the simplest class of 
extensions of the SM. It is represented by models with ``minimal flavour 
violation'' (MFV), which we may characterize as follows 
\cite{Buras-Cracow,B-MFV} (for alternative definitions, see 
\cite{MFV-DGIA,MFV-BEKU}):
\begin{itemize}
\item All flavour-changing transitions are still governed by the
CKM matrix, in particular no new phases.
\item The only relevant operators are those already present in the SM.
\end{itemize}
Important examples are the Two-Higgs-Doublet Model II, the constrained MSSM 
(if $\tan\overline{\beta}=v_2/v_1$ is not too large), and models with 
universal extra dimensions \cite{Buras-Cracow}. As was pointed out in 
\cite{B-MFV}, a ``universal unitarity triangle'' can be constructed for 
such MFV models with the help of those quantities that are not affected 
by the corresponding NP contributions. Following these lines, the ``true'' 
values of $\bar\rho$ and $\bar\eta$ can still be determined in a 
transparent manner, despite the presence of NP. 

Because of the items listed above, all SM expressions for decay 
amplitudes, as well as for particle--antiparticle mixing, can be 
generalized to the MFV models through a straightforward replacement 
of the initial Wilson coefficients for the renormalization-group 
evolution from $\mu={\cal O}(M_W)$ down to appropriate ``low-energy'' 
scales $\mu$ through characteristic NP coefficients. If we consider,
for example, $B^0_d$--$\bar B^0_d$ mixing, we just have to make the 
following substitution for the Inami--Lim function $S_0(x_t)$:
\begin{equation}\label{S-MFV}
S_0(x_t)\to S (v),
\end{equation}
where $v$, which equals $x_t=m_t^2/M_W^2$ in the SM, denotes collectively 
the parameters of a given MFV model. Note that the {\it same} 
short-distance function governs also 
$B^0_s$--$\bar B^0_s$ mixing, as well as $K^0$--$\bar K^0$ mixing,
so that it also enters the expression for the CP-violating 
observable $\varepsilon$.

Since no new phases appear in MFV models, one may think that the 
$B^0_d$--$\bar B^0_d$ mixing phase introduced in (\ref{phiq-def})
would not be affected in such scenarios. However, because of a 
subtlety, this is actually not the case \cite{BF-MFV}. If we look
at (\ref{M12-calc}), we observe that the sign of $S_0(x_t)$ enters
implicitly $\phi_d$; in (\ref{theta-def}) and (\ref{phiq-def}), we 
have actually used the fact that $S_0(x_t)$ is {\it positive}. However, since 
$S_0(x_t)$ is now replaced by $S(v)$, which needs no longer be positive, 
the expression for $\phi_d$ in (\ref{phiq-def}) is generalized as follows:
\begin{equation}\label{phid-def}
\phi_d=2\beta+\mbox{arg}(S (v)),
\end{equation}
so that $\phi_d^{\rm NP}$ in (\ref{phid-NP}) is either $0^\circ$ or 
$180^\circ$ for $S(v)>0$ or $S(v)<0$, respectively. Consequently, 
in the most general MFV case, the mixing-induced CP asymmetry of 
$B_d\to J/\psi K_{\rm S}$ is given by
\begin{equation}\label{apsiK-def}
-{\cal A}_{\rm CP}^{\rm mix}(B_d\to J/\psi K_{\rm S})
\equiv a_{\psi K_{\rm S}}={\rm sgn}(S (v))\sin2\beta.
\end{equation}
On the other hand, $\Delta M_d^{\rm NP}$ in (\ref{DMd-NP}) may have a 
significant impact on $\Delta M_d$. Similarly, also $\varepsilon$ may 
be affected. However, since the NP effects enter $\Delta M_d$ and 
$\varepsilon$ through the same generalized Inami--Lim function $S(v)$, 
we obtain correlations between these observables. In fact, the interplay 
between $B^0_d$--$\bar B^0_d$ mixing and $\varepsilon$ in the CKM fits
implies bounds on $\sin2\beta$ \cite{BuBu01}. Using (\ref{apsiK-def}), 
we may cancel the sign ambiguity due to ${\rm sgn}(S (v))$, and obtain 
the following lower bounds for $a_{\psi K_{\rm S}}$:
\begin{equation}
\left(a_{\psi K_{\rm S}}\right)_{\rm min}=\left\{
\begin{array}{ll}
0.42 & \mbox{($S(v)>0$ \cite{BuBu01})}\\
0.69 & \mbox{($S(v)<0$ \cite{BF-MFV}).}
\end{array}\right.
\end{equation}
Although these bounds were very exciting immediately after the first 
$B$-factory data for $a_{\psi K_{\rm S}}$ were announced, which favoured
rather small values, they are now not effective because of the world 
average given in (\ref{s2b-average}). We shall come back to NP scenarios 
with MFV in Subsections~\ref{ssec:rare-gen}--\ref{ssec:Kpinunu-detail}. 
For a very comprehensive discussion, we refer the reader to 
\cite{Buras-Cracow}.

\section{AMPLITUDE RELATIONS}\label{sec:A-REL}
\setcounter{equation}{0}
As we have noted in Subsection~\ref{To-CP}, amplitude relations 
offer another important tool to explore CP violation. Let us now have 
a closer look at the corresponding strategies, where we distinguish 
between the use of theoretically clean and flavour-symmetry 
relations.

\subsection{Theoretically Clean Relations}\label{subsec:clean-rel}
\subsubsection{$B^\pm\to K^\pm D$}
\begin{figure}
\begin{center}
\leavevmode
\epsfysize=3.8truecm 
\epsffile{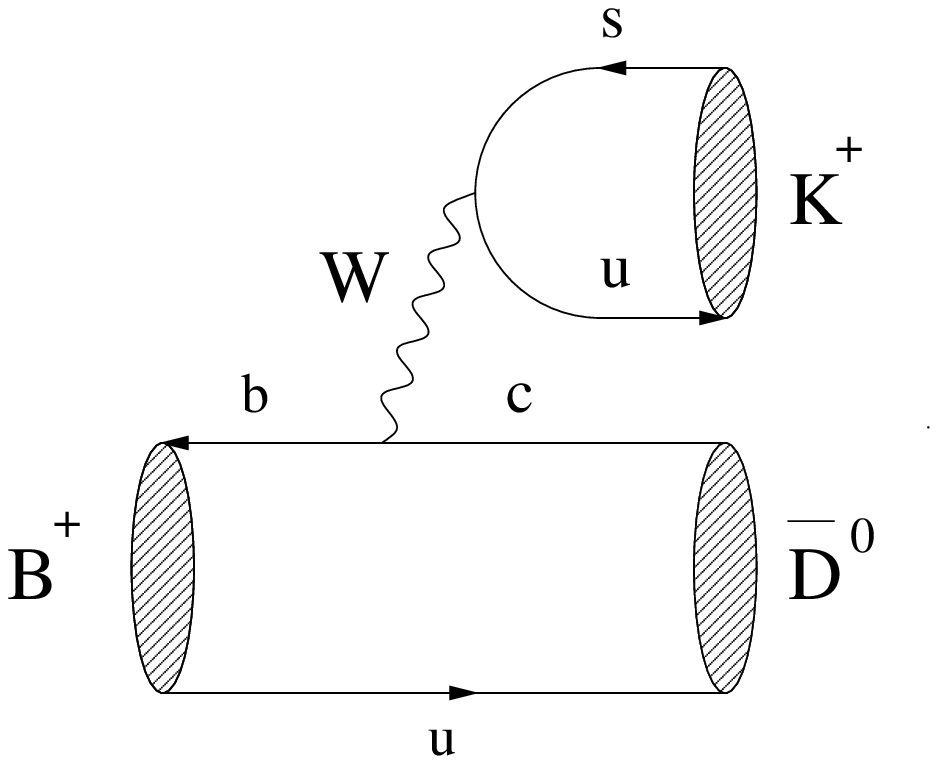} \hspace*{1truecm}
\epsfysize=4.5truecm 
\epsffile{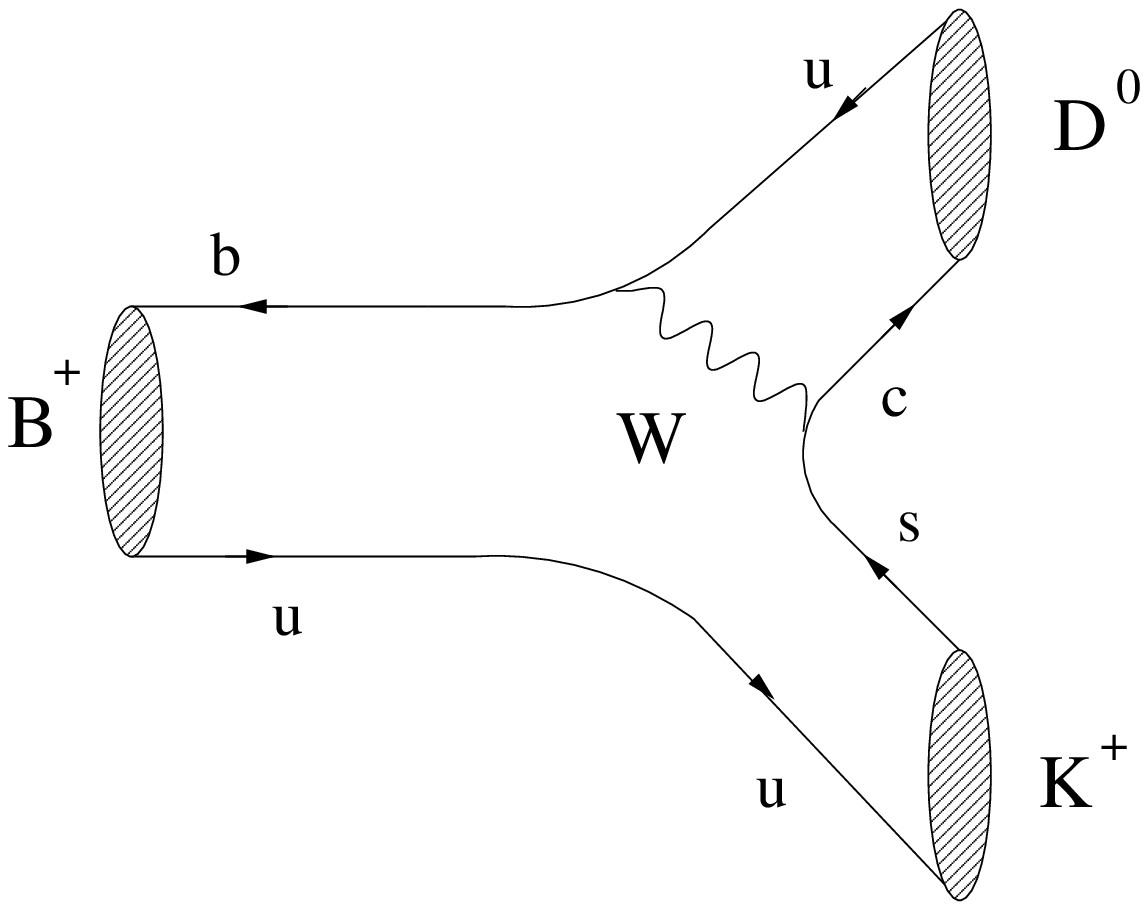}
\end{center}
\vspace*{-0.2truecm}
\caption{Feynman diagrams contributing to $B^+\to K^+\bar D^0$ and 
$B^+\to K^+D^0$. }\label{fig:BDK}
\end{figure}

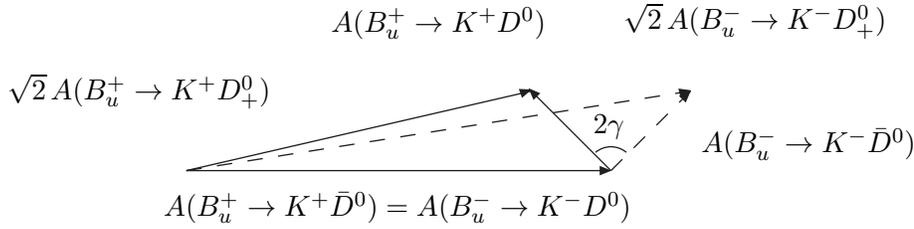
\begin{figure}
\vspace*{0.4truecm}
\begin{center}
\begin{picture}(320,75)(0,0)
\Line(50,10)(208,10) \ArrowLine(207,10)(209,10)
\DashLine(50,10)(240,40){6}\ArrowLine(237,39.5)(239,40)
\DashLine(210,10)(240,40){6}\ArrowLine(238,38)(239,39)
\Line(50,10)(180,40)\ArrowLine(177,39.5)(179,40.2)
\Line(210,10)(180,40)\ArrowLine(181,39)(180,40)
\Text(130,2)[t]{$A(B^+_u\to K^+\bar D^0)=A(B^-_u\to K^-D^0)$}
\Text(82,40)[r]{$\sqrt{2}\,A(B^+_u\to K^+ D_+^0)$}
\Text(245,20)[l]{$A(B^-_u\to K^-\bar D^0)$}
\Text(315,60)[br]{$\sqrt{2}\,A(B^-_u\to K^- D_+^0)$}
\Text(105,60)[lb]{$A(B^+_u\to K^+D^0)$}
\Text(210,26.5)[c]{$2\gamma$}\CArc(210,10)(9,50,130)
\end{picture}
\end{center}
\caption{The extraction of $\gamma$ from 
$B^\pm\to K^\pm\{D^0,\bar D^0,D^0_+\}$ 
decays.}\label{fig:BDK-triangle}
\end{figure}

The prototype of the strategies using theoretically clean amplitude 
relations is provided by $B^\pm \to K^\pm D$ decays \cite{gw}. Looking at 
Fig.~\ref{fig:BDK}, we observe that $B^+\to K^+\bar D^0$ and $B^+\to K^+D^0$ 
are pure ``tree'' decays. If we consider, in addition, the transition 
$B^+\to D^0_+K^+$, where $D^0_+$ denotes the CP 
eigenstate of the neutral $D$-meson system with eigenvalue $+1$,
\begin{equation}\label{ED85}
|D^0_+\rangle=\frac{1}{\sqrt{2}}\left[|D^0\rangle+
|\bar D^0\rangle\right],
\end{equation}
we obtain interference effects, which are described by
\begin{eqnarray}
\sqrt{2}A(B^+\to K^+D^0_+)&=&A(B^+\to K^+D^0)+
A(B^+\to K^+\bar D^0)\\
\sqrt{2}A(B^-\to K^-D^0_+)&=&A(B^-\to K^-\bar D^0)+
A(B^-\to K^-D^0).
\end{eqnarray}
These relations can be represented as two triangles in 
the complex plane. Since we have only to deal with tree-diagram-like 
topologies, we have moreover
\begin{eqnarray}
A(B^+\to K^+\bar D^0)&=&A(B^-\to K^-D^0)\\
A(B^+\to K^+D^0)&=&A(B^-\to K^-\bar D^0)\times e^{2i\gamma},
\end{eqnarray}
allowing a {\it theoretically clean} extraction of $\gamma$, as shown 
in Fig.~\ref{fig:BDK-triangle}. Unfortunately, these triangles are 
very squashed, since $B^+\to K^+D^0$ is colour-suppressed 
with respect to $B^+\to K^+\bar D^0$:
\begin{equation}\label{BDK-suppr}
\left|\frac{A(B^+\to K^+D^0)}{A(B^+\to K^+\bar D^0}\right|=
\left|\frac{A(B^-\to K^-\bar D^0)}{A(B^-\to K^-D^0}\right|\approx
\frac{1}{\lambda}\frac{|V_{ub}|}{|V_{cb}|}\times\frac{a_2}{a_1}
\approx 0.4\times0.3={\cal O}(0.1),
\end{equation}
where the phenomenological ``colour'' factors were introduced in
Subsection~\ref{ssec:ME-fact}. 

Another -- more subtle -- problem is related to the measurement of
$\mbox{BR}(B^+\to K^+D^0)$. From the theoretical point of view, 
$D^0\to K^-\ell^+\nu$ would be ideal to measure this tiny 
branching ratio. However, because of the huge background from 
semileptonic $B$ decays, we must rely on Cabibbo-allowed hadronic 
$D^0\to f_{\rm NE}$ decays, such as $f_{\rm NE}=\pi^+K^-$, $\rho^+K^-$,
$\ldots$, i.e.\ have to measure 
\begin{equation}\label{chain1}
B^+\to K^+D^0 \,[\to f_{\rm NE}].
\end{equation}
Unfortunately, we then encounter another decay path into the {\it same} 
final state $K^+ f_{\rm NE}$ through 
\begin{equation}\label{chain2}
B^+\to K^+\bar D^0 \,[\to f_{\rm NE}], 
\end{equation}
where BR$(B^+\to K^+\bar D^0)$ is {\it larger} than BR$(B^+\to K^+D^0)$
by a factor of ${\cal O}(10^2)$, while $\bar D^0\to f_{\rm NE}$ is doubly 
Cabibbo-suppressed, i.e.\ the corresponding branching ratio is suppressed
with respect to the one of $D^0\to f_{\rm NE}$ by a factor of 
${\cal O}(10^{-2})$. Consequently, we obtain interference effects of 
${\cal O}(1)$ between the decay chains in (\ref{chain1}) and (\ref{chain2}). 
If two different final states $f_{\rm NE}$ are considered, 
$\gamma$ could -- in principle -- be extracted \cite{ADS}, although
this determination would then be more involved than the original
triangle approach presented in \cite{gw}.

\subsubsection{$B_c^\pm\to D_s^\pm D$}
\begin{figure}
\begin{center}
\leavevmode
\epsfysize=4.4truecm 
\epsffile{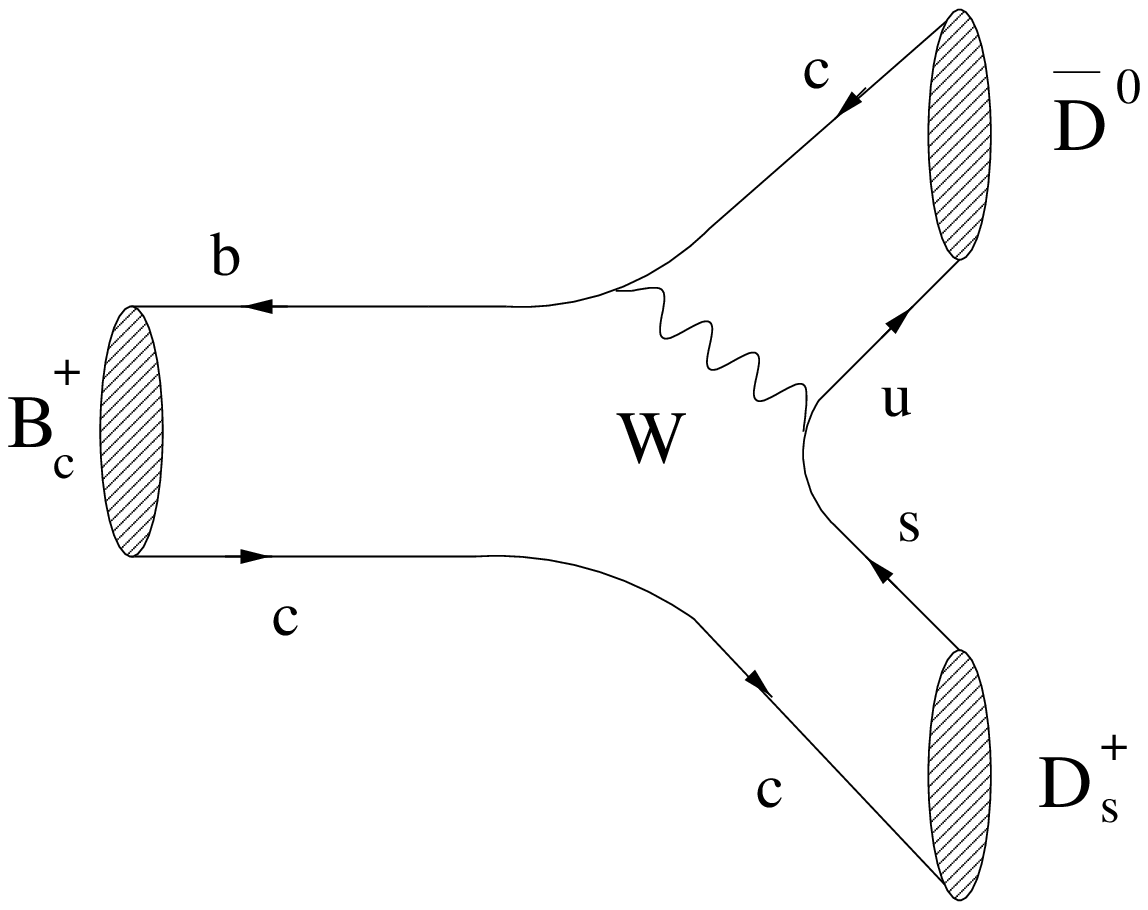} \hspace*{1.3truecm}
\epsfysize=4.0truecm 
\epsffile{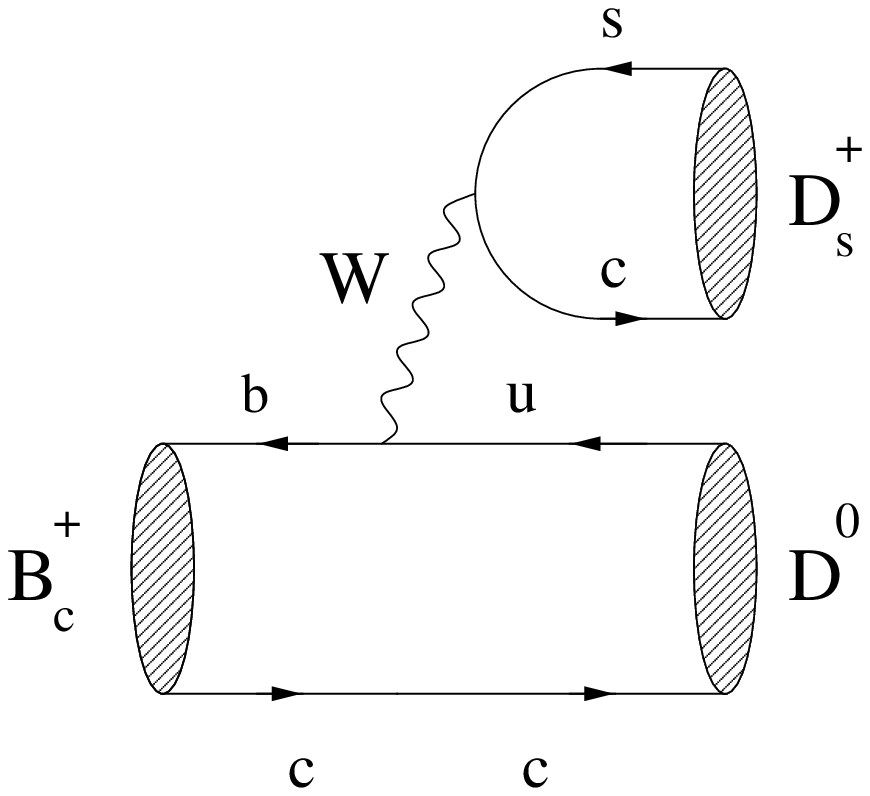}
\end{center}
\vspace*{-0.4truecm}
\caption{Feynman diagrams contributing to $B^+_c\to D_s^+\bar D^0$ and 
$B^+\to D_s^+D^0$. }\label{fig:BcDsD}
\end{figure}

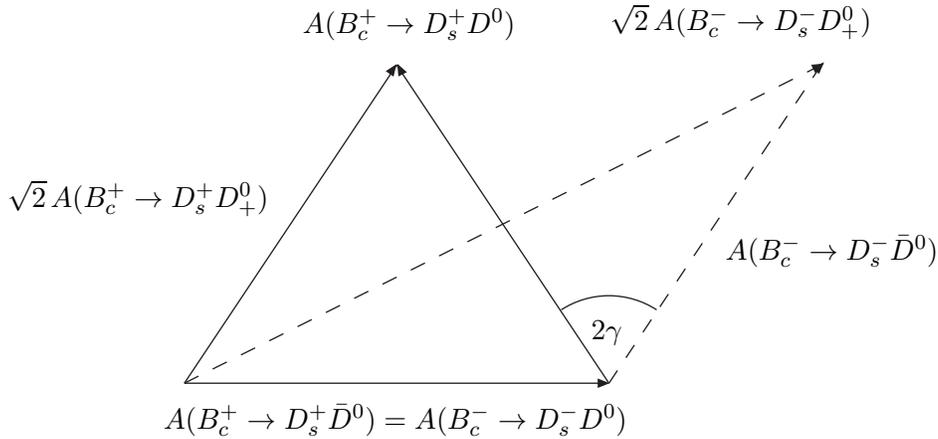
\begin{figure}
\vspace*{0.4truecm}
\begin{center}
\begin{picture}(320,150)(0,0)
\Line(50,10)(208,10) \ArrowLine(207,10)(209,10)
\DashLine(50,10)(290,130){6}\ArrowLine(288,129)(290,130)
\DashLine(210,10)(290,130){6}\ArrowLine(289,129)(290,130)
\Line(50,10)(130,130)\ArrowLine(128.3,128)(129,129)
\Line(210,10)(130,130)\ArrowLine(131.5,128)(131,129)
\Text(130,2)[t]{$A(B^+_c\to D_s^+\bar D^0)=A(B^-_c\to D^-_sD^0)$}
\Text(82,80)[r]{$\sqrt{2}\,A(B^+_c\to D_s^+ D_+^0)$}
\Text(255,60)[l]{$A(B^-_c\to D^-_s\bar D^0)$}
\Text(310,140)[br]{$\sqrt{2}\,A(B^-_c\to D_s^- D_+^0)$}
\Text(95,140)[lb]{$A(B^+_c\to D^+_sD^0)$}
\Text(210,29)[c]{$2\gamma$}\CArc(210,10)(33,57,123)
\end{picture}
\end{center}
\caption{The extraction of $\gamma$ from 
$B_c^\pm\to D^\pm_s\{D^0,\bar D^0,D^0_+\}$ decays.}\label{fig:triangles}
\end{figure}

In addition to the ``conventional'' $B_u^\pm$ mesons, there is yet another 
species of charged $B$ mesons, the $B_c$-meson system, which consists of
$B_c^+\sim c\overline{b}$ and $B_c^-\sim b\overline{c}$. These mesons were 
observed by the CDF collaboration through their decay 
$B_c^+\to J/\psi \ell^+ \nu$, with the following mass and lifetime 
\cite{CDF-Bc}:
\begin{equation}
M_{B_c}=(6.40\pm0.39\pm0.13)\,\mbox{GeV}, \quad
\tau_{B_c}=(0.46^{+0.18}_{-0.16}\pm 0.03)\,\mbox{ps}.
\end{equation}
Since a huge number of $B_c$ mesons ($\sim 10^{10}$/year) will 
be produced at LHCb \cite{LHC-Book}, the natural question of
whether also the charged $B_c$-meson system provides a triangle approach 
to determine $\gamma$ arises. Such a determination is actually offered by 
the decays $B_c^\pm\to D_s^\pm D$, which are the $B_c$-meson counterparts 
of the $B_u^\pm\to K^\pm D$ modes (see Fig.\ \ref{fig:BcDsD}), 
and satisfy the following amplitude relations \cite{masetti}:
\begin{eqnarray}
\sqrt{2}A(B_c^+\to D_s^+D^0_+)&=&A(B_c^+\to D_s^+D^0)+
A(B_c^+\to D_s^+\bar D^0)\\
\sqrt{2}A(B_c^-\to D_s^-D^0_+)&=&A(B_c^-\to D_s^-\bar D^0)+
A(B_c^-\to D_s^-D^0),
\end{eqnarray}
with
\begin{eqnarray}
A(B^+_c\to D_s^+\bar D^0)&=&A(B^-_c\to D_s^-D^0)\\
A(B_c^+\to D_s^+D^0)&=&A(B_c^-\to D_s^-\bar D^0)\times e^{2i\gamma}.
\end{eqnarray}
At first sight, everything is completely analogous to the $B_u^\pm\to K^\pm D$
case. However, there is an important difference \cite{fw}, 
which becomes obvious by comparing the Feynman diagrams shown in 
Figs.~\ref{fig:BDK} and \ref{fig:BcDsD}: in the $B_c^\pm\to D_s^\pm D$ 
system, the amplitude with the rather small CKM matrix element $V_{ub}$ 
is not colour-suppressed, while the larger element $V_{cb}$ comes with 
a colour-suppression factor. Therefore, we obtain
\begin{equation}\label{Bc-ratio1}
\left|\frac{A(B^+_c\to D_s^+ D^0)}{A(B^+_c\to D_s^+ 
\bar D^0)}\right|=\left|\frac{A(B^-_c\to D_s^-\bar D^0)}{A(B^-_c\to D_s^- 
D^0)}\right|\approx\frac{1}{\lambda}\frac{|V_{ub}|}{|V_{cb}|}
\times\frac{a_1}{a_2}\approx0.4\times 3 = {\cal O}(1),
\end{equation}
and conclude that the two amplitudes are similar in size. In contrast 
to this favourable situation, in the decays $B_u^{\pm}\to K^{\pm}D$, 
the matrix element $V_{ub}$ comes with the colour-suppression factor, 
resulting in a very stretched triangle. The extraction of $\gamma$ from 
the $B_c^\pm\to D_s^\pm D$ triangles is illustrated in 
Fig.~\ref{fig:triangles}, which should be compared with the
squashed $B^\pm_u\to K^\pm D$ triangles shown in 
Fig.\ \ref{fig:BDK-triangle}. Another important advantage is that 
the interference effects arising from $D^0,\bar D^0\to\pi^+K^-$ are 
practically unimportant for the measurement of BR$(B^+_c\to D_s^+ D^0)$ 
and BR$(B^+_c\to D_s^+ \bar D^0)$ since the $B_c$-decay amplitudes are 
of the same order of magnitude. Consequently, the $B_c^\pm\to D_s^\pm D$
decays provide -- from the theoretical point of view -- the ideal
realization of the ``triangle'' approach to determine $\gamma$. On
the other hand, the practical implementation still appears to 
be challenging, although detailed experimental feasibility studies for
LHCb are strongly encouraged. The corresponding branching ratios
were recently estimated in \cite{IKP}, with a pattern in accordance 
with (\ref{Bc-ratio1}).

\boldmath\subsection{Flavour-Symmetry Relations: 
$B\to\pi K$}\unboldmath\label{ssec:BpiK}
Let us now turn to amplitude relations that follow from the
flavour symmetries of the strong interactions, which are -- in contrast
to the relations discussed in Subsection~\ref{subsec:clean-rel} -- 
not theoretically clean, but are nevertheless very useful to explore
CP violation and to obtain insights into hadron dynamics. Here the 
prototype is provided by $B\to\pi K$ decays, which received a lot of 
attention in the $B$-physics community. Since a detailed discussion of 
the corresponding strategies is beyond the scope of these lectures, we 
address only their most important features and refer the interested 
reader to \cite{RF-Phys-Rep}, where also a comprehensive list of 
references can be found.

\subsubsection{General Features}
\begin{figure}
\begin{center}
\leavevmode
\epsfysize=4.5truecm 
\epsffile{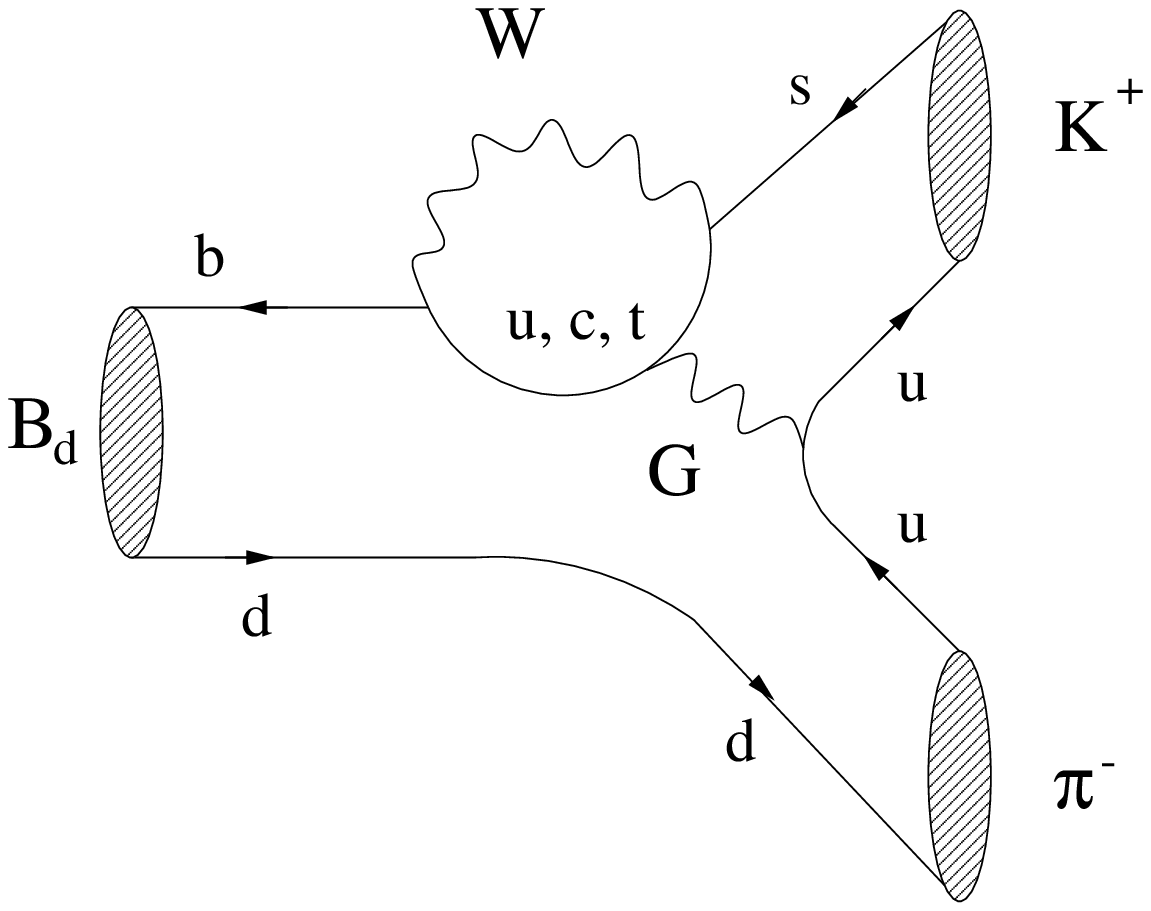} \hspace*{1.4truecm}
\epsfysize=4.0truecm 
\epsffile{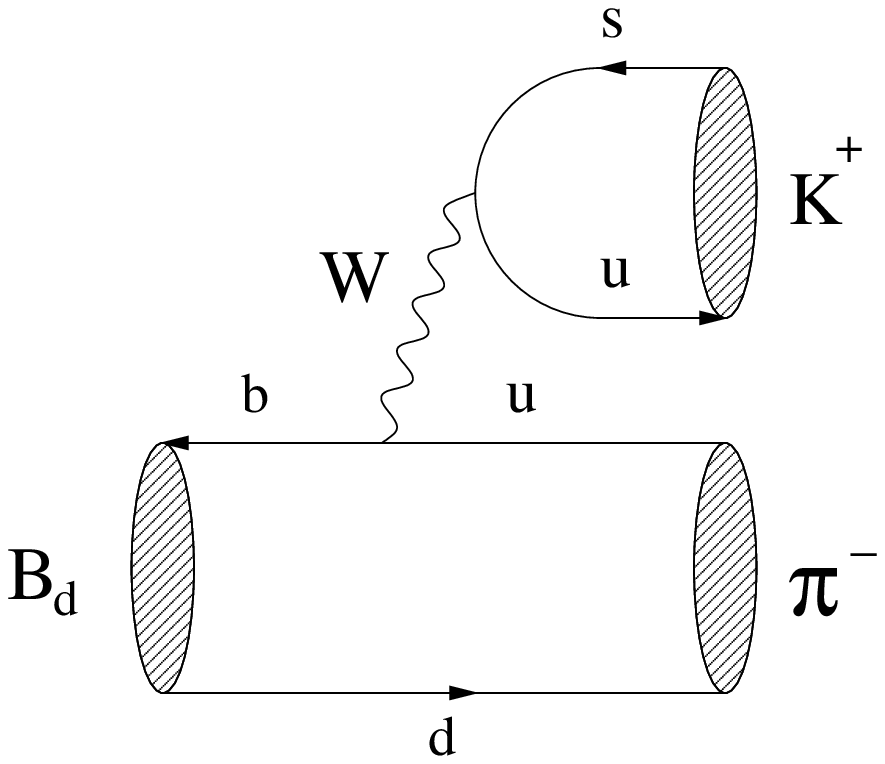}
\end{center}
\vspace*{-0.3truecm}
\caption{Feynman diagrams contributing to 
$B^0_d\to\pi^-K^+$.}\label{fig:BpiK-neutral}
\end{figure}

In order to get more familiar with the $B\to\pi K$ modes, let us 
consider the decay $B^0_d\to\pi^-K^+$. As can be seen in 
Fig.~\ref{fig:BpiK-neutral}, this channel receives contributions from 
penguin and tree topologies. Consequently, $B^0_d\to\pi^-K^+$ exhibits 
interference effects between the penguin and tree amplitudes, where the 
latter brings the angle $\gamma$ of the UT into the game. Because of 
the small ratio $|V_{us}V_{ub}^\ast/(V_{ts}V_{tb}^\ast)|\approx0.02$, 
the QCD penguin topologies play the dominant r\^ole in this decay, 
despite their loop suppression. The ratio of the tree to the penguin 
amplitudes is generically expected at the $20\%$ level. Interestingly, 
all $B\to\pi K$ modes are governed by their QCD penguin contributions. 
Because of the large top-quark mass, we have also to care about 
EW penguins:
\begin{itemize}
\item In the case of $B^0_d\to\pi^-K^+$ and $B^+\to\pi^+K^0$, these 
topologies contribute only in colour-suppressed form and are hence 
expected to play a minor r\^ole, thereby leading to contributions to 
the decay amplitudes of ${\cal O}(1\%)$.
\item On the other hand, EW penguins may also contribute to 
$B^+\to\pi^0K^+$ and $B^0_d\to\pi^0K^0$ in colour-allowed form, and 
may here even compete with the tree-diagram-like topologies, thereby 
leading to contributions to the decay amplitudes of ${\cal O}(20\%)$.
\end{itemize}
It can be shown that the isospin flavour symmetry of strong 
interactions implies the relation
\begin{displaymath}
\sqrt{2}A(B^+\to\pi^0K^+)+A(B^+\to\pi^+K^0)=
\sqrt{2}A(B^0_d\to\pi^0K^0)+A(B^0_d\to\pi^-K^+)
\end{displaymath}
\begin{equation}\label{BpiK-iso}
=-\Bigl[\underbrace{|T+C|e^{i\delta_{T+C}}e^{i\gamma}}_{\mbox{tree
topologies}}+
\underbrace{(P_{\rm ew}+P_{\rm ew}^{\rm C})}_{\mbox{EW penguins}}\Bigr]
\propto\left[e^{i\gamma}-q\right],
\end{equation}
where the $T$ ($P_{\rm ew}$) and $C$ ($P_{\rm ew}^{\rm C}$) denote the 
amplitudes of the colour-allowed and colour-suppressed tree (EW penguin) 
topologies, respectively, $\delta_{T+C}$ is a CP-conserving strong 
phase, and the factors of $\sqrt{2}$ originate from the wave functions
of the neutral pions. Note that the QCD penguin contributions cancel in 
this expression. A relation with an analogous phase structure can also 
be derived for the $B^+\to\pi^+ K^0$, $B_d^0\to\pi^- K^+$ system.

\subsubsection{Extraction of $\gamma$ and Strong Phases}
The $B\to\pi K$ observables allow us to determine the angle $\gamma$ 
of the UT. Because of the isospin relation in (\ref{BpiK-iso}), 
we may separately consider the following decay combinations to this 
end:
\begin{itemize}
\item The ``mixed'' system of the charged $B^\pm\to\pi^\pm K$ and
neutral $B_d\to\pi^\mp K^\pm$ modes \cite{PAPIII}--\cite{defan}.
\item The system of the charged $B^\pm\to\pi^\pm K$, 
$B^\pm\to\pi^0 K^\pm$ modes \cite{NR}--\cite{BF-neutral1}.
\item The system of the neutral $B_d\to\pi^0 K$, $B_d\to\pi^\mp K^\pm$
modes \cite{BF-neutral1,BF-neutral2}. 
\end{itemize}
Correspondingly, we introduce the following sets of observables 
\cite{BF-neutral1}:
\begin{equation}\label{mixed-obs}
\mbox{}\hspace*{0.4truecm}\left\{\begin{array}{c}R\\A_0\end{array}\right\}
\equiv\left[\frac{\mbox{BR}(B^0_d\to\pi^-K^+)\pm
\mbox{BR}(\bar B^0_d\to\pi^+K^-)}{\mbox{BR}(B^+\to\pi^+K^0)+
\mbox{BR}(B^-\to\pi^-\bar K^0)}\right]\frac{\tau_{B^+}}{\tau_{B^0_d}}
\end{equation}
\begin{equation}\label{charged-obs}
\left\{\begin{array}{c}R_{\rm c}\\A_0^{\rm c}\end{array}\right\}
\equiv2\left[\frac{\mbox{BR}(B^+\to\pi^0K^+)\pm
\mbox{BR}(B^-\to\pi^0K^-)}{\mbox{BR}(B^+\to\pi^+K^0)+
\mbox{BR}(B^-\to\pi^-\bar K^0)}\right]
\end{equation}
\begin{equation}\label{neut-obs}
\left\{\begin{array}{c}R_{\rm n}\\A_0^{\rm n}\end{array}\right\}
\equiv\frac{1}{2}\left[\frac{\mbox{BR}(B^0_d\to\pi^-K^+)\pm
\mbox{BR}(\bar B^0_d\to\pi^+K^-)}{\mbox{BR}(B^0_d\to\pi^0K^0)+
\mbox{BR}(\bar B^0_d\to\pi^0\bar K^0)}\right],
\end{equation}
where the $R_{\rm (c,n)}$ and $A_0^{\rm (c,n)}$ refer to the plus 
and minus signs, respectively; the factors of $2$ and $1/2$ are due 
to the wave functions of the neutral pions. In contrast to the
observables in (\ref{mixed-obs}), those in (\ref{charged-obs}) and 
(\ref{neut-obs}) are significantly affected by EW penguins. We will
return to this important feature below.

\begin{figure}
\vspace*{-0.0cm}
$$\hspace*{-0.3cm}
\epsfysize=0.18\textheight
\epsfxsize=0.20\textheight
\epsffile{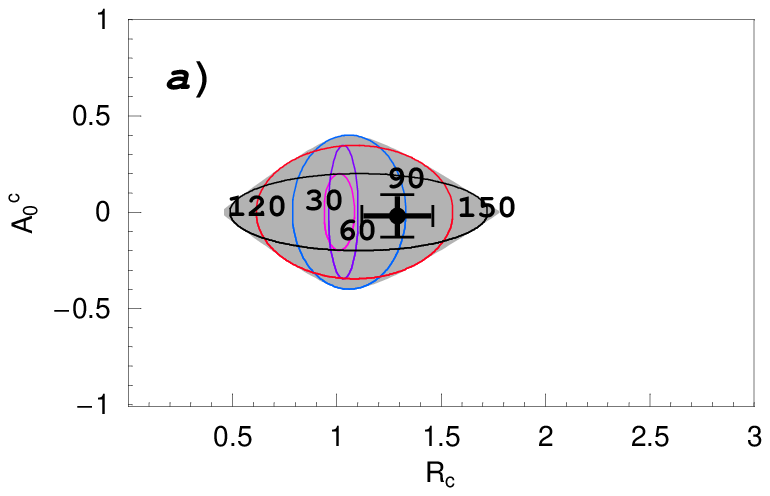} \hspace*{0.3cm}
\epsfysize=0.18\textheight
\epsfxsize=0.20\textheight
\epsffile{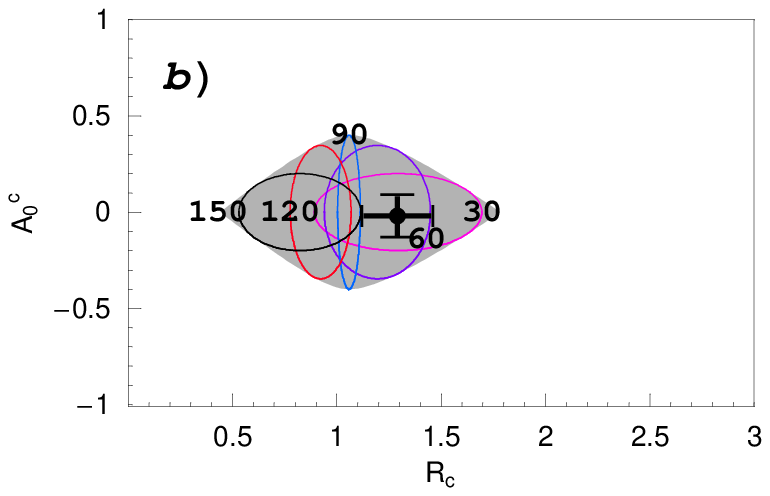}
$$
\vspace*{-0.6cm}
$$\hspace*{-0.3cm}
\epsfysize=0.18\textheight
\epsfxsize=0.20\textheight
\epsffile{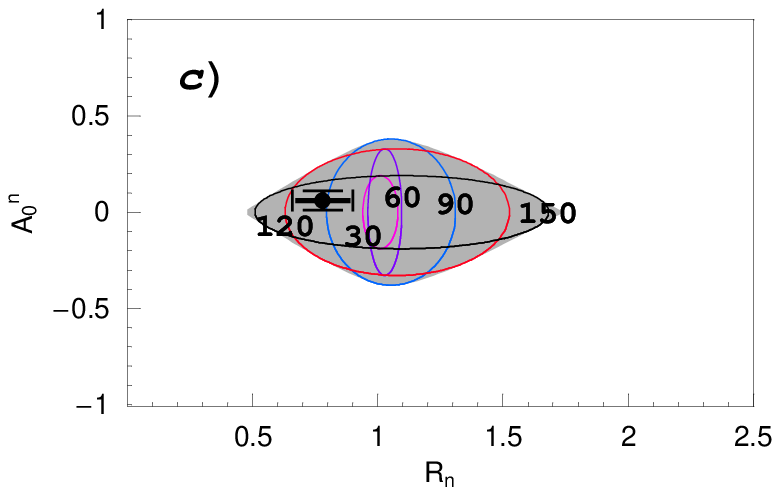} \hspace*{0.3cm}
\epsfysize=0.18\textheight
\epsfxsize=0.20\textheight
\epsffile{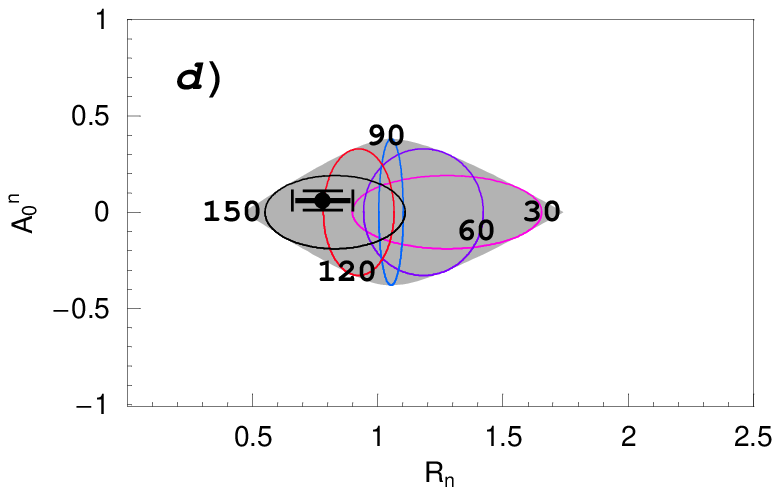}
$$
\vspace*{-0.8truecm}
\caption{The allowed regions in observable space of the charged
($r_{\rm c}=0.20$; (a), (b)) and neutral ($r_{\rm n}=0.19$; (c), (d))
$B\to \pi K$ systems for $q=0.69$: in (a) and (c), we show also the 
contours for fixed values of $\gamma$, whereas we give the curves 
arising for fixed values of $|\delta_{\rm c}|$ and $|\delta_{\rm n}|$ 
in (b) and (d), respectively.}\label{fig:BpiK-OS}
\end{figure}

As noted in \cite{BF-neutral1}, all three $B\to\pi K$ systems can be 
described by the same set of formulae, just making straightforward 
replacements of variables. Let us first focus on the charged and neutral 
$B\to\pi K$ systems. For the parametrization of their observables, we 
employ the isospin relation mentioned above, and assume that certain 
rescattering effects are small; large rescattering processes would be 
indicated by large direct CP violation in $B^\pm\to\pi^\pm K$, which
is not supported by the current $B$-factory average \cite{HFAG}:
\begin{equation}\label{ACPdirBpPipK}
{\cal A}_{\rm CP}^{\rm dir}(B^\pm\to\pi^\pm K)=-0.02\pm0.06,
\end{equation}
and by an enhancement of the $B\to KK$ branching ratios, which 
are already strongly constrained by the $B$-factory data as well 
(for detailed discussions, see \cite{RF-Phys-Rep,BFRS3}). Following these 
lines, we may write
\begin{equation}
R_{\rm c,n}=\mbox{function}(q,r_{\rm c,n},\delta_{\rm c,n},\gamma), \quad
A_0^{\rm c,n}=\mbox{function}(r_{\rm c,n},\delta_{\rm c,n},\gamma),
\end{equation}
where the parameters $q$, $r_{\rm c,n}$ and $\delta_{\rm c,n}$ have 
the following meaning:
\begin{itemize}
\item $q$ describes the ratio of the EW penguin to tree contributions
(see (\ref{BpiK-iso})), which can be determined with the help of $SU(3)$ 
flavour-symmetry arguments, yielding the following SM result \cite{BFRS3,NR}:
\begin{equation}\label{q-SM}
\left. q \right|_{\rm SM}=0.69\times\left[\frac{0.086}{|V_{ub}/V_{cb}|}
\right].
\end{equation}
\item The parameters $r_{\rm c,n}$ measure the ratios of the tree to 
QCD penguin topologies, and can be fixed through $SU(3)$ arguments 
and the data for BR$(B^\pm\to\pi^\pm\pi^0)$ \cite{GRL}, yielding
$r_{\rm c,n}\sim0.2$. 
\item The $\delta_{\rm c,n}$ are the CP-conserving strong phases 
between the tree and QCD penguin amplitudes.
\end{itemize}
Let us now consider either the charged or the neutral 
$B\to\pi K$ system. Since we may fix $q$ and the corresponding 
$r_{\rm c,n}$ with the help of $SU(3)$ flavour-symmetry relations, 
the observables $R_{\rm c,n}$ and $A_0^{\rm c,n}$ depend only 
on the two ``unknown'' parameters $\delta_{\rm c,n}$ and $\gamma$. If 
we vary them within their allowed ranges, 
i.e.\ $-180^\circ\leq \delta_{\rm c,n}\leq+180^\circ$ and 
$0^\circ\leq \gamma \leq180^\circ$, we obtain an allowed region in the 
$R_{\rm c,n}$--$A_0^{\rm c,n}$ plane \cite{FlMa2,FlMa1}. Should the
measured values of $R_{\rm c,n}$ and $A_0^{\rm c,n}$ fall outside this 
region, we would have an immediate signal for NP. On the 
other hand, should the measurements lie inside the allowed range, 
$\gamma$ and $\delta_{\rm c,n}$ could be extracted. The value of $\gamma$ 
thus obtained could then be compared with the results of other
strategies, whereas the strong phase $\delta_{\rm c,n}$ would offer
interesting insights into hadron dynamics. 

In Fig.~\ref{fig:BpiK-OS}, we show the allowed regions in the 
$R_{\rm c,n}$--$A_0^{\rm c,n}$ planes following \cite{FlMa2}, where 
the crosses represent the averages of the $B$-factory data. As can be read 
off from the contours in these figures, both the charged and the 
neutral $B\to \pi K$ data favour $\gamma\gsim90^\circ$, which would 
be in conflict with the results of the usual CKM fits, as summarized
in (\ref{UT-range}). Moreover, we observe that the charged modes point 
towards $|\delta_{\rm c}|\lsim90^\circ$ (QCD factorization predicts 
$\delta_{\rm c}$ to be close to $0^\circ$ \cite{BBNS,Be-Ne}), whereas 
the neutral decays prefer $|\delta_{\rm n}|\gsim90^\circ$. Since we 
do not expect $\delta_{\rm c}$ to differ significantly from $\delta_{\rm n}$, 
we arrive at a ``puzzling'' picture of the kind that was already pointed
out in the year 2000 \cite{BF-neutral2}, and was recently
reconsidered in \cite{BFRS3,BFRS2,Be-Ne,yoshikawa,GR-EWP,BFRS1}. 
In the experimental values
\begin{equation}\label{Rcn-exp}
R_{\rm c}=1.17\pm0.12,\quad 
R_{\rm n}=0.76\pm0.10,
\end{equation}
this puzzle is reflected in particular by $R_{\rm n}<1$, while $R_{\rm c}>1$,
as is now consistently favoured by the separate BaBar, Belle and CLEO data 
\cite{HFAG}. Concerning the mixed $B\to\pi K$ system, the data fall well into 
the SM region in observable space and do not indicate any ``anomalous'' 
behaviour \cite{FlMa2}.

\subsubsection{The ``$B\to\pi K$ Puzzle'' and Recent 
Developments}\label{ssec:BpiK-puzzle}
Since $R_{\rm c}$ and $R_{\rm n}$ are affected significantly by 
colour-allowed EW penguins, whereas such topologies may only 
contribute to $R$ in colour-suppressed form, the experimental pattern 
for these observables discussed above may be a manifestation of NP in 
the EW penguin sector \cite{Be-Ne,BF-neutral2,yoshikawa,GR-EWP,BFRS1}, 
offering an attractive avenue for physics beyond the SM to enter the 
$B\to\pi K$ system \cite{FM-NP,trojan}. In order to deal with
these effects quantitatively, we have to replace the parameter
in (\ref{q-SM}), which characterizes the EW penguins in the SM, through 
a generalized parameter $q$, which may, in particular, also be associated 
with a CP-violating NP phase $\phi$. 

A detailed analysis of the $B\to\pi K$ puzzle was recently performed 
in \cite{BFRS3,BFRS2}. The starting point is the $B\to\pi\pi$ puzzle
addressed in Subsection~\ref{ssec:Bpipi-puzzle}, which indicates that
another hadronic parameter of the neutral $B\to\pi K$ system, 
$\rho_{\rm n}e^{i\theta_{\rm n}}$, is not as small as na\"\i vely 
expected. However, using the $SU(3)$ flavour symmetry and plausible 
dynamical assumptions, it can be shown that we may fix all relevant 
hadronic $B\to\pi K$ parameters -- including CP-conserving strong 
phases -- through their $B\to\pi\pi$ counterparts, i.e.\ with the
help of the $B$-factory data. Moreover, if we complement $B_d\to\pi^+\pi^-$ 
with $B_d\to\pi^\mp K^\pm$, we may also extract $\gamma$ (see 
Subsection~\ref{ssec:gam-Bpipi-piK}), with a result in excellent 
accordance with the range for $\gamma$ in (\ref{BFRS-gam}). 
Since EW penguins play a 
very minor r\^ole in $B\to\pi\pi$ and $B_d\to\pi^\mp K^\pm$ decays, these 
modes -- and the parameters extracted from their observables -- are 
essentially unaffected by NP in the EW penguin sector. Having all 
$B\to\pi K$ parameters at hand, we may then analyse the $B\to\pi K$ 
system in the SM. 

As far as the ``mixed'' $B\to\pi K$ system is concerned, we obtain 
\begin{equation}\label{R-pred}
\left.R\right|_{\rm SM}=0.943^{+0.033}_{-0.026},
\end{equation}
which agrees well with the experimental result $R=0.91\pm0.07$
following from the averages compiled in \cite{HFAG}.
Additional information is provided by direct CP violation. Whereas
the direct CP asymmetry of $B^\pm\to\pi^\pm K$ vanishes within our 
working assumptions, in accordance with the experimental value 
in (\ref{ACPdirBpPipK}), we find
\begin{equation}
\left.{\cal A}_{\rm CP}^{\rm dir}(B_d\to\pi^\mp K^\pm)\right|_{\rm SM}
=0.140^{+0.139}_{-0.087},
\end{equation}
which is in agreement with the current $B$-factory average
${\cal A}_{\rm CP}^{\rm dir}(B_d\to\pi^\mp K^\pm)=+0.095\pm0.028$.

\begin{figure}
\vspace*{0.3truecm}
\begin{center}
\psfrag{Rn}{$R_{\rm n}$}\psfrag{Rc}{$R_{\rm c}$}
\psfrag{f}{\small $\!\!\phi$}\psfrag{expRegion}{exp.\ region}\psfrag{SM}{SM}
\psfrag{q069}{$q=0.69$}\psfrag{q122}{$q=1.22$}\psfrag{q175}{$q=1.75$}
\includegraphics[width=10.2cm]{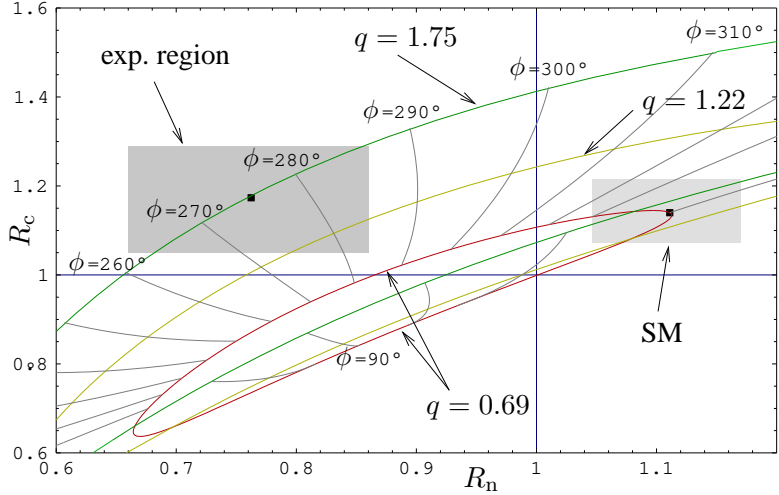}
\end{center}
\caption{The situation in the $R_{\rm n}$--$R_{\rm c}$ plane, 
where the current experimental and SM ranges are indicated in grey. 
We show also contours for the EW penguin parameters $q=0.69$, 
$q=1.22$ and $q=1.75$, with a NP phase 
$\phi \in [0^\circ,360^\circ]$.}\label{fig:Rn-Rc}
\end{figure}

In order to discuss the observables $R_{\rm n}$ and $R_{\rm c}$,
it is convenient to consider the $R_{\rm n}$--$R_{\rm c}$ plane. 
Since all hadronic parameters are fixed through the $B\to\pi\pi$ 
data, these observables now depend only on the EW penguin parameters $q$ 
and $\phi$, where the SM is described by (\ref{q-SM}), corresponding
to $\phi=0^\circ$. As can nicely be seen in 
Fig.~\ref{fig:Rn-Rc}, the pattern of the SM predictions
\begin{equation}\label{Rcn-SM}
\left.R_{\rm c}\right|_{\rm SM}=1.14^{+0.08}_{-0.07},\quad
\left.R_{\rm n}\right|_{\rm SM}=1.11^{+0.06}_{-0.07}
\end{equation}
is {\it not} in accordance with the current experimental picture
(\ref{Rcn-exp}), so that we are actually back at the $B\to\pi K$ 
puzzle described above. In this figure, we have also included 
various contours corresponding to different fixed values of $q$, 
where each point is parametrized through the value of
$\phi\in[0^\circ,360^\circ]$. We observe that we may in fact 
move to the experimental region for an enhanced value of $q\sim 1.8$ 
and $\phi\sim -90^\circ$, where in particular the large CP-violating 
phase is in stark contrast to the SM. In order to put these observations 
on a more quantitative level, we may convert the experimental values 
of $R_{\rm c}$ and $R_{\rm n}$ into values of $q$ and $\phi$, with 
the following result:
\begin{equation}\label{q-det}
q=1.75^{+1.27}_{-0.99},\quad  \phi=-(85^{+11}_{-14})^\circ.
\end{equation}
Because of the large, non-vanishing value of $\phi$, this scenario
of NP would require new sources for CP violation, i.e.\ would {\it not}
belong to the simple class of MFV models specified in 
Subsection~\ref{ssec:MFV}. As far as direct CP violation in 
$B^\pm\to\pi^0 K^\pm$ is concerned, we obtain
\begin{equation}
{\cal A}_{\rm CP}^{\rm dir}(B^\pm\to\pi^0 K^\pm)=0.04^{+0.37}_{-0.28}
\end{equation}
in our NP scenario, in accordance with the experimental 
number ${\cal A}_{\rm CP}^{\rm dir}(B^\pm\to\pi^0 K^\pm)=0.00\pm0.07$.
As was pointed out in \cite{PAPIII}, also the CP asymmetries of 
$B_d\to\pi^0 K_{\rm S}$ are an important tool to explore the KM mechanism 
of CP violation, where the SM corresponds (for $\rho_{\rm n}=0$)
to the relations
\begin{equation}
{\cal A}_{\rm CP}^{\rm dir}(B_d\to\pi^0 K_{\rm S})=0, \quad
{\cal A}_{\rm CP}^{\rm mix}(B_d\to\pi^0 K_{\rm S})=-\sin\phi_d=
{\cal A}_{\rm CP}^{\rm mix}(B_d\to J/\psi K_{\rm S}),
\end{equation}
in analogy to (\ref{Bd-phiKS-SM-rel}).
Recently, the BaBar collaboration reported the following results 
\cite{BABAR-Bdpi0KS}:
\begin{equation}\label{Adir-Bdpi0KS}
{\cal A}_{\rm CP}^{\rm dir}(B_d\to\pi^0 K_{\rm S})=
0.40^{+0.27}_{-0.28} \pm 0.09,\quad
{\cal A}_{\rm CP}^{\rm mix}(B_d\to\pi^0 K_{\rm S})=
-0.48^{-0.38}_{+0.47} \pm 0.06.
\end{equation}
Moreover, also a measurement of the direct CP asymmetry of 
the $B_d^0\to \pi^0K^0$ channel is available \cite{HFAG}:
\begin{equation}
{\cal A}_{\rm CP}^{\rm dir}(B_d^0\to\pi^0 K^0)=-0.03\pm0.36\pm0.09,
\end{equation}
which is supposed to agree with the direct CP asymmetry in
(\ref{Adir-Bdpi0KS}). Consequently, these experimental numbers are
expected to change significantly in the future. On the other hand,
the $B\to\pi\pi,\pi K$ analysis described above yields the predictions
\begin{equation}
{\cal A}_{\rm CP}^{\rm dir}(B_d\to\pi^0 K_{\rm S})=
+0.05^{+0.24}_{-0.29},\quad 
{\cal A}_{\rm CP}^{\rm mix}(B_d\to\pi^0 K_{\rm S})=
-0.99^{+0.04}_{-0.01}.
\end{equation}
The measurement of these CP asymmetries will allow an interesting 
test of the NP scenario of enhanced EW penguins with a large 
CP-violating phase that is suggested by the $B\to\pi K$ puzzle.
In this respect, it is important to consider also rare $B$ and $K$ decays,
which offer particularly sensitive probes for the exploration
of this kind of NP. We shall return to the corresponding NP effects 
in Subsection~\ref{ssec:NP-rare}, where we will also briefly 
address the impact on $\mbox{Re}(\varepsilon'/\varepsilon)$,
$B_d\to J/\psi K_{\rm S}$ and $B_d\to\phi K_{\rm S}$.

\section{THE {\boldmath$B_s$-MESON\unboldmath} SYSTEM}\label{sec:Bs}
\setcounter{equation}{0}
\subsection{General Features}
\subsubsection{Comparison of the $B_d$ and $B_s$ Systems}
At the $e^+e^-$ $B$ factories operating at the $\Upsilon(4S)$ resonance
(BaBar and Belle), the $B_s$-meson system is not accessible since 
$\Upsilon(4S)$ states decay only into $B_{u,d}$ but not into 
$B_s$ mesons.\footnote{Operating
these machines on the $\Upsilon(5S)$ resonance would also allow the
production of $B_s$ mesons.} On the other hand, plenty of $B_s$ mesons will 
be produced at hadron colliders. Consequently, these particles are the 
``El Dorado'' for $B$-decay studies at run II of the Tevatron 
\cite{TEV-Book}, and later on at the LHC \cite{LHC-Book}. 
There are important differences between the 
$B_d$ and $B_s$ systems:
\begin{itemize}
\item The $B^0_s$--$\bar B^0_s$ mixing phase is 
negligibly small in the SM, 
\begin{equation}
\phi_s\equiv2\,\mbox{arg}(V_{ts}^\ast V_{tb})=-2\delta\gamma=
-2\lambda^2\eta={\cal O}(-2^\circ),
\end{equation}
whereas $\phi_d\equiv2\,\mbox{arg}(V_{td}^\ast V_{tb})=
2\beta={\cal O}(50^\circ)$.

\item A large mixing parameter $x_s$ is expected in the SM,
\begin{equation}
x_s\equiv\frac{\Delta M_s}{\Gamma_s}={\cal O}(20),
\end{equation}
whereas $x_d=0.771\pm0.012$. Consequently, we have to deal with 
rapid $B^0_s$--$\bar B^0_s$ oscillations. The current experimental 
lower bound for the mass difference of the $B_s$ mass eigenstates 
is given by $\Delta M_s>14.5\,\mbox{ps}^{-1}$, corresponding to
$x_s>20.8$ (95\% C.L.) \cite{LEPBOSC,HFAG}.

\item There may be a sizeable difference between the decay widths of
the $B_s$ mass eigenstates,
\begin{equation}
\frac{\Delta\Gamma_s}{\Gamma_s}={\cal O}(-10\%),
\end{equation}
whereas $\Delta\Gamma_d/\Gamma_d$ is negligibly small, as we have seen
in Subsection~\ref{ssec:Mix-Par}. The current
CDF and LEP average is given by
$\Delta\Gamma_s/\Gamma_s=-0.16^{+0.16}_{-0.15}$, 
$|\Delta\Gamma_s|/\Gamma_s<0.54$ (95\% C.L.) \cite{LEPBOSC,HFAG}.

\end{itemize}

\subsubsection{Impact of $\Delta M_s$ on the 
Unitarity Triangle}\label{ssec:DMs-UT}
As we discussed in Subsection~\ref{ssec:BBbar-mix}, the mass differences
of the $B_q$ mass eigenstates satisfy 
\begin{equation}\label{DMq-simple}
\Delta M_q\propto M_{B_q}\hat B_{B_q}f_{B_q}^2 |V_{tq}^\ast V_{tb}|^2.
\end{equation}
In the $B_d$-meson case, this particular structure leads
to (\ref{Rt-det}), allowing us to determine the side $R_t$ of the UT.
To this end, in addition to the CKM parameter $A$ (see
(\ref{Vcb-det})), also the non-perturbative quantity 
\begin{equation}\label{np-Bd-mix}
\sqrt{\hat B_{B_d}}f_{B_d}=(235\pm33^{+0}_{-24})\,\mbox{MeV}
\end{equation}
has to be known, where the numerical value follows from lattice QCD 
studies \cite{CKM-Book}; QCD sum rules give a similar picture \cite{SR-calc}.
On the other hand, if we apply the expressions for $V_{cb}$ and 
$V_{ts}$ in (\ref{NLO-wolf}), 
we obtain
\begin{equation}\label{Rt-simple-rel}
R_t\equiv\frac{1}{\lambda}\left|\frac{V_{td}}{V_{cb}}\right|=
\frac{1}{\lambda}\left|\frac{V_{td}}{V_{ts}}\right|
\left[1+{\cal O}(\lambda^2)\right].
\end{equation}
Consequently, we may -- up to corrections entering at the $\lambda^2$ 
level -- determine $R_t$ through the ratio $|V_{td}/V_{ts}|$. Using now 
(\ref{DMq-simple}) yields the following expression \cite{Buras-Schladming}:
\begin{equation}\label{Rt-SU3}
R_t=0.90\left[\frac{\xi}{1.24}\right]
\sqrt{\frac{18.4\,{\rm ps}^{-1}}{\Delta M_s}}
\sqrt{\frac{\Delta M_d}{0.5\,{\rm ps}^{-1}}},
\end{equation}
where
\begin{equation}\label{xi-SU3}
\xi\equiv\frac{\sqrt{\hat B_s}f_{B_s}}{\sqrt{\hat B_d}f_{B_d}}
\end{equation}
is an $SU(3)$-breaking parameter; lattice QCD studies give
\begin{equation}\label{xi-lat}
\xi=1.18\pm0.04^{+0.12}_{-0},
\end{equation}
where $\xi=1.24\pm0.08$ should be used for analyses of the UT, 
as discussed in \cite{CKM-Book}. In comparison with the quantity in
(\ref{np-Bd-mix}) entering (\ref{Rt-det}), the ratio in (\ref{xi-SU3}) 
is more favourable and represents an important aspect of current 
non-perturbative research \cite{CKM-Book}. Another advantage 
of (\ref{Rt-SU3}) is that $A$, the Inami--Lim function 
$S_0(x_t)$, and the short-distance QCD correction factor $\eta_B$ cancel
in this expression. Interestingly, it allows us also to convert the lower 
experimental bound $\Delta M_s>14.5\,{\rm ps}^{-1}$ into the 
upper bound $R_t<1.0\times[\xi/1.24]$, which implies $\gamma\lsim90^\circ$,
thereby excluding a large fraction of the $\bar\rho$--$\bar\eta$ plane.

\subsubsection{$\Delta\Gamma_s$ and ``Untagged'' $B_s$ Rates}
The width difference of the $B_s$-meson system may provide interesting 
studies of CP violation through ``untagged'' $B_s$ rates 
\cite{dun}--\cite{FD-NCP}, which are defined as 
\begin{equation}
\langle\Gamma(B_s(t)\to f)\rangle
\equiv\Gamma(B^0_s(t)\to f)+\Gamma(\bar B^0_s(t)\to f),
\end{equation}
and are characterized by the feature that we do not distinguish between
initially, i.e.\ at time $t=0$, present $B^0_s$ or $\bar B^0_s$ mesons. 
If we consider a final state $f$ to which both a $B^0_s$ and a $\bar B^0_s$ 
may decay, and use the expressions in (\ref{rates}), we find
\begin{equation}\label{untagged-rate}
\langle\Gamma(B_s(t)\to f)\rangle
\propto \left[\cosh(\Delta\Gamma_st/2)-{\cal A}_{\Delta\Gamma}(B_s\to f)
\sinh(\Delta\Gamma_st/2)\right]e^{-\Gamma_s t},
\end{equation}
where ${\cal A}_{\Delta\Gamma}(B_s\to f)\propto \mbox{Re}\,\xi_f$ was 
introduced in (\ref{ADGam}). We observe that the rapidly oscillating 
$\Delta M_st$ terms cancel, and that we may obtain information on the 
phase structure of the observable $\xi_f$, thereby providing valuable
insights into CP violation. For instance, the untagged observables 
offered by the angular distribution of the 
$B_s\to K^{*+}K^{*-}, K^{*0}\bar K^{*0}$ decay products allow
the determination of the UT angle $\gamma$, provided $\Delta\Gamma_s$ is 
actually sizeable \cite{FD-CP}. Although $B$-decay experiments at hadron 
colliders should be able to resolve the $B^0_s$--$\bar B^0_s$ oscillations, 
untagged $B_s$ rates are interesting in terms of efficiency, 
acceptance and purity.

\boldmath\subsection{$B_s\to J/\psi\phi$}\unboldmath
This particularly promising channel is the $B_s$-meson counterpart of the 
``golden'' mode $B_d\to J/\psi K_{\rm S}$, as can be seen from the
diagrams shown in Fig.~\ref{fig:BdPsiKS}, where we just have to replace
the down spectator quark by a stange quark in order to obtain the
$B_s\to J/\psi\phi$ diagrams. Consequently, this decay is described 
by a transition amplitude with a structure that is completely 
analogous to that of (\ref{BdpsiK-ampl2}). On the other hand, in 
contrast to $B_d\to J/\psi K_{\rm S}$, the final state of 
$B_s\to J/\psi\phi$ is an admixture of different CP eigenstates, 
which can, however, be disentangled through an angular analysis of the 
$J/\psi [\to\ell^+\ell^-]\phi [\to\ K^+K^-]$ decay products \cite{DDLR,DDF}. 
Their angular distribution exhibits tiny direct CP violation, whereas 
mixing-induced CP-violating effects allow the extraction of
\begin{equation}\label{sinphis}
\sin\phi_s+{\cal O}(\overline{\lambda}^3)=\sin\phi_s+{\cal O}(10^{-3}).
\end{equation}
Since we have $\phi_s=-2\lambda^2\eta={\cal O}(10^{-2})$ in the SM, the 
determination of this phase from (\ref{sinphis}) is affected by
generic hadronic uncertainties of ${\cal O}(10\%)$, which may become an
important issue for the LHC era. These uncertainties can be controlled with
the help of flavour-symmetry arguments through the decay 
$B_d\to J/\psi \rho^0$ \cite{RF-ang}. Needless to note, the big hope is 
that experiments will find a {\it sizeable} value of $\sin\phi_s$, which 
would immediately signal the presence of NP contributions to 
$B^0_s$--$\bar B^0_s$ mixing.

Other interesting aspects of the $B_s\to J/\psi\phi$ angular distribution
are the determination of the width difference $\Delta\Gamma_s$ from untagged 
data samples \cite{DDF} (for recent LHC feasibility studies, see 
\cite{belkov}), and the extraction of $\cos\delta_f\cos\phi_s$ terms, where 
the $\delta_f$ are CP-conserving strong phases. If we fix the signs of 
$\cos\delta_f$ through factorization, we may extract the sign of $\cos\phi_s$, 
allowing an {\it unambiguous} determination of $\phi_s$ \cite{DFN}. 
In this context, $B_s\to D_\pm\eta^{(')}$, $ D_\pm\phi$, ...\ decays 
offer also interesting methods \cite{RF-gam-eff-03,RF-gam-det-03}.

\boldmath\subsection{$B_s\to K^+K^-$}\unboldmath\label{ssec:BsKK}
As can be seen from Fig.~\ref{fig:bpipi}, the decay $B_d\to\pi^+\pi^-$ is
related to the $B_s\to K^+K^-$ channel through an interchange of {\it all} 
down and strange quarks. Because of this feature, the $U$-spin flavour
symmetry of strong interactions, which connects the down and strange quarks 
through $SU(2)$ transformations in the same manner as the ordinary isospin 
symmetry connects the down and up quarks, allows us to relate the hadronic 
$B_d\to\pi^+\pi^-$ parameters to their $B_s\to K^+K^-$ counterparts. It can 
then be shown that these quantities -- and the angle $\gamma$ of the UT -- 
can be extracted from the measured CP asymmetries of the $B_d\to\pi^+\pi^-$, 
$B_s\to K^+K^-$ system \cite{RF-BsKK}. Also other $U$-spin strategies were 
developed, using $B_{s(d)}\to J/\psi K_{\rm S}$ or 
$B_{d(s)}\to D_{d(s)}^+D_{d(s)}^-$ \cite{RF-BdsPsiK}, 
$B_{d(s)}\to K^{0(*)}\bar K^{0(*)}$ \cite{RF-Phys-Rep,RF-ang}, 
$B_{(s)}\to \pi K$ \cite{GR-BspiK}, or $B_{s(d)}\to J/\psi \eta$
modes \cite{skands}. Since the $B_s\to K^+K^-$, $B_d\to\pi^+\pi^-$ 
system is particularly promising from an experimental point of view, 
thereby providing an interesting playground for CDF-II \cite{TEV-Book} 
and LHCb \cite{LHC-Book,LHCb-analyses}, let us now have a closer look 
at the corresponding strategy \cite{RF-BsKK}.

\subsubsection{Amplitude Structure and CP Asymmetries}
If we follow Subsection~\ref{ssec:Bpipi}, we may write the 
$B_d\to\pi^+\pi^-$, $B_s\to K^+K^-$ amplitudes as 
\begin{eqnarray}
A(B_d^0\to\pi^+\pi^-)&=&{\cal C}\left[e^{i\gamma}-
d e^{i\theta}\right]\label{Bpipi-ampl2}\\
A(B_s^0\to K^+K^-)&=&\left(\frac{\lambda}{1-\lambda^2/2}\right){\cal C}'
\left[e^{i\gamma}+\left(\frac{1-\lambda^2}{\lambda^2}\right)
d'e^{i\theta'}\right],\label{BsKK-ampl}
\end{eqnarray}
where $de^{i\theta}$ was introduced in (\ref{D-DEF}), $d'e^{i\theta'}$ is 
the $B_s\to K^+K^-$ counterpart of this quantity, and the overall
normalization factors ${\cal C}$ and ${\cal C}'$ are CP-conserving 
strong amplitudes. Using these general parametrizations, we may write
the corresponding CP-violating observables in the following generic 
form:
\begin{eqnarray}
{\cal A}_{\rm CP}^{\rm dir}(B_d\to\pi^+\pi^-)&=&
\mbox{fct}(d,\theta,\gamma), \qquad
{\cal A}_{\rm CP}^{\rm mix}(B_d\to\pi^+\pi^-)=
\mbox{fct}(d,\theta,\gamma,\phi_d) \label{Bpipi-obs}\\
{\cal A}_{\rm CP}^{\rm dir}(B_s\to K^+K^-)&=&
\mbox{fct}(d',\theta',\gamma), \quad
{\cal A}_{\rm CP}^{\rm mix}(B_s\to K^+K^-)=
\mbox{fct}(d',\theta',\gamma,\phi_s).\label{BsKK-obs}
\end{eqnarray}
The explicit expressions for the direct and mixing-induced 
CP asymmetries of $B_d\to\pi^+\pi^-$ are given in (\ref{Adir-Bpipi-gen}) 
and (\ref{Amix-Bpipi-gen}), respectively, whereas those for their
$B_s\to K^+K^-$ counterparts can be found in \cite{RF-BsKK}. Fortunately, 
these rather complicated expressions are not required for the 
following discussion.

\begin{figure}
\centerline{
\rotate[r]{
\epsfxsize=6.6truecm
{\epsffile{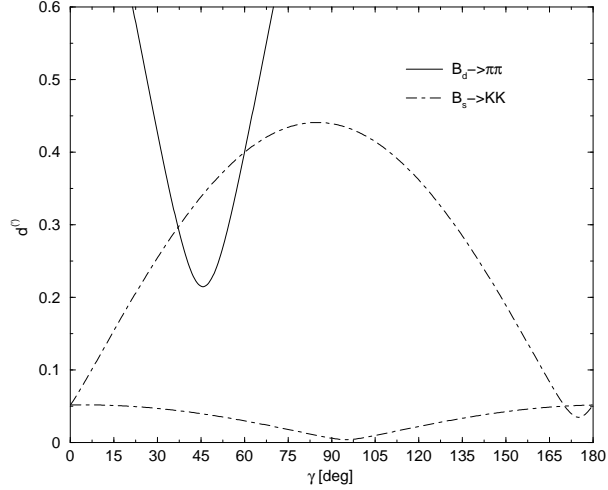}}}}
\caption{The contours in the $\gamma$--$d^{(')}$ plane for a 
specific example with $d=d'=0.4$, $\theta=\theta'=140^\circ$, 
$\phi_d=47^\circ$, $\phi_s=0^\circ$, $\gamma=60^\circ$, corresponding to
${\cal A}_{\rm CP}^{\rm dir}(B_d\to\pi^+\pi^-)=-0.30$,
${\cal A}_{\rm CP}^{\rm mix}(B_d\to\pi^+\pi^-)=+0.63$,
${\cal A}_{\rm CP}^{\rm dir}(B_s\to K^+K^-)=+0.16$ and
${\cal A}_{\rm CP}^{\rm mix}(B_s\to K^+K^-)=-0.17$. 
}\label{fig:gam-d}
\end{figure}

\subsubsection{Extraction of $\gamma$ and Hadronic 
Parameters}\label{ssec:BsKK-gam}
As we saw in Subsection~\ref{subsec:BpsiK}, $\phi_d$ can be extracted 
through the ``golden'' mode $B_d\to J/\psi K_{\rm S}$, with the result 
in (\ref{phid-det}). On the other hand, $\phi_s$ can be assumed to be 
negligibly small in the SM, or can be fixed through $B_s\to J/\psi \phi$, 
as we discussed above. These experimental determinations work also in 
the presence of NP contributions to $B^0_q$--$\bar B^0_q$ mixing, as is 
obvious from the discussion in Subsection~\ref{ssec:NP}.

Looking at (\ref{Bpipi-obs}), we observe that a measurement of
${\cal A}_{\rm CP}^{\rm dir}(B_d\to\pi^+\pi^-)$ 
and ${\cal A}_{\rm CP}^{\rm mix}(B_d\to\pi^+\pi^-)$ allows us to eliminate 
the strong phase $\theta$, thereby yielding $d$ as a function of 
$\gamma$ in a {\it theoretically clean} way. In complete analogy, we
may use the general parametrizations of the form in (\ref{BsKK-obs})
to eliminate $\theta'$, and to determine $d'$ in a 
{\it theoretically clean} manner as a function of $\gamma$ from 
the measured values of ${\cal A}_{\rm CP}^{\rm dir}(B_s\to K^+K^-)$ 
and ${\cal A}_{\rm CP}^{\rm mix}(B_s\to K^+K^-)$. Since $B_d\to\pi^+\pi^-$ 
and $B_s\to K^+K^-$ are related to each other by interchanging all down 
and strange quarks, the $U$-spin flavour symmetry of strong interactions 
implies the following relations:
\begin{equation}\label{U-spin-rel}
d'=d, \quad \theta'=\theta.
\end{equation}
Applying the former, we may extract $\gamma$ and $d$ from the 
theoretically clean $\gamma$--$d$ and $\gamma$--$d'$ contours,
which we have illustrated for a specific example in Fig.~\ref{fig:gam-d}.
As discussed in \cite{RF-BsKK}, it is also possible to resolve the 
twofold ambiguity for $(\gamma,d)$ arising from the intersections
of the solid and dot-dashed curves in Fig.~\ref{fig:gam-d}. Moreover, 
we may determine $\theta$ and $\theta'$, thereby allowing an interesting 
internal consistency check of the second $U$-spin relation in 
(\ref{U-spin-rel}).\footnote{Alternatively, we may eliminate $d$ and 
$d'$, and may then extract these parameters and $\gamma$ through the 
relation $\theta'=\theta$.} 

This strategy is very promising from an experimental point of view: 
at run II of the Tevatron and at the LHC, experimental accuracies for 
$\gamma$ of ${\cal O}(10^\circ)$ and ${\cal O}(1^\circ)$, 
respectively, are expected \cite{TEV-Book,LHCb-analyses}. As far as 
the $U$-spin-breaking corrections to $d'=d$ are concerned, they enter 
the determination of $\gamma$ through a relative shift of the 
$\gamma$--$d$ and $\gamma$--$d'$ contours; their impact on the 
extracted value of $\gamma$ therefore depends on the form of these curves, 
which is fixed through the measured observables. In the examples discussed 
in \cite{RF-Phys-Rep,RF-BsKK}, as well as in the one shown in 
Fig.~\ref{fig:gam-d}, the extracted value of $\gamma$ would be very 
stable under such corrections. Let us also note that the $U$-spin
relations in (\ref{U-spin-rel}) appear to be quite robust, since
the relevant form factors and decay constants cancel within factorization,
so that they do not receive $U$-spin-breaking corrections in this
approach \cite{RF-BsKK}. On the other hand, the ratio $|{\cal C}'/
{\cal C}|$, which equals 1 in the strict $U$-spin limit and enters 
the $U$-spin relation
\begin{equation}
\frac{{\cal A}_{\rm CP}^{\rm mix}
(B_s\to K^+K^-)}{{\cal A}_{\rm CP}^{\rm dir}(B_d\to\pi^+\pi^-)}=
-\left|\frac{{\cal C}'}{{\cal C}}\right|^2
\left[\frac{\mbox{BR}(B_d\to\pi^+\pi^-)}{\mbox{BR}(B_s\to K^+K^-)}\right]
\frac{\tau_{B_s}}{\tau_{B_d}},
\end{equation}
is affected by $U$-spin-breaking effects within factorization. An 
estimate of the corresponding form factors was recently performed
in \cite{KMM}, and certain non-factorizable effects were addressed 
in \cite{beneke}.

\subsubsection{Replacing $B_s\to K^+K^-$ by 
$B_d\to\pi^\mp K^\pm$}\label{ssec:gam-Bpipi-piK}
\begin{figure}
\begin{center}
\psfrag{eta}{$\bar\eta$}\psfrag{rho}{$\bar\rho$}
\includegraphics[width=8.4cm]{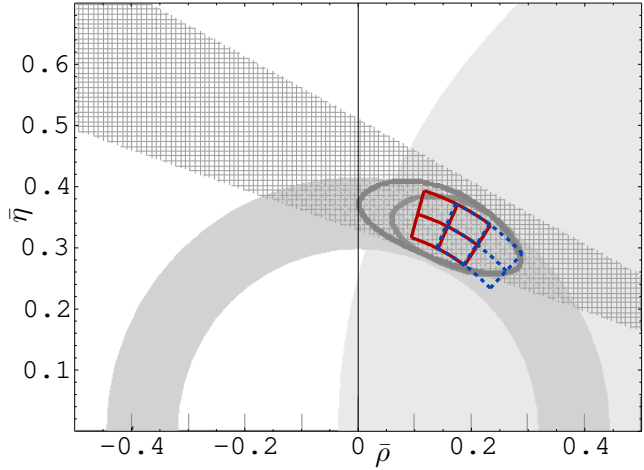}
\end{center}
\vspace*{-0.3truecm}
\caption{Comparison of the determination of $\gamma$ from the 
$B\to\pi\pi,\pi K$ data with the CKM fits, as discussed in the 
text.}\label{fig:ut-compare}
\end{figure}

Since $B_s\to K^+K^-$ is not accessible at the $e^+e^-$ $B$ factories
operating at the $\Upsilon(4S)$ resonance, we may not yet implement the 
strategy discussed above. However, as can easily be seen by looking at 
the corresponding Feynman diagrams, $B_s\to K^+K^-$ is related to 
$B_d\to\pi^\mp K^\pm$ through an interchange of spectator quarks. 
Consequently, we may approximately replace $B_s\to K^+K^-$ through 
$B_d\to\pi^\mp K^\pm$ in order to deal with the penguin problem in 
$B_d\to\pi^+\pi^-$ \cite{RF-Bpipi}. The utility of $B_d\to\pi^\mp K^\pm$ 
decays to control the penguin effects in $B_d\to\pi^+\pi^-$ was also 
emphasized in \cite{SiWo}. In order to explore the implications of the 
$B$-factory data, the following quantity plays a key r\^ole:
\begin{equation}\label{H-det}
H=\frac{1}{\epsilon}\left(\frac{f_K}{f_\pi}\right)^2
\left[\frac{\mbox{BR}(B_d\to\pi^+\pi^-)}{\mbox{BR}(B_d\to\pi^\mp K^\pm)}
\right]=7.17\pm0.75.
\end{equation}
Here $\epsilon\equiv\lambda^2/(1-\lambda^2)$, the ratio
$f_K/f_\pi=160/131$ describes factorizable $SU(3)$-breaking corrections, 
and the numerical value refers to the averages compiled in \cite{HFAG}. 
Applying (\ref{U-spin-rel}), we obtain
\begin{equation}\label{H-expr}
H=\frac{1-2 d \cos\theta\cos\gamma+d^2}{\epsilon^2+
2\epsilon d \cos\theta\cos\gamma+d^2}. 
\end{equation}
If we now combine the CP asymmetries
${\cal A}_{\rm CP}^{\rm dir}(B_d\to\pi^+\pi^-)$ 
and ${\cal A}_{\rm CP}^{\rm mix}(B_d\to\pi^+\pi^-)$ with $H$, we have
sufficient information available to determine $\gamma$, as well as $d$ and 
$\theta$ \cite{RF-BsKK,RF-Bpipi}. In practice, this can be
done with the help of the expressions in (\ref{Adir-Bpipi-gen}),
(\ref{Amix-Bpipi-gen}) and (\ref{H-expr}). A detailed discussion of
this strategy was given in \cite{FIM,FlMa2}, where also the impact
of NP contributions to $B^0_d$--$\bar B^0_d$ mixing was explored. 
Using additional information from the $B\to\pi K$ analysis 
discussed in Subsection~\ref{ssec:BpiK-puzzle}, the corresponding
determination of $\gamma$ was recently refined in \cite{BFRS3}, where
in particular a twofold ambiguity for $\gamma$ could be resolved, yielding
\begin{equation}\label{gam-fin}
\gamma=(64.7^{+6.3}_{-6.9})^\circ,
\end{equation}
which is in excellent agreement with the SM picture summarized in
(\ref{BFRS-gam}). If we complement this result with the experimental 
range for $R_b$ and apply the simple relations 
\begin{equation}
\bar\rho=R_b\cos\gamma,\quad 
\bar\eta=R_b\sin\gamma,
\end{equation}
which follow straightforwardly from Fig.~\ref{fig:cont-scheme},
we may also determine $\alpha$ and $\beta$:
\begin{equation}\label{alpha-beta}
\alpha=(93.6^{+10.3}_{-9.1})^\circ,  \quad  
\beta=(21.7^{+2.5}_{-2.6})^\circ.
\end{equation}
In Fig.~\ref{fig:ut-compare}, we compare these results with the allowed 
region for the apex of the UT that follows from the CKM fits, as implemented 
in \cite{BurasParodiStocchi}.\footnote{The small and large ellipses in
Fig.~\ref{fig:ut-compare} refer to the analyses of the SM and NP scenarios 
with MFV, respectively, as obtained in a recent update 
\cite{Buras-Schladming} of \cite{BurasParodiStocchi}.} 
Here the solid window corresponds to 
the range for $\gamma$ in (\ref{gam-fin}), whereas the dashed window 
indicates how the results change when the recently reported new Belle 
data \cite{Belle-New-Bpipi} are used. Needless to note, the consistency 
of the overall picture is very remarkable.

In the analysis leading to (\ref{gam-fin}) and (\ref{alpha-beta}),
it has been assumed that $\phi_d\sim47^\circ$, as in the SM. 
However, as discussed in \cite{FIM,FlMa2}, it is interesting to consider 
also the second, unconventional solution of 
$\phi_d\sim133^\circ$ in (\ref{phid-det}). There are simple relations
to go from one solution to the other. In particular, if  
$\phi_d$, $\gamma$, $d$ and $\theta$ are solutions of (\ref{Adir-Bpipi-gen}),
(\ref{Amix-Bpipi-gen}) and (\ref{H-expr}), then
\begin{equation}\label{phid-rel}
\pi-\phi_d,\quad \pi-\gamma,\quad d,\quad \pi-\theta
\end{equation}
are solutions as well. Consequently, (\ref{phid-rel}) allows us to go 
easily from the $\phi_d\sim47^\circ$ to the $\phi_d\sim133^\circ$ case. 
Interestingly, for the value of $\theta$ in (\ref{d-det}), we obtain 
$\cos\theta\sim-0.7<0$, having the same sign as in factorization, 
where $\left.\cos\theta\right|_{\rm fact}=-1$. On the other hand, the value 
of $\theta$ corresponding to $\phi_d\sim133^\circ$ yields 
$\cos\theta\sim+0.7>0$, i.e.\ the opposite sign, thereby disfavouring 
the $\phi_d\sim133^\circ$ solution~\cite{BFRS3}. 

Let us finally note that the results for $d$ and $\theta$ in 
(\ref{d-det}) following from the $B\to\pi\pi$ analysis discussed 
in Subsection~\ref{ssec:Bpipi-puzzle} allow us also to obtain
SM predictions for the CP-violating $B_s\to K^+K^-$ observables 
with the help of (\ref{U-spin-rel}) \cite{BFRS3}:
\begin{equation}
\left.{\cal A}_{\rm CP}^{\rm dir}(B_s\to K^+K^-)\right|_{\rm SM}=
0.14^{+0.14}_{-0.09}, \quad
\left.{\cal A}_{\rm CP}^{\rm mix}(B_s\to K^+K^-)\right|_{\rm SM}=
-0.18^{+0.08}_{-0.07}.
\end{equation}
On the other hand, the prediction of $\mbox{BR}(B_s\to K^+K^-)$
requires information on the $SU(3)$-breaking form-factor ratios 
entering $|{\cal C}'/{\cal C}|$, where the estimates of \cite{KMM}
correspond to a branching ratio at the $3.5\times 10^{-5}$ level.
It will be very interesting to see the first data for the $B_s\to K^+K^-$ 
channel from run II of the Tevatron, and to fully expoit its physics
potential at LHCb and BTeV. The decay $B_s\to\pi^\pm K^\mp$ offers 
also various ways to complement the $B\to\pi\pi,\pi K$ strategy 
discussed in Subsection~\ref{ssec:BpiK-puzzle}.

\boldmath\subsection{$B_s\to D_s^{(\ast)\pm} K^\mp$}\unboldmath
\label{ssec:BsDsK}
Decays of the kind $B_s\to D_s^{(\ast)\pm} K^\mp, ...$\ and their 
counterparts $B_d\to D^{(\ast)\pm} \pi^\mp, ...$\ provide another
important tool to explore CP violation \cite{BsDsK,BdDpi}. 
Since these transitions can be described on the same theoretical 
basis, we will consider them simultaneously in this subsection, 
following \cite{RF-BdDpi}.

\begin{figure}
\begin{center}
\leavevmode
\epsfysize=4.0truecm 
\epsffile{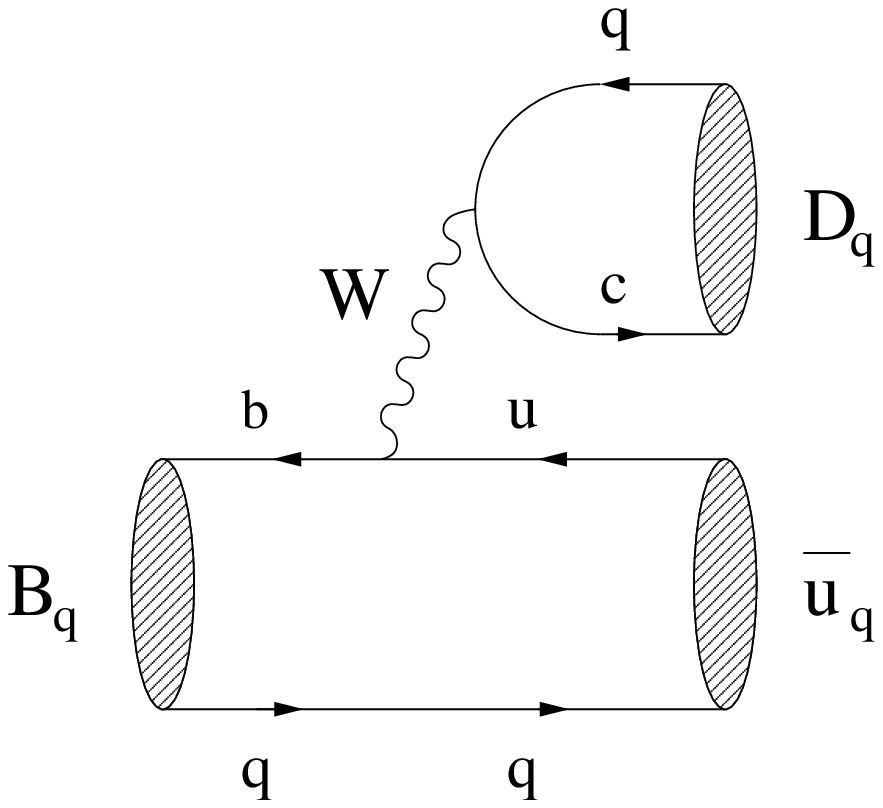} \hspace*{1truecm}
\epsfysize=4.0truecm 
\epsffile{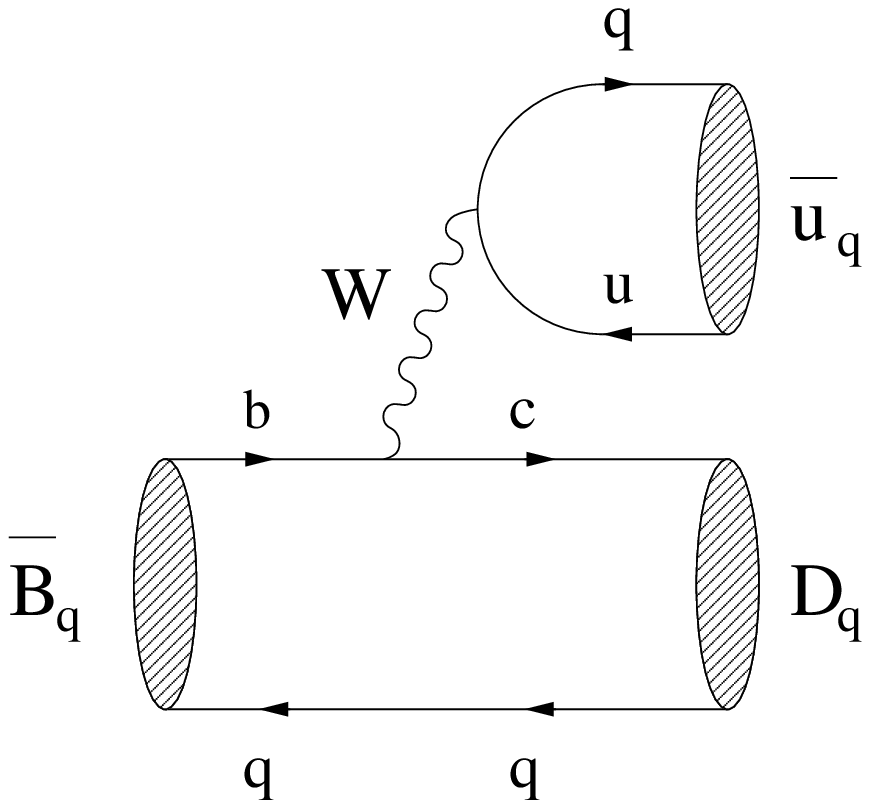}
\end{center}
\vspace*{-0.3truecm}
\caption{Feynman diagrams contributing to $B_q^0\to D_q \bar u_q$ and 
$\bar B_q^0\to D_q \bar u_q$.}\label{fig:BqDu}
\end{figure}

\subsubsection{Basic Features}
It is convenient to write $B_s\to D_s^{(\ast)\pm} K^\mp, ...$\ and 
$B_d\to D^{(\ast)\pm} \pi^\mp, ...$\ decays generically as 
$B^0_q\to D_q\bar u_q$, so that we may easily distinguish 
between the following cases: 
\begin{itemize}
\item $q=s$: $D_s\in\{D_s^+, D_s^{\ast+}, ...\}$, 
$u_s\in\{K^+, K^{\ast+}, ...\}$.
\item $q=d$: $D_d\in\{D^+, D^{\ast+}, ...\}$, 
$u_d\in\{\pi^+, \rho^+, ...\}$.
\end{itemize}
In the discussion given below, we shall only consider those 
$B^0_q\to D_q\bar u_q$ decays where at least one of the 
$D_q$, $\bar u_q$ states is a pseudoscalar meson. 
In the opposite case, for example $B^0_s\to D_s^{\ast+}K^{\ast-}$, the 
extraction of weak phases would require a complicated angular analysis. 
If we look at Fig.~\ref{fig:BqDu}, we observe that 
$B^0_q\to D_q\bar u_q$ originates from colour-allowed tree topologies, 
and that also a $\bar B^0_q$ meson may decay into the same final state 
$D_q\bar u_q$. The latter feature leads to interference effects between 
$B^0_q$--$\bar B^0_q$ mixing and decay processes, providing valuable
information about the CP-violating phase $\phi_q+\gamma$.

\subsubsection{Rate Asymmetries}
Let us first consider $B_q$ decays into $D_q\bar u_q$. Since both a 
$B^0_q$ and a $\bar B^0_q$ meson may decay into this state, 
we obtain a time-dependent rate asymmetry of the following form: 
\begin{eqnarray}
\lefteqn{\frac{\Gamma(B^0_q(t)\to D_q\bar u_q)-
\Gamma(\bar B^0_q(t)\to D_q\bar u_q)}{\Gamma(B^0_q(t)\to 
D_q\bar u_q)+\Gamma(\bar B^0_q(t)\to D_q\bar u_q)}}
\nonumber\\
&&=\left[\frac{C(B_q\to D_q\bar u_q)\cos(\Delta M_q t)+
S(B_q\to D_q\bar u_q)\sin(\Delta M_q t)}{\cosh(\Delta\Gamma_qt/2)-
{\cal A}_{\rm \Delta\Gamma}(B_q\to D_q\bar u_q)
\sinh(\Delta\Gamma_qt/2)}\right],\label{rate-asym}
\end{eqnarray}
having a structure that is completely analogous to the one of (\ref{ee6}).
Applying the formalism discussed in Section~\ref{sec:mix}, we find
that these observables are given by
\begin{equation}\label{Obs-expr}
C(B_q\to D_q\bar u_q)\equiv C_q=\frac{1-|\xi_q|^2}{1+|\xi_q|^2}, \quad
S(B_q\to D_q\bar u_q)\equiv S_q=\frac{2\,\mbox{Im}\,\xi_q}{1+
|\xi_q|^2},
\end{equation}
where
\begin{equation}\label{xi-BDpi-def}
\xi_q\equiv-e^{-i\phi_q}\left[e^{i\phi_{\rm CP}(B_q)}
\frac{A(\bar B_q^0\to D_q\bar u_q)}{A(B_q^0\to D_q\bar u_q)}
\right]
\end{equation}
measures the strength of the interference effects between the
$B^0_q$--$\bar B^0_q$ mixing and decay processes. 

If we take the Feynman diagrams shown in Fig.~\ref{fig:BqDu} into account
and use an appropriate low-energy effecitve Hamiltonian of the kind
discussed in Subsection~\ref{subsec:ham}, we may write
\begin{equation}\label{A-BbarDubar}
A(\bar B^0_q\to D_q\bar u_q)=\langle \bar u_qD_q|
{\cal H}_{\rm eff}(\bar B^0_q\to D_q\bar u_q)|
\bar B^0_q\rangle=\frac{G_{\rm F}}{\sqrt{2}}\bar v_q
\bar M_q,
\end{equation}
where the hadronic matrix element
\begin{equation}\label{Mbar-def}
\bar M_q\equiv 
\langle \bar u_qD_q|\bar {\cal O}_1^{\, q}\,{ C}_1(\mu)+
\bar {\cal O}_2^{\, q}\,{ C}_2(\mu)|\bar B^0_q\rangle
\end{equation}
involves the current--current operators
\begin{equation}
\bar {\cal O}_1^{\, q}\equiv
(\bar q_\alpha u_\beta)_{\mbox{{\scriptsize 
V--A}}}\left(\bar c_\beta b_\alpha\right)_{\mbox{{\scriptsize V--A}}},
\quad
\bar {\cal O}_2^{\, q}\equiv
(\bar q_\alpha u_\alpha)_{\mbox{{\scriptsize 
V--A}}}\left(\bar c_\beta b_\beta\right)_{\mbox{{\scriptsize V--A}}},
\end{equation}
and the CKM factors $\bar v_q$ are given by
\begin{equation}
\bar v_s\equiv V_{us}^\ast V_{cb}= A\lambda^3, \quad 
\bar v_d\equiv V_{ud}^\ast V_{cb}= A\lambda^2(1-\lambda^2/2).
\end{equation}
On the other hand, the $B^0_q\to D_q\bar u_q$ decay amplitude
takes the following form:
\begin{equation}\label{ampl-BDubar1}
A(B^0_q\to D_q\bar u_q)=\langle\bar u_qD_q|
{\cal H}_{\rm eff}(B^0_q\to D_q\bar u_q)|B^0_q\rangle
=\frac{G_{\rm F}}{\sqrt{2}}v_q^\ast \langle\bar u_qD_q|
{\cal O}_1^{q\dagger}\,{ C}_1(\mu)+{\cal O}_2^{q\dagger}\,{ C}_2(\mu)
|B^0_q\rangle,
\end{equation}
where we have to deal with the current--current operators 
\begin{equation}
{\cal O}_1^{q}\equiv(\bar q_\alpha c_\beta)_{\mbox{{\scriptsize V--A}}}
\left(\bar u_\beta b_\alpha\right)_{\mbox{{\scriptsize V--A}}},
\quad
{\cal O}_2^{q}\equiv(\bar q_\alpha c_\alpha)_{\mbox{{\scriptsize V--A}}}
\left(\bar u_\beta b_\beta\right)_{\mbox{{\scriptsize V--A}}},
\end{equation}
and the CKM factors $v_q$ are defined as 
\begin{equation}
v_s\equiv V_{cs}^\ast V_{ub}=A\lambda^3R_be^{-i\gamma}, 
\quad 
v_d\equiv V_{cd}^\ast V_{ub}=
-\left(\frac{A\lambda^4R_b}{1-\lambda^2/2}\right)e^{-i\gamma}.
\end{equation}
If we introduce CP phases for the $D_q$ and $u_q$ mesons in 
analogy to (\ref{CP-def}), we obtain
\begin{equation}
({\cal CP})|D_q\bar u_q\rangle=(-1)^L e^{i[\phi_{\rm CP}(D_q)-
\phi_{\rm CP}(u_q)]}|\bar D_q u_q\rangle,
\end{equation}
where $L$ denotes the angular momentum of the $D_q\bar u_q$ state.
Using now the relations $({\cal CP})^\dagger ({\cal CP})=\hat 1$
and $({\cal CP}){\cal O}_k^{q\dagger}({\cal CP})^\dagger={\cal O}_k^{q}$
as in Subsection~\ref{subsec:CPasym}, we may rewrite 
(\ref{ampl-BDubar1}) as 
\begin{equation}\label{A-BDubar}
A(B^0_q\to D_q\bar u_q)=(-1)^L e^{i[\phi_{\rm CP}(B_q)-
\phi_{\rm CP}(D_q)+\phi_{\rm CP}(u_q)]}
\frac{G_{\rm F}}{\sqrt{2}}v_q^\ast M_q, 
\end{equation}
with
\begin{equation}\label{M-def}
M_{q}\equiv \langle u_q \bar D_q|{\cal O}_1^{q}\,{ C}_1(\mu)+
{\cal O}_2^{q}\,{ C}_2(\mu)|\bar B^0_q\rangle.
\end{equation}
An analogous calculation for the $\bar B^0_q\to \bar D_qu_q$ 
and $B^0_q\to\bar D_qu_q$ transitions yields
\begin{equation}\label{ampl-BbarDbaru}
A(\bar B^0_q\to \bar D_qu_q)=\frac{G_{\rm F}}{\sqrt{2}}v_q M_q
\end{equation}
\begin{equation}\label{ampl-BDu}
A(B^0_q\to \bar D_q u_q)=(-1)^L e^{i[\phi_{\rm CP}(B_q)+
\phi_{\rm CP}(D_q)-\phi_{\rm CP}(u_q)]}\frac{G_{\rm F}}{\sqrt{2}}
\bar v_q^\ast \bar M_q,
\end{equation}
where the same hadronic matrix elements as in the
$B^0_q\to D_q\bar u_q$ and $\bar B^0_q\to D_q\bar u_q$
modes arise. 

If we now insert (\ref{A-BbarDubar}) and (\ref{A-BDubar}) into 
(\ref{xi-BDpi-def}), we observe that the convention-dependent 
phase $\phi_{\rm CP}(B_q)$ is cancelled through the amplitude ratio, 
and arrive at
\begin{equation}\label{xi-BDpi-expr}
\xi_q=-(-1)^Le^{-i(\phi_q+\gamma)}\left[\frac{1}{x_qe^{i\delta_q}}\right],
\end{equation}
where
\begin{equation}\label{xq-def}
x_{s}\equiv R_b a_{s}, \quad 
x_{d}\equiv-\left(\frac{\lambda^2R_b}{1-\lambda^2}\right)a_{d},
\end{equation}
with
\begin{equation}\label{aq-def}
a_qe^{i\delta_q}\equiv e^{-i[\phi_{\rm CP}(D_q)-\phi_{\rm CP}(u_q)]}
\frac{M_q}{\bar M_q}.
\end{equation}
It should be noted that the convention-dependent phases 
$\phi_{\rm CP}(D_q)$ and $\phi_{\rm CP}(u_q)$ in (\ref{aq-def}) 
are cancelled through the ratio of hadronic matrix elements, so 
that $a_qe^{i\delta_q}$ is actually a physical observable 
(this is shown explicitly in \cite{RF-BdDpi}). Applying now
(\ref{Obs-expr}), we finally arrive at 
\begin{equation}
C_q=-\left[\frac{1-x_q^2}{1+x_q^2}\right], \quad
S_q=(-1)^L\left[\frac{2\,x_q\sin(\phi_q+\gamma+\delta_q)}{1+x_q^2}\right].
\end{equation} 

An analogous calculation for the decays into the CP-conjugate 
final state $\bar D_q u_q$ yields
\begin{equation}
\bar\xi_q=-e^{-i\phi_q}\left[e^{i\phi_{\rm CP}(B_q)}
\frac{A(\bar B_q^0\to \bar D_q u_q)}{A(B_q^0\to\bar D_qu_q)}
\right]=-(-1)^Le^{-i(\phi_q+\gamma)}\left[x_qe^{i\delta_q}\right],
\end{equation}
which implies
\begin{equation}
\bar C_q=+\left[\frac{1-x_q^2}{1+x_q^2}\right], \quad
\bar S_q=(-1)^L\left[\frac{2\,x_q
\sin(\phi_q+\gamma-\delta_q)}{1+x_q^2}\right],
\end{equation} 
where $\bar C_q\equiv C(B_q\to \bar D_q u_q)$ and
$\bar S_q\equiv S(B_q\to \bar D_q u_q)$. Note that 
$\bar \xi_q$ and $\xi_q$ satisfy the relation
\begin{equation}\label{xi-rel}
\bar\xi_q\times\xi_q=e^{-i2(\phi_q+\gamma)},
\end{equation}
where the hadronic parameter $x_qe^{i\delta_q}$ {\it cancels}. 
Consequently, we may extract $\phi_q+\gamma$ in a {\it theoretically clean} 
way from the corresponding observables.

\subsubsection{Conventional Extraction of $\phi_q+\gamma$}
It is convenient to introduce the following combinations of
observables:
\begin{equation}\label{Cp-def}
\langle C_q\rangle_+\equiv\frac{\bar C_q+C_q}{2}=0
\end{equation}
\begin{equation}\label{Cm-def}
\langle C_q\rangle_-\equiv\frac{\bar C_q-C_q}{2}=\frac{1-x_q^2}{1+x_q^2}
\end{equation}
\begin{equation}\label{Sp-def}
\langle S_q\rangle_+\equiv\frac{\bar S_q+S_q}{2}=+(-1)^L
\left[\frac{2\,x_q\cos\delta_q}{1+x_q^2}\right]\sin(\phi_q+\gamma)
\end{equation}
\begin{equation}\label{Sm-def}
\langle S_q\rangle_-\equiv\frac{\bar S_q-S_q}{2}=-(-1)^L
\left[\frac{2\,x_q\sin\delta_q}{1+x_q^2}\right]\cos(\phi_q+\gamma).
\end{equation}

We observe that (\ref{Cm-def}) allows us -- in principle -- to 
determine $x_q$ from $\langle C_q\rangle_-$. However, to this end,
terms entering at the $x_q^2$ level have to be resolved experimentally. 
In the case of $q=s$, we have $x_s={\cal O}(R_b)$, implying 
$x_s^2={\cal O}(0.16)$, so that this may actually be possible, although 
challenging \cite{BsDsK}. On the other hand, $x_d={\cal O}(-\lambda^2R_b)$ 
is doubly Cabibbo-suppressed. Although it should be possible to resolve 
terms of ${\cal O}(x_d)$, this will be impossible for the vanishingly 
small $x_d^2={\cal O}(0.0004)$ terms, so that alternative approaches 
to fix $x_d$ are required \cite{BdDpi}. 

In contrast to the observables associated with the $\cos(\Delta M_qt)$
terms, the mixing-induced observables entering the rate asymmetries
with $\sin(\Delta M_qt)$ provide information on $\phi_q+\gamma$. Let 
us now assume that $x_q$ is known. We may then consider
\begin{eqnarray}
s_+&\equiv& (-1)^L
\left[\frac{1+x_q^2}{2 x_q}\right]\langle S_q\rangle_+
=+\cos\delta_q\sin(\phi_q+\gamma)\\
s_-&\equiv&(-1)^L
\left[\frac{1+x_q^2}{2x_q}\right]\langle S_q\rangle_-
=-\sin\delta_q\cos(\phi_q+\gamma),
\end{eqnarray}
yielding
\begin{equation}\label{conv-extr}
\sin^2(\phi_q+\gamma)=\frac{1}{2}\left[(1+s_+^2-s_-^2) \pm
\sqrt{(1+s_+^2-s_-^2)^2-4s_+^2}\right].
\end{equation}
This expression implies an eightfold solution for $\phi_q+\gamma$. 
If we fix the sign of $\cos\delta_q$ with the help of factorization, 
a fourfold discrete ambiguity emerges. Since we may determine
$\phi_d$ and $\phi_s$ through analyses of $B_d\to J/\psi K_{\rm S}$
and $B_s\to J/\psi\phi$ decays, respectively, we may extract
$\gamma$ from $\phi_q+\gamma$.

\subsubsection{New Strategies and Recent Developments}
Let us now discuss new strategies to explore the 
$B_q\to D_q \bar u_q$ modes \cite{RF-BdDpi}. If the
width difference $\Delta\Gamma_s$ is sizeable, the time-dependent untagged 
rates (see (\ref{untagged-rate}))
\begin{equation}\label{untagged}
\langle\Gamma(B_q(t)\to D_q\bar u_q)\rangle=
\langle\Gamma(B_q\to D_q\bar u_q)\rangle 
\left[\cosh(\Delta\Gamma_qt/2)-{\cal A}_{\rm \Delta\Gamma}
(B_q\to D_q\overline{u}_q)\,\sinh(\Delta\Gamma_qt/2)\right]e^{-\Gamma_qt}
\end{equation}
and their CP conjugates provide 
${\cal A}_{\rm \Delta\Gamma}(B_s\to D_s\bar u_s)
\equiv {\cal A}_{\rm \Delta\Gamma_s}$ and 
${\cal A}_{\rm \Delta\Gamma}(B_s\to \bar D_s u_s)\equiv 
\bar {\cal A}_{\rm \Delta\Gamma_s}$. It can be shown that these ``untagged''
observables can be combined with their ``tagged'' counterparts 
$\langle S_s\rangle_\pm$ in the form of the following simple relation: 
\begin{equation}\label{untagged-extr}
\tan(\phi_s+\gamma)=
-\left[\frac{\langle S_s\rangle_+}{\langle{\cal A}_{\rm \Delta\Gamma_s}
\rangle_+}\right]
=+\left[\frac{\langle{\cal A}_{\rm \Delta\Gamma_s}
\rangle_-}{\langle S_s\rangle_-}\right],
\end{equation}
where $\langle{\cal A}_{\rm \Delta\Gamma_s}\rangle_+$ 
and $\langle{\cal A}_{\rm \Delta\Gamma_s}\rangle_-$ are defined
in analogy to (\ref{Sp-def}) and (\ref{Sm-def}), respectively. Obviously, 
(\ref{untagged-extr}) offers an elegant extraction of $\phi_s+\gamma$,
up to a twofold ambiguity. If we fix again the sign of $\cos\delta_q$ 
through factorization, we may determine $\phi_s+\gamma$ in an 
{\it unambiguous} manner, which should be compared with the fourfold 
ambiguity arising in this case from (\ref{conv-extr}). In particular,
we may decide whether $\gamma\in[0^\circ,180^\circ]$, as 
in the SM, or $\gamma\in[180^\circ,360^\circ]$. Another important 
advantage of (\ref{untagged-extr}) is that we do {\it not} 
have to rely on the resolution of ${\cal O}(x_s^2)$ terms, as 
$\langle S_s\rangle_\pm$ and 
$\langle {\cal A}_{\rm \Delta\Gamma_s}\rangle_\pm$ are both proportional 
to $x_s$. On the other hand, we need a sizeable value of $\Delta\Gamma_s$. 
Measurements of untagged rates are also very useful in the case of 
a vanishingly small $\Delta\Gamma_q$, since the ``unevolved'' 
(i.e.\ time-independent) untagged rates in (\ref{untagged}) offer 
various interesting strategies to determine $x_q$ from the ratio of 
$\langle\Gamma(B_q\to D_q\bar u_q)\rangle+
\langle\Gamma(B_q\to \bar D_q u_q)\rangle$ to CP-averaged rates of
appropriate $B^\pm$ or flavour-specific $B_q$ decays.

If we keep the hadronic parameter $x_q$ and the associated strong phase
$\delta_q$ as ``unknown'', free parameters in the expressions for the
$\langle S_q\rangle_\pm$, we may derive the relations 
\begin{equation}
|\sin(\phi_q+\gamma)|\geq|\langle S_q\rangle_+|, \quad
|\cos(\phi_q+\gamma)|\geq|\langle S_q\rangle_-|,
\end{equation}
which can straightforwardly be converted into bounds on $\phi_q+\gamma$. 
If $x_q$ is known, stronger constraints are implied by 
\begin{equation}\label{bounds}
|\sin(\phi_q+\gamma)|\geq|s_+|, \quad
|\cos(\phi_q+\gamma)|\geq|s_-|.
\end{equation}
Once $s_+$ and $s_-$ are known, we may of course determine
$\phi_q+\gamma$ through the ``conventional'' approach, using 
(\ref{conv-extr}). However, the bounds following from (\ref{bounds})
provide essentially the same information and are much simpler to 
implement. Moreover, as discussed in detail in \cite{RF-BdDpi}
for several examples, the bounds following from the $B_s$ and $B_d$ 
modes may be highly complementary, thereby providing particularly narrow, 
theoretically clean ranges for $\gamma$. 
Whereas the $B_s$ decays are not yet accessible, first results for 
the $B_d\to D^{(\ast)\pm}\pi^\mp$ modes obtained by BaBar give
$|\sin(\phi_d+\gamma)|>0.87$ ($0.58$) at the 68\% (95\%) C.L.
\cite{BaBar-BDpi}. Looking at (\ref{Sp-def}), we observe that 
we may extract the sign of $\sin(\phi_q+\gamma)$ from $\langle S_q\rangle_+$ 
if we assume that the sign of $\cos\delta_q$ is as in factorization.
To this end, the factor $(-1)^L$ has to be properly taken into account.
The information on the sign of $\sin(\phi_d+\gamma)$ is very useful,
as it allows us to distinguish directly between the two solutions for
$(\phi_d,\gamma)$ discussed in Subsection~\ref{ssec:gam-Bpipi-piK}.
If we apply (\ref{phid-rel}), the analysis of CP violation in 
$B_d\to\pi^+\pi^-$ gives $(\phi_d,\gamma)\sim (47^\circ,65^\circ)$ or 
$(133^\circ,115^\circ)$ \cite{FIM,FlMa2}, corresponding to 
$\sin(\phi_q+\gamma)\sim+0.9$ or $-0.9$, respectively. The BaBar 
analysis favours the former case \cite{RF-BdDpi}, i.e.\ the picture 
of the SM, in accordance with the discussion after (\ref{phid-rel}). 
The exploration of $B_d\to D^{(\ast)\pm}\pi^\mp$ modes is also in 
progress at Belle \cite{Belle-BDpi}. Unfortunately, the
current Belle results for (fully reconstructed) 
$B_d\to D^{(\ast)\pm}\pi^\mp$ decays favour the sign opposite 
to the one obtained by BaBar (see also \cite{HFAG}), so that
the experimental picture is not yet conclusive. 

Let us now further exploit the complementarity between the
$B_s^0\to D_s^{(\ast)+}K^-$ and $B_d^0\to D^{(\ast)+}\pi^-$ modes.
If we look at their decay topologies, we observe that these channels are 
related to each other through an interchange of all down and strange quarks. 
Consequently, the $U$-spin flavour symmetry of strong interactions implies 
$a_s=a_d$ and $\delta_s=\delta_d$. There are various possibilities 
to implement these relations. A particularly simple picture emerges if we 
assume that $a_s=a_d$ {\it and} $\delta_s=\delta_d$, 
which yields
\begin{equation}
\tan\gamma=-\left[\frac{\sin\phi_d-S
\sin\phi_s}{\cos\phi_d-S\cos\phi_s}
\right]\stackrel{\phi_s=0^\circ}{=}
-\left[\frac{\sin\phi_d}{\cos\phi_d-S}\right].
\end{equation}
Here we have introduced
\begin{equation}
S\equiv-R\left[\frac{\langle S_d\rangle_+}{\langle S_s\rangle_+}\right]
\end{equation}
with
\begin{equation}
R\equiv\left(\frac{1-\lambda^2}{\lambda^2}\right)
\left[\frac{1}{1+x_s^2}\right],
\end{equation}
which can be fixed from untagged $B_s$ rates through
\begin{equation}
R=\left(\frac{f_K}{f_\pi}\right)^2
\left[\frac{\Gamma(\bar B^0_s\to D_s^{(\ast)+}\pi^-)+
\Gamma(B^0_s\to D_s^{(\ast)-}\pi^+)}{\langle\Gamma(B_s\to D_s^{(\ast)+}K^-)
\rangle+\langle\Gamma(B_s\to D_s^{(\ast)-}K^+)\rangle}\right].
\end{equation}
Alternatively, we may {\it only} assume that $\delta_s=\delta_d$ {\it or} 
that $a_s=a_d$. Apart from features related to multiple discrete ambiguities, 
the most important advantage with respect to the ``conventional'' approach 
is that the experimental resolution of the $x_q^2$ terms is not required. In 
particular, $x_d$ does {\it not} have to be fixed, and $x_s$ may only enter 
through a $1+x_s^2$ correction, which can straightforwardly be determined 
through untagged $B_s$ rate measurements. In the most refined implementation 
of this strategy, the measurement of $x_d/x_s$ would {\it only} be interesting 
for the inclusion of $U$-spin-breaking effects in $a_d/a_s$.

\section{RARE DECAYS}\label{sec:rare}
\setcounter{equation}{0}
\subsection{General Features and Impact of New Physics in Models with
Minimal Flavour Violation}\label{ssec:rare-gen}
In order to complement the exploration of flavour physics through 
the CP-violating phenomena discussed above, also various rare 
decays of $B$ and $K$ mesons offer very interesting strategies.
As we have already noted, by ``rare'' decays we mean transitions 
that do {\it not} arise at the tree level in the SM, but may originate
through loop effects. Consequently, rare $B$ decays are mediated by 
FCNC processes of the kind $\bar b\to \bar s$ or $\bar b\to \bar d$, 
whereas rare $K$ decays originate from their $\bar s\to \bar d$
counterparts. Prominent examples of rare $B$ decays are the
following exclusive decay modes:
\begin{itemize}
\item $B\to K^\ast\gamma$, $B\to \rho\gamma$, $...$ 
\item $B\to K\mu^+\mu^-$, $B\to \pi\mu^+\mu^-$, $...$ 
\item $B_{s,d}\to \mu^+\mu^-$. 
\end{itemize}
While the $B_{s,d}\to \mu^+\mu^-$ transitions are very clean, the former 
two decay classes suffer from theoretical uncertainties that are related to
hadronic form factors and long-distance contributions. On the other hand, the 
hadronic uncertainties are much smaller in the corresponding inclusive 
decays, $B\to X_{s,d}\gamma$ and $B\to X_{s,d}\mu^+\mu^-$, which are 
therefore more promising from the theoretical point of view, but are 
unfortunately more difficult to measure; the cleanest rare $B$ decays 
are given by $B\to X_{s,d} \nu\bar \nu$ processes. Let us note that a 
tremendous amount of work went into the calculation of the branching ratio 
of the prominent $B\to X_s\gamma$ channel (for an overview, see \cite{BuMi}); 
the agreement of the experimental value with the SM expectation implies 
important constraints for the allowed parameter space of popular NP 
scenarios. The phenomenology of the kaon system includes also interesting 
rare decays:
\begin{itemize}
\item $K_{\rm L}\to\mu^+\mu^-$
\item $K_{\rm L}\to\pi^0 e^+ e^-$
\item $K_{\rm L}\to\pi^0\nu\bar\nu$, $K^+\to\pi^+\nu\bar\nu$,
\end{itemize}
where the ``golden'' modes are given by the $K\to\pi\nu\bar\nu$ 
processes, which are essentially theoretically clean, as we have
already noted in Subsection~\ref{subsec:rare-kaon-brief}.

In order to deal with rare decays theoretically, appropriate low-energy 
effective Hamiltonians are used, in analogy to the analysis of non-leptonic 
$B$ decays. The structure of the corresponding transition amplitudes is 
similar to the one of (\ref{ee2}), i.e.\ the short-distance physics is 
described by perturbatively calculable Wilson coefficient functions, 
whereas the long-distance dynamics is encoded in non-perturbative
hadronic matrix elements of local operators. It is useful to follow 
\cite{PBE,BH92}, and to rewrite the rare-decay implementation of 
(\ref{ee2}) as
\begin{equation}\label{mmaster}
{A({\rm decay})}= 
P_0({\rm decay}) + \sum_r P_r({\rm decay} ) F_r(v).
\end{equation}
For the derivation of this expression, we choose $\mu=\mu_0={\cal O}(M_W)$, 
and rewrite the corresponding Wilson coefficients $C_k(\mu_0)$ as linear 
combinations of ``master functions'' $F_r(v)$, which follow from the 
evaluation of penguin and box diagrams with heavy particle exchanges. 
Expression (\ref{mmaster}) does not only apply to the SM, but also to 
NP scenarios with MFV (see Subsection~\ref{ssec:MFV}), where the 
parameters involved are collectively denoted by $v$. In the SM, the 
functions $F_r(v)$ reduce to the well-known Inami--Lim functions \cite{IL}, 
with $v=x_t=m_t^2/M_W^2$. The term $P_0$ summarizes the contributions 
that originate from light internal quarks, such as the charm and up quarks, 
and the sum takes the remaining contributions into account. For a detailed 
discussion of this formalism and the general features of the $P_0$, $P_r$ 
and $F_r$, we refer the reader to \cite{Buras-Cracow}. Let us here just 
emphasize the following important points:
\begin{itemize}
\item The $F_r(v)$ are {\it process-independent, universal} 
functions that depend on the particular model considered. NP enters
the decay amplitudes only through these functions.
\item The $P_0$ and $P_r$ are {\it process-dependent} quantities. In 
particular, they depend on the hadronic matrix elements of the operators 
$Q_k$.
\end{itemize}
In models with MFV, the set of the $F_r(v)$ consists of seven functions
\begin{equation}\label{masterf}
S(v),~X(v),~Y(v),~Z(v),~E(v),~D'(v),~E'(v),
\end{equation}
which are discussed in detail in \cite{Buras-Cracow}. In (\ref{S-MFV}),
we encountered already one of them, the function $S(v)$, which  
governs $B^0_q$--$\bar B^0_q$ and $K^0$--$\bar K^0$ mixing; below, we
will come across $X(v)$ and $Y(v)$, which characterize rare $K$, $B$ 
decays with $\nu\bar\nu$ and $\ell^+\ell^-$ in the final states, 
respectively. The important property of the functions in (\ref{masterf}) 
is that they do not -- within the framework of MFV --  contain complex 
phases, so that the CP-violating effects are governed {\it entirely} by 
the KM phase hiding in the parameters $P_r$.

For detailed discussions of the many interesting aspects of rare $B$
and $K$ decays and recent developments, we refer the reader to 
\cite{BF-rev,B-LH98,BuMi,rare}. Let us here choose 
$B_{s,d}\to\mu^+\mu^-$ and $K\to\pi\nu\bar\nu$ processes as 
representative examples, which are particularly clean from 
the theoretical point of view; the former channels are also an 
important element of the $B$-physics programme of the LHC \cite{LHC-Book}. 
Finally, we shall illustrate the impact of NP that does {\it not} belong 
to the class of MFV models on rare decays. To this end, we consider a 
NP scenario that is suggested by the ``$B\to\pi K$ puzzle'' discussed in 
Subsection~\ref{ssec:BpiK-puzzle}.

\boldmath\subsection{$B_{s,d}\to\mu^+\mu^-$}
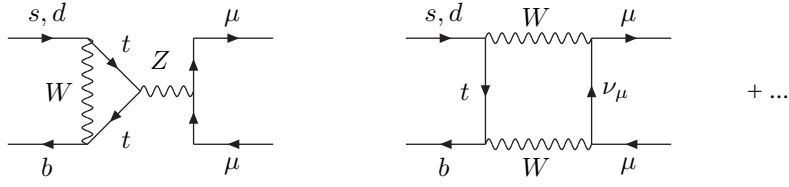
\begin{figure}
\begin{center}
{\small
\hspace*{-6.0truecm}\begin{picture}(250,70)(0,45)
\ArrowLine(60,100)(90,100)
\ArrowLine(130,100)(160,100)
\ArrowLine(90,60)(60,60)
\ArrowLine(160,60)(130,60)
\ArrowLine(90,100)(110,80)
\ArrowLine(110,80)(90,60)
\ArrowLine(130,60)(130,80)
\ArrowLine(130,80)(130,100)
\Photon(90,100)(90,60){2}{8}
\Photon(110,80)(130,80){2}{4}
\Text(75,105)[b]{$s,d$}\Text(105,95)[b]{$t$}\Text(145,105)[b]{$\mu$}
\Text(75,55)[t]{$b$}\Text(105,65)[t]{$t$}\Text(145,55)[t]{$\mu$}
\Text(85,80)[r]{$W$}\Text(114,92)[l]{$Z$}
\ArrowLine(210,100)(240,100)\Photon(240,100)(280,100){2}{8}
\ArrowLine(280,100)(310,100)
\ArrowLine(240,60)(210,60)\Photon(240,60)(280,60){2}{8}
\ArrowLine(310,60)(280,60)
\ArrowLine(240,100)(240,60)\ArrowLine(280,60)(280,100)
\Text(225,105)[b]{$s,d$}\Text(260,105)[b]{$W$}\Text(295,105)[b]{$\mu$}
\Text(225,55)[t]{$b$}\Text(260,55)[t]{$W$}\Text(295,55)[t]{$\mu$}
\Text(235,80)[r]{$t$}\Text(285,80)[l]{$\nu_\mu$}
\Text(340,80)[l]{+ ...}
\end{picture}}
\end{center}
\vspace*{-0.3truecm}
\caption{Decay processes contributing to $B_{s,d}\to\mu^+\mu^-$
in the SM.}\label{fig:Bsd-mumu-diag}
\end{figure}

\unboldmath
As can be seen in Fig.~\ref{fig:Bsd-mumu-diag}, within the framework
of the SM, the decays $B_{s,d}\to\mu^+\mu^-$ originate from $Z^0$ penguins 
and box diagrams. These transitions belong to the cleanest 
modes in the category of rare $B$ decays, since they involve only 
the hadronic matrix element of a quark current between a $B_q$-meson
and the vacuum state, i.e.\ the decay constant $f_{B_q}$ that we introduced 
in (\ref{ME-rel2}), NLO QCD corrections were calculated, and 
long-distance contributions are expected to play a negligible 
r\^ole \cite{BB-Bmumu}. The low-energy effective Hamiltonian 
describing $B_q\to\mu^+\mu^-$ decays is given as follows ($q\in\{s,d\}$):
\begin{equation}\label{Heff-Bmumu}
{\cal H}_{\rm eff}=-\frac{G_{\rm F}}{\sqrt{2}}\left[
\frac{\alpha}{2\pi\sin^2\Theta_{\rm W}}\right]
V_{tb}^\ast V_{tq} \eta_Y Y_0(x_t)(\bar b q)_{\rm V-A}(\bar\mu\mu)_{\rm V-A} 
\,+\, {\rm h.c.},
\end{equation}
where $\alpha$ denotes the QED coupling and $\Theta_{\rm W}$ is the
Weinberg angle. Here the short-distance physics is described by 
\begin{equation}
Y(x_t)=\eta_Y Y_0(x_t),
\end{equation}
where $\eta_Y=1.012$ is a perturbative QCD 
correction factor \cite{BB-Bmumu,eta-Y,MiU}, and $Y_0(x_t)$, which is another 
Inami--Lim function \cite{IL}, describes the top-quark mass dependence 
of the Feynman diagrams shown in Fig.~\ref{fig:Bsd-mumu-diag}. In the SM, we 
may write $Y_0(x_t)$ -- to a very good approximation -- as follows
\cite{Buras-Cracow}:
\begin{equation}
Y_0(x_t)=0.98\times\left[\frac{m_t}{167 \, \mbox{GeV}}\right]^{1.56}.
\end{equation}

We observe that the matrix element of (\ref{Heff-Bmumu}) between a 
$\langle\mu^-\mu^+|$ final state and a $|B_q\rangle$ initial state indeed 
involves the decay constant $f_{B_q}$. The corresponding SM branching ratios 
then take the following form \cite{Brev01}:
\begin{equation}\label{BR-Bsmumu}
\mbox{BR}( B_s \to \mu^+ \mu^-) = 4.1 \times 10^{-9} \left[
\frac{f_{B_s}}{0.24 \, \mbox{GeV}} \right]^2 \left[
\frac{|V_{ts}|}{0.040}           \right]^2 \left[
\frac{\tau_{B_s}}{1.5 \, \mbox{ps}} \right] \left[ \frac{m_t}{167 
\, \mbox{GeV} } \right]^{3.12}
\end{equation}
\begin{equation}\label{BR-Bdmumu}
\mbox{BR}(B_d \to \mu^+ \mu^-) = 1.1 \times 10^{-10}
\left[ \frac{f_{B_d}}{0.20 \, \mbox{GeV}} \right]^2 \left[
\frac{|V_{td}|}{0.008}           \right]^2
\left[ \frac{\tau_{B_d}}{1.5 \, \mbox{ps}} \right] \left[
\frac{m_t}{167 \, \mbox{GeV} } \right]^{3.12},
\end{equation}
which should be compared with the experimental $90\%$ C.L.\ bounds
\begin{equation}\label{Bmumu-exp}
\mbox{BR}( B_s \to \mu^+ \mu^-)< 5.8 \times 10^{-7}, \quad
\mbox{BR}( B_d \to \mu^+ \mu^-)< 1.5 \, (1.6) \times 10^{-7}
\end{equation}
obtained by the CDF (Belle) collaboration \cite{Bmumu-bounds}.
Looking at (\ref{BR-Bsmumu}) and (\ref{BR-Bdmumu}), we see that
a measurement of these branching ratios would allow clean
determinations of $|V_{ts}|$ and $|V_{td}|$, respectively, provided 
the non-perturbative decay constants $f_{B_s}$ and $f_{B_d}$ were 
known reliably. The current status following from lattice QCD studies
is given as follows \cite{CKM-Book}:
\begin{equation}
f_{B_d}=(203\pm27^{+0}_{-20})\,\mbox{MeV}, \quad
f_{B_s}=(238\pm31)\,\mbox{MeV};
\end{equation}
similar results were obtained with the help of QCD sum rules 
\cite{SR-calc}. If we consider the ratio
\begin{equation}\label{RT1-rare}
\frac{\mbox{BR}(B_d\to\mu^+\mu^-)}{\mbox{BR}(B_s\to\mu^+\mu^-)}=
\left[\frac{\tau_{B_d}}{\tau_{B_s}}\right]
\left[\frac{M_{B_d}}{M_{B_s}}\right]
\left[\frac{f_{B_d}}{f_{B_s}}\right]^2
\left|\frac{V_{td}}{V_{ts}}\right|^2,
\end{equation}
these parameters enter only in the form of the following $SU(3)$-breaking
ratio (see also (\ref{xi-lat})):
\begin{equation}
\frac{f_{B_s}}{f_{B_d}}=1.18\pm0.04^{+0.12}_{-0}.
\end{equation}
Using now (\ref{Rt-simple-rel}), the relation in (\ref{RT1-rare}) allows a 
determination of the side $R_t$ of the UT. On the other hand, 
we may also write (see (\ref{DMq-simple}))
\begin{equation}\label{RT2-DM}
\frac{\Delta M_d}{\Delta M_s}=
\left[\frac{M_{B_d}}{M_{B_s}}\right]
\left[\frac{\hat B_{B_d}}{\hat B_{B_s}}\right]
\left[\frac{f_{B_d}}{f_{B_s}}\right]^2
\left|\frac{V_{td}}{V_{ts}}\right|^2,
\end{equation}
allowing us to fix $R_t$ with the help of (\ref{Rt-SU3}). Consequently,
(\ref{RT1-rare}) and (\ref{RT2-DM}) provide complementary determinations 
of the UT side $R_t$. Moreover, these expressions imply also the 
following relation:
\begin{equation}\label{Bmumu-DM-rel}
\frac{\mbox{BR}(B_s\to\mu^+\mu^-)}{\mbox{BR}(B_d\to\mu^+\mu^-)}=
\left[\frac{\tau_{B_s}}{\tau_{B_d}}\right]
\left[\frac{\hat B_{B_d}}{\hat B_{B_s}}\right]
\left[\frac{\Delta M_s}{\Delta M_d}\right],
\end{equation}
which suffers from theoretical uncertainties that are smaller than
those affecting (\ref{RT1-rare}) and (\ref{RT2-DM}), since the 
dependence on $(f_{B_d}/f_{B_s})^2$ cancels, and 
$\hat B_{B_d}/\hat B_{B_s}=1$ up to tiny $SU(3)$-breaking 
corrections \cite{Buras-rel}. In particular, QCD lattice simulations
give the following numbers \cite{CKM-Book}:
\begin{equation}
\hat B_{B_d}=1.34\pm0.12, \quad \hat B_{B_s}=1.34\pm0.12, \quad 
\frac{\hat B_{B_s}}{\hat B_{B_d}}=1.00\pm0.03.
\end{equation}
Moreover, we may also use the (future) experimental data for 
$\Delta M_{(s)d}$ to reduce the hadronic uncertainties in the 
SM predictions for the $B_q\to\mu^+\mu^-$ branching ratios 
\cite{Buras-rel}, yielding
\begin{equation}
\left.\mbox{BR}(B_s \to \mu^+ \mu^-)\right|_{\rm SM} = (3.42 \pm 0.53)\times
\left[\frac{\Delta M_s}{18.0\, {\rm ps}^{-1}}\right]
\times 10^{-9}
\end{equation}
\begin{equation}
\left.\mbox{BR}(B_d\to \mu^+ \mu^-)\right|_{\rm SM} = 
(1.00 \pm 0.14)\times 10^{-10}.
\end{equation}

Since these branching ratios are very small, we could only hope 
to observe the $B_q \to \mu^+ \mu^-$ decays at the LHC, should they 
actually be governed by their SM contributions \cite{LHC-Book}. 
However, as these transitions are mediated by rare FCNC processes,
they are sensitive probes for NP. In particular, as was recently
reviewed in \cite{Buras-NP-rev}, the $B_q \to \mu^+ \mu^-$ branching 
ratios may be dramatically enhanced in specific NP (SUSY) scenarios.
Should this actually be the case, these decays may be seen at run II 
of the Tevatron, and the $e^+e^-$ $B$ factories could observe 
$B_d\to \mu^+ \mu^-$. In the case of models with MFV, we just have 
to make the replacement 
\begin{equation}
Y(x_t) \to Y(v)
\end{equation}
in order to take the NP contributions to the $B_q \to \mu^+ \mu^-$ decays 
into account. In particular, the {\it same} $Y(v)$ enters the 
$B_s \to \mu^+ \mu^-$ and $B_d \to \mu^+ \mu^-$ channels 
(see (\ref{masterf})). In analogy, the {\it same} generalized function 
$S(v)$ governs the mass differences $\Delta M_s$ and $\Delta M_d$, as we 
have seen in Subsection~\ref{ssec:MFV}. Consequently, within 
MFV scenarios, the NP effects cancel in 
(\ref{RT1-rare}), (\ref{RT2-DM}) and (\ref{Bmumu-DM-rel}), where
in particular the latter relation offers an interesting test of
this picture.

\boldmath\subsection{$K\to\pi\nu\bar\nu$}\unboldmath\label{ssec:Kpinunu-detail}
As we discussed in Subsection~\ref{subsec:rare-kaon-brief},
$K\to\pi\nu\bar\nu$ decays originate from $Z^0$ penguins and box diagrams.
Let us first have a closer look at the charged mode $K^+\to\pi^+\nu\bar\nu$. 
The low-energy effective Hamiltonian describing this decay is
given as follows \cite{Brev01}:
\begin{equation}\label{Heff-Kpinunu}
{\cal H}_{\rm eff}=\frac{G_{\rm F}}{\sqrt{2}}\left[
\frac{\alpha}{2\pi\sin^2\Theta_{\rm W}}\right]\sum_{\ell=e,\mu,\tau}
\left[\lambda_c X^\ell_{\rm NL}+
\lambda_t X(x_t)\right]
(\bar s d)_{\rm V-A}(\bar\nu_\ell \nu_\ell)_{\rm V-A},
\end{equation}
where 
\begin{equation}
\lambda_c\equiv V_{cs}^\ast V_{cd}=-\lambda\left(1-\frac{\lambda^2}{2}\right)
\end{equation}
and $\lambda_t\equiv V_{ts}^\ast V_{td}$ with
\begin{equation}\label{RE-IM}
{\rm Im}\,\lambda_t=\eta A^2 \lambda^5,\quad 
{\rm Re}\,\lambda_t=-\left(1-\frac{\lambda^2}{2}\right)A^2\lambda^5
(1-\bar \rho)
\end{equation}
are CKM factors, and
\begin{equation}
X(x_t)=\eta_X X_0(x_t)
\end{equation}
describes the top-quark mass dependence originating from the $Z^0$ 
penguin and box diagrams, where $X_0(x_t)$ is another Inam--Lim 
function \cite{IL}, and $\eta_X=0.994$ is a perturbative NLO QCD 
correction factor \cite{BB-Bmumu,eta-Y,MiU,eta-X}. Within the SM, we may 
write $X_0(x_t)$ -- to a very good approximation -- as follows
\cite{Buras-Cracow}:
\begin{equation}
X_0(x_t)=1.53\times\left[\frac{m_t}{167 \, \mbox{GeV}}\right]^{1.15}.
\end{equation}
The counterpart of $X(x_t)$ in the charm sector is given by 
$X^\ell_{\rm NL}$. For the analysis of $K^+\to\pi^+\nu\bar\nu$, 
the following combination is relevant:\footnote{The small numerical
difference of $P_c(\nu\bar\nu)$ with respect to the value given in 
\cite{Brev01}, where $\lambda=0.2205$ was used, is related to the very 
recent value of $\lambda=0.2240$ \cite{CKM-Book}. A similar comment 
applies to the quantities $\kappa_+$ and $\kappa_{\rm L}$, to be 
introduced below.}
\begin{equation}
P_c(\nu\bar\nu)=\frac{1}{\lambda^4}\left[\frac{2}{3}X^e_{\rm NL}+
\frac{1}{3}X^\tau_{\rm NL}\right]=0.39\pm0.06.
\end{equation}
If we calculate the matrix element of (\ref{Heff-Kpinunu}) between
the $\langle\bar\nu\nu\pi^+|$ final state and the $|K^+\rangle$ initial 
state, we encounter a hadronic matrix element of the 
$(\bar s d)_{\rm V-A}$  current
that can be extracted -- with the help of the isospin flavour
symmetry of strong interactions -- from the semileptonic decay
$K^+\to\pi^0e^+\nu$, which is a tree decay that is described by
the following Hamiltonian \cite{Brev01}:
\begin{equation}
{\cal H}_{\rm eff}(K^+\to\pi^0e^+\nu)=\frac{G_{\rm F}}{\sqrt{2}}V_{us}^\ast
(\bar s u)_{\rm V-A}(\bar\nu_e e)_{\rm V-A}.
\end{equation}
Using the isospin relation
\begin{equation}
\langle\pi^+|(\bar s d)_{\rm V-A}|K^+\rangle=
\sqrt{2}\langle\pi^0|(\bar s u)_{\rm V-A}|K^+\rangle,
\end{equation}
and neglecting the phase-space differences due to $M_{\pi^+}\not=M_{\pi^0}$
and $M_e\not=0$, we obtain
\begin{equation}\label{rare-K-iso}
\frac{\mbox{BR}(K^+\to\pi^+\nu\bar\nu)}{\mbox{BR}(K^+\to\pi^0e^+\nu)}
=\frac{\alpha^2}{|V_{us}|^22\pi^2\sin^4\Theta_{\rm W}}
\sum_{\ell=e,\mu,\tau}|\lambda_c X^\ell_{\rm NL}+
\lambda_t X(x_t)|^2.
\end{equation}
Consequently, we may determine the hadronic matrix element 
relevant to the rare decay $K^+\to\pi^+\nu\bar\nu$ through
the experimental data for the (non-rare) decay $K^+\to\pi^0e^+\nu$.
Because of this important feature, $K^+\to\pi^+\nu\bar\nu$ is 
a very {\it clean} decay. 

It is useful to write the $K^+\to\pi^+\nu\bar\nu$ branching ratio as 
\begin{equation}\label{b1}
B_1\equiv\frac{1}{\kappa_+}{\rm BR}(K^+\to\pi^+\nu\bar\nu),
\end{equation}
with
\begin{equation}\label{kappa-p}
\kappa_+=r_{K^+}\left[\frac{3\alpha^2{\rm BR}(K^+\to\pi^0e^+\nu)}{2\pi^2
\sin^4\Theta_{\rm W}}\right]\lambda^8=4.78\times 10^{-11},
\end{equation}
where $r_{K^+}=0.901$ describes the isospin-breaking
corrections that arise in relating $K^+\to\pi^+\nu\bar\nu$ to 
$K^+\to\pi^0e^+\nu$. Let us now consider the general MFV case, where
\begin{equation}
X(x_t)\to X(v). 
\end{equation}
The ``reduced'' $K^+\to\pi^+\nu\bar\nu$ branching 
ratio $B_1$ can then be expressed as follows \cite{BF-MFV}:
\begin{equation}\label{bkpn}
B_1=\left[{{\rm Im}\lambda_t\over\lambda^5}|X(v)|\right]^2+
\left[{{\rm Re}\lambda_c\over\lambda}{\rm sgn}(X(v)) P_c(\nu\bar \nu) +
{{\rm Re}\lambda_t\over\lambda^5}|X(v)|\right]^2;
\end{equation}
the corresponding SM prediction following from the very recent update
in \cite{BFRS3} was given in (\ref{Kpinunu-SM}). It is now an easy 
exercise to show that the measured $K^+ \to \pi^+ \nu \bar\nu$ branching 
ratio determines an ellipse in the $\bar\rho$--$\bar\eta$ plane,
\begin{equation}\label{ellipse}
\left(\frac{\bar\rho-\rho_0}{\bar \rho_1}\right)^2+
\left(\frac{\bar\eta}{\bar \eta_1}\right)^2=1,
\end{equation}
centred at $(\rho_0,0)$ with 
\begin{equation}\label{110}
\rho_0 = 1 + {\rm sgn}(X(v)) \frac{P_c(\nu\bar\nu)}{A^2 |X(v)|},
\end{equation}
and having the squared axes
\begin{equation}\label{110a}
\bar\rho_1^2 = r^2_0, \quad \bar\eta_1^2 = 
\left(\frac{r_0}{\sigma}\right)^2,
\end{equation}
with
\begin{equation}
r^2_0 = \frac{\sigma B_1}{A^4 |X(v)|^2}, \quad
\sigma = \frac{1}{(1-\lambda^2/2)^2}.
\end{equation}

Concerning $K_{\rm L}\to \pi^0\nu\bar\nu$, we may
introduce -- in analogy to (\ref{b1}) -- the reduced branching ratio
\begin{equation}\label{b2}
B_2\equiv\frac{1}{\kappa_{\rm L}}{\rm BR}(K_{\rm L}\to\pi^0\nu\bar\nu),
\end{equation}
which is characterized by
\begin{equation}\label{kappa-L}
\kappa_{\rm L}=\left[\frac{r_{K_{\rm L}}}{r_{K^+}}
\frac{\tau_{K_{\rm L}}}{\tau_{K^+}}\right]\kappa_+=2.09\times 10^{-10},
\end{equation}
where $r_{K_{\rm L}}=0.944$ describes the isospin-breaking
corrections that arise in relating $K_{\rm L}\to \pi^0\nu\bar\nu$ to 
$K^+\to\pi^0e^+\nu$. As discussed in detail in \cite{Brev01},
the decay $K_{\rm L}\to \pi^0\nu\bar\nu$ is dominated in the SM
by direct CP violation, and is completely governed by the short-distance
loop diagrams with internal top-quark exchanges. Since the charm contribution
can be fully neglected, the decay $K_{\rm L}\to \pi^0\nu\bar\nu$ is
{\it even cleaner} than $K^+\to\pi^+\nu\bar\nu$. In models with MFV, the 
reduced $K_{\rm L}\to \pi^0\nu\bar\nu$ branching ratio is given as follows:
\begin{equation}\label{bklpn}
B_2=\left[{{\rm Im}\lambda_t\over\lambda^5}|X(v)|\right]^2;
\end{equation}
the SM corresponds to (\ref{Kpinunu-SM}). 
If we now follow \cite{BBSIN}, but admit both signs of $X(v)$ and $S(v)$, 
we obtain
\begin{equation}\label{rhetb}
\bar\rho=1+\left[{\pm\sqrt{\sigma(B_1-B_2)}+
{\rm sgn}(X(v)) P_c(\nu\bar\nu)\over A^2 |X(v)|}\right],\quad
\bar\eta= {\rm sgn}(S(v)){\sqrt{B_2}\over\sqrt{\sigma} A^2 |X(v)|}.
\end{equation}
The dependence on $|X(v)|$ cancels in the following quantity \cite{BF-MFV}:
\begin{equation}\label{cbbnew}
r_s\equiv{1-\overline{\rho}\over\overline{\eta}}={\rm ctg}\beta=
{\rm sgn}(S(v))\,\sqrt{\sigma}\left[{\mp\sqrt{\sigma(B_1-B_2)}-
{\rm sgn}(X(v))P_c(\nu\bar\nu)\over\sqrt{B_2}}\right],
\end{equation}
which allows the determination of $(\sin2\beta)_{\pi\nu\bar\nu}$ 
in (\ref{s2b-K-B}) through
\begin{equation}
\sin2\beta=\frac{2r_s}{1+r_s^2}.
\end{equation}
Note that (\ref{cbbnew}) reduces to 
\begin{equation}
r_s=\sqrt{\sigma}\left[\frac{\sqrt{\sigma(B_1-B_2)}-
P_c(\nu\bar\nu)}{\sqrt{B_2}}\right]
\end{equation}
in the case of positive values of $S(v)$ and $X(v)$ \cite{BBSIN}. 
Because of the relation in (\ref{apsiK-def}), it is actually more 
appropriate to consider the CP-violating observable $a_{\psi K_{\rm S}}$
instead of $\sin2\beta$. Consequently, we obtain a very interesting link 
between the mixing-induced CP violation in the ``golden'' mode 
$B_d\to J/\psi K_{\rm S}$ and the branching ratios of the rare
$K\to\pi\nu\bar\nu$ decays. 

Since $a_{\psi K_{\rm S}}$ has already been measured with 
impressive accuracy and BR$(K^+\to\pi^+\nu\bar \nu)$ 
will be known rather accurately prior to the measurement of 
BR$(K_{\rm L}\to\pi^0\nu\bar\nu)$, it is of particular interest to 
calculate BR$(K_{\rm L}\to\pi^0\nu\bar\nu)$ as a function of 
BR$(K^+\to\pi^+\nu\bar\nu)$ for a given value of $a_{\psi K_{\rm S}}$
\cite{BF-MFV}. To this end, it is useful to introduce the quantity 
\begin{equation}
f(\beta)\equiv \mbox{sgn}(S(v))\,{\rm ctg}\beta=
\frac{1-\bar\rho}{|\bar\eta|},
\end{equation}
which can be determined {\it unambiguously} through 
\begin{equation}\label{rs-unambig}
f(\beta)=\frac{1+\sqrt{1-a_{\psi K_{\rm S}}^2}}{a_{\psi K_{\rm S}}}
=2.279^{+0.235}_{-0.215};
\end{equation}
the numerical value corresponds to $a_{\psi K_{\rm S}}=0.736\pm0.049$.
We then obtain the following expression:
\begin{equation}\label{B1B2}
B_1=B_2+\left[\frac{f(\beta)\sqrt{B_2}+
{\rm sgn}(X(v))\sqrt{\sigma}P_c(\nu\bar\nu)}
{\sigma}\right]^2.
\end{equation}
In comparison with (\ref{cbbnew}), the advantage of (\ref{B1B2}) 
is the absence of the sign ambiguities due to ${\rm sgn}(S(v))$ and
the $\mp$ in front of $\sqrt{\sigma(B_1-B_2)}$. Consequently, 
for given values of $a_{\psi K_{\rm S}}$ and BR$(K^+\to\pi^+\nu\bar\nu)$, 
only two values of BR$(K_{\rm L}\to\pi^0\nu\bar\nu)$ are allowed for 
the full class of MFV models, independently of any new parameter
present in these models. These two values of the $K_{\rm L}\to\pi^0\nu\bar\nu$
branching ratio correspond to the two possible signs of $X(v)$. 
The measurement of BR$(K_{\rm L}\to\pi^0\nu\bar\nu)$ will therefore 
either select one of these two possible values or will rule out all 
MFV models.

\boldmath\subsection{New Physics Beyond Minimal Flavour
Violation: An Example}\unboldmath\label{ssec:NP-rare}
As we have seen in Subsection~\ref{ssec:BpiK-puzzle}, the pattern of
the current $B$-factory data for the $B\to\pi K$ system suggests an
enhancement of the corresponding EW penguin parameter $q$, and the
presence of a CP-violating NP phase $\phi$ in the EW penguin sector,
as summarized in (\ref{q-det}). Since we encounter here CP-violating
effects that are {\it not} associated with the CKM matrix, the 
corresponding NP does {\it not} belong to the category of MFV models 
considered above. In order to explore the implications for rare $B$ and 
$K$ decays, let us follow \cite{BFRS3,BFRS2}, and consider a specific 
scenario, where the NP effects enter through enhanced $Z^0$ penguins, 
which are described by a short-distance function $C$. 

The implications of enhanced $Z^0$ penguins with a large new complex 
phase for rare and CP-violating $K$ and $B$ decays were already 
discussed in \cite{Buras:1998ed}--\cite{Buchalla:2000sk}, where 
model-independent analyses and studies within particular 
supersymmetric scenarios were presented. Here we determine the 
size of the enhancement of the $Z^0$-penguin function $C$ and 
the magnitude of its complex phase through the $B\to\pi K$ data.
As was pointed out in \cite{BFRS1}, a connection between 
rare decays and the $B\to\pi K$ system can be established by 
relating the EW penguin parameter $q$ to the $Z^0$-penguin function 
$C$, which can be properly done with the help of a renormalization-group
analysis. In the case of a complex EW penguin parameter, with
a non-vanishing weak phase $\phi$, we obtain the following
relation \cite{BFRS3,BFRS2}:
\begin{equation}\label{RG}
C\equiv|C|e^{i\theta_C}= 2.35~ \bar q e^{i\phi} -0.82,\quad 
\bar q= q \left[\frac{|V_{ub}/V_{cb}|}{0.086}\right].
\end{equation}
This quantity enters the short-distance functions
$X$ and $Y$, which govern the rare $K$, $B$ decays with $\nu\bar\nu$ 
and $\mu^+\mu^-$ in the final states, respectively, in the linear combinations
\begin{equation}
X\equiv|X|e^{i\theta_X}=C+B^{\nu\bar\nu}, \quad
Y\equiv|Y|e^{i\theta_Y}=C+B^{\mu^+\mu^-},
\end{equation}
where $B^{\nu\bar\nu}$ and $B^{\mu^+\mu^-}$ describe the
box diagrams with $\nu\bar\nu$ and $\mu^+\mu^-$, respectively. 
If we evaluate, in the spirit of \cite{BFRS1,Buras:1998ed,Buras:1999da}, 
these box-diagram contributions in the SM and use (\ref{RG}), we obtain
\begin{equation}\label{X-C-rel}
|X|e^{i\theta_X}=|C|e^{i\theta_C}+0.73 \quad \mbox{and} \quad 
|Y|e^{i\theta_Y}=|C|e^{i\theta_C}+0.18.
\end{equation}
While the analysis described here does not rely on a particular model, 
concrete models with enhanced CP-violating $Z^0$-mediated FCNC 
couplings, generated either at the one-loop level or even at the tree 
level, were discussed in the literature (see, for instance,
\cite{GNR97,trojan,Buras:1998ed,Buras:1999da,Buchalla:2000sk}).
Let us also note that models with $Z^\prime$-mediated FCNCs could 
be put in this class, provided their contributions can effectively 
be absorbed in the function $C$ (for a recent analysis, 
see~\cite{BCLL03}).

If we now insert the numerical values in (\ref{q-det}) into 
(\ref{X-C-rel}), we obtain
a central value for $|Y|$ that violates the upper bound $|Y|\le 2.2$ 
following from the BaBar and Belle data on $B\to X_s\mu^+\mu^-$ 
\cite{Kaneko:2002mr}, and the upper bound on 
$\mbox{BR}(K_{\rm L}\to \pi^0 e^+e^-)$ of $2.8\times 10^{-10}$ from KTeV 
\cite{KTEVKL}. However, we may still encounter significant deviations from 
the SM. In order to illustrate this exciting feature, we consider
only the subset of those values of $(q,\phi)$ in (\ref{q-det}) that 
satisfy the constraint of $|Y|= 2.2$. If we then use (\ref{RG})
and (\ref{X-C-rel}), we obtain  
\begin{equation}\label{CXY}
\begin{array}{rclcrcl}
|C|&=&2.24\pm 0.04,&&\theta_C&=&-(105\pm 12)^\circ,\\
|X|&=&2.17\pm0.12,&&\theta_X&=&-(87\pm12)^\circ,\\
|Y|&=&2.2\,\, {\rm (input)}, &&\theta_Y&=&-(103\pm12)^\circ,
\end{array}
\end{equation}
which should be compared with the SM values $C(x_t)=0.79$, $X(x_t)=1.53$ 
and $Y(x_t)=0.98$, corresponding to $m_t=167~{\rm GeV}$.

Going now back to the $B_q\to\mu^+\mu^-$ decays, we find
\begin{equation}\label{BqLL-enhance}
\frac{\mbox{BR}(B_{s}\to\mu^+\mu^-)}{\mbox{BR}(B_{s}\to\mu^+\mu^-)_{\rm SM}}
=\frac{\mbox{BR}(B_{d}\to\mu^+\mu^-)}{\mbox{BR}(B_{d}\to\mu^+\mu^-)_{\rm SM}}
=\left|\frac{Y}{Y_{\rm SM}}\right|^2 \approx 5.0.
\end{equation}
This significant enhancement corresponds to the branching ratios
\begin{equation}
\mbox{BR}(B_{s}\to\mu^+\mu^-) \approx 17 \times 10^{-9}, \quad 
\mbox{BR}(B_{d}\to\mu^+\mu^-) \approx 5 \times 10^{-10},
\end{equation}
which are still well below the experimental bounds summarized in
(\ref{Bmumu-exp}).

\begin{figure}
\vspace*{0.3truecm}
\begin{center}
\psfrag{K+}{BR($K^+\rightarrow\pi^+\nu\bar\nu$)}
\psfrag{KL}{BR($K_{\rm L}\rightarrow\pi^0\nu\bar\nu$)}
\psfrag{SM}{SM}\psfrag{expRange}{exp.\ range}
\psfrag{b25}{$\beta_X=25^\circ$}\psfrag{b50}{$50^\circ$}\psfrag{b70}{$70^\circ$}
\psfrag{b111}{$111^\circ$}\psfrag{b130}{$130^\circ$}\psfrag{b150}{$150^\circ$}
\psfrag{GN}{GN bound}\psfrag{E949}{E949 result}
\includegraphics[width=11cm]{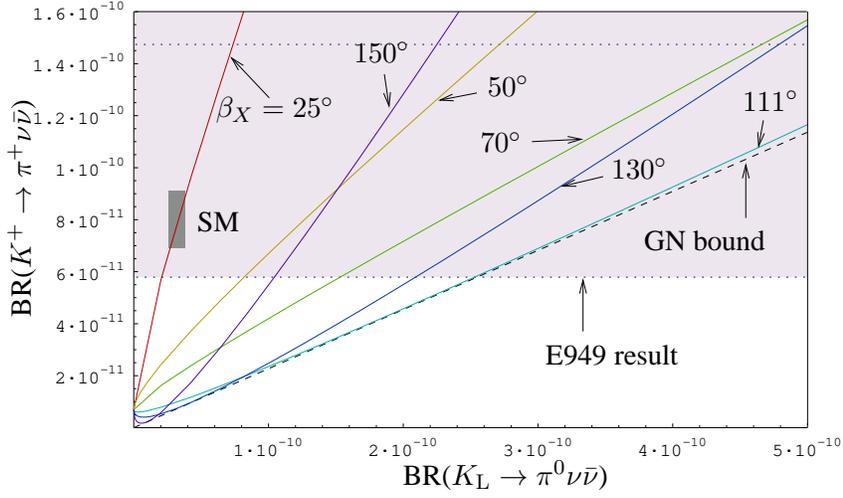}
\end{center}
\caption{$\mbox{BR}(K^+\to\pi^+\nu\bar\nu)$ as a function of 
$\mbox{BR}(K_{\rm L}\to\pi^0\nu\bar\nu)$ for various values of 
$\beta_X$. The dotted horizontal lines indicate the experimental 
range (\ref{Brook}) and the grey area the SM prediction. We also show the 
bound in (\ref{BOUND}).  \label{fig:KpKl}}
\end{figure}

As far as the $K\to\pi\nu\bar\nu$ decays are concerned, this NP 
analysis implies 
\begin{equation}\label{NPkpnr}
\mbox{BR}(K^+\to\pi^+\nu\bar\nu)=
(7.5 \pm 2.1)\times 10^{-11},\quad \mbox{BR}(K_{\rm L}\to\pi^0\nu\bar\nu)=
(3.1 \pm 1.0)\times 10^{-10},
\end{equation}
which should be compared with the SM predictions in (\ref{Kpinunu-SM}).
We observe that the impact of NP on the $K^+\to\pi^+\nu\bar\nu$
branching ratio would be small, whereas 
$\mbox{BR}(K_{\rm L}\to\pi^0\nu\bar\nu)$ would be dramatically enhanced. 
If we introduce 
\begin{equation}
\beta_X\equiv\beta-\beta_s-\theta_X \quad \mbox{with} \quad
\beta_s\equiv -\delta\gamma=-\lambda^2\eta,
\end{equation}
we see that this exciting pattern is dominantly the 
consequence of $\beta_X\approx 111^\circ$, as
\begin{equation}
\frac{\mbox{BR}(K_{\rm L}\to\pi^0
\nu\bar\nu)}{\mbox{BR}(K_{\rm L}\to\pi^0\nu\bar\nu)_{\rm SM}}=
\left|\frac{X}{X_{\rm SM}}\right|^2
\left[\frac{\sin\beta_X}{\sin(\beta-\beta_s)}\right]^2
\end{equation}
and
\begin{equation}
\frac{\mbox{BR}(K_{\rm L}\to\pi^0\nu\bar\nu)}{\mbox{BR}
(K^+\to\pi^+\nu\bar\nu)}\approx 4.4\times (\sin\beta_X)^2
\approx 4.2\pm 0.2.  
\end{equation}
It is interesting to note that $\mbox{BR}(K_{\rm L}\to\pi^0\nu\bar\nu)$
is very close to its model-independent
upper bound~\cite{GN-bound}:
\begin{equation}\label{BOUND}
\mbox{BR}(K_{\rm L}\to\pi^0\nu\bar\nu)\le 4.4 \times 
\mbox{BR}(K^+\to\pi^+\nu\bar\nu).
\end{equation}
A spectacular implication of these findings is a strong violation of 
the relation in (\ref{s2b-K-B}). Indeed, 
\begin{equation}
(\sin 2 \beta)_{\pi \nu\bar\nu}=\sin 2\beta_X =-(0.69^{+0.23}_{-0.41}),
\end{equation}
in striking disagreement with $(\sin 2 \beta)_{\psi K_{\rm S}}= 
0.736\pm0.049$. In Fig.~\ref{fig:KpKl}, we plot -- in the spirit of 
\cite{BF-MFV} -- $\mbox{BR}(K^+\to\pi^+\nu\bar\nu)$ as a function of 
$\mbox{BR}(K_{\rm L}\to\pi^0\nu\bar\nu)$ for fixed values of
$\beta_X$. As this plot is independent of $|X|$, it offers a direct
measurement of the phase $\beta_X$. The first line on the left 
represents the MFV models with $\beta_X=\beta-\beta_s$, whereas the 
first line on the right corresponds to the model-independent Grossman--Nir 
bound given in (\ref{BOUND}). The central value $\beta_X=111^\circ$ found 
in \cite{BFRS3,BFRS2} is very close to this bound. As can be seen in 
Fig.~\ref{fig:KpKl}, the measured $K\to\pi\nu\bar\nu$ branching ratios 
allow us to determine $\beta_X$ up to discrete ambiguities, which can be 
resolved by considering other rare decays simultaneously. The corresponding 
plot for different values of $\beta_X$ that are close to $\beta$ can be 
found in \cite{BF-MFV}.

In addition to the significant and -- in the case of 
$K_{\rm L}\to\pi^0\nu\bar\nu$ and $(\sin 2 \beta)_{\pi \nu\bar\nu}$ -- 
even spectacular NP effects discussed above, there are further 
interesting implications of this scenario \cite{BFRS3,BFRS2}:
\begin{itemize}
\item The branching ratio
\begin{equation}
\mbox{BR}(K_{\rm L}\to\pi^0e^+e^-)= (7.8\pm 1.6)\times 10^{-11}
\end{equation}
is significantly enhanced and governed by direct CP violation. On
the other hand, the SM result $(3.2^{+1.2}_{-0.8})\times 10^{-11}$ 
\cite{BI03} is dominated by indirect CP violation. In a very 
recent analysis \cite{ISU}, the same NP scenario was considered 
as well, addressing also the decay $K_{\rm L}\to\pi^0\mu^+\mu^-$.

\item The integrated forward--backward CP asymmetry for 
$B_d\to K^*\mu^+\mu^-$ \cite{Buchalla:2000sk}, which is 
given by
\begin{equation}
A^{\rm CP}_{\rm FB}=(0.03\pm0.01)\times \tan\theta_Y, 
\end{equation}
can be very large in view of $\theta_Y\approx -100^\circ$. The corresponding
NP effects for the lepton polarization asymmetries of 
$B \to X_s \ell^+ \ell^-$ decays were recently studied in \cite{CCG}.

\item The $B\to X_{s,d}\nu\bar\nu$ branching ratios are enhanced 
by a factor of $2$ with respect to the SM.

\item Enhanced $Z^0$ penguins may also play an important r\^ole in 
$\mbox{Re}(\varepsilon'/\varepsilon)$ \cite{Buras:1998ed}. As far as
the enhancement of $|C|$ and its large negative phase suggested by the 
$B\to\pi K$ analysis are concerned, the consistency with (\ref{epspeps-WA}) 
requires a significant enhancement of the hadronic matrix element of the 
relevant QCD penguin operator with respect to that of the relevant 
EW penguin operator. The corresponding large hadronic uncertainties leave 
sufficient room for such effects. 

\item It is also interesting to explore the implications for 
$B_d\to J/\psi K_{\rm S}$ and $B_d\to \phi K_{\rm S}$. As far as the 
former channel is concerned, the NP corrections to the determination 
of $\sin2\beta$ from the mixing-induced $B_d\to J/\psi K_{\rm S}$ 
CP asymmetry are at the 0.05 level, corresponding to a shift of 
$\beta$ by at most $\pm2^\circ$. Such small effects are still 
beyond the current experimental and theoretical accuracy, but could 
be reinvestigated in the LHC era. 
Concerning the decay $B_d\to \phi K_{\rm S}$, large hadronic uncertainties 
preclude a precise prediction. However, if we assume that the sign of the 
cosine of a strong phase agrees with factorization, we find 
\begin{equation}\label{NP-pattern}
\underbrace{(\sin 2 \beta)_{\phi K_{\rm S}}}_{-
{\cal A}_{\rm CP}^{\rm mix}(B_d\to \phi K_{\rm S})} \, > \quad
\underbrace{(\sin 2 \beta)_{\psi K_{\rm S}}=0.736\pm0.049}_{-
{\cal A}_{\rm CP}^{\rm mix}(B_d\to J/\psi K_{\rm S})},
\end{equation}
where $(\sin 2 \beta)_{\phi K_{\rm S}}\sim1$ may well be possible.
This pattern is qualitatively different from the present $B$-factory 
data summarized in (\ref{aCP-Bd-phiK-mix}), which are, however, not 
yet conclusive. In particular, we could easily accommodate a value of  
$(\sin 2 \beta)_{\phi K_{\rm S}}$ of the same magnitude as the
central value found by Belle but of {\it opposite} sign. On the other 
hand, a future confirmation of the pattern in (\ref{NP-pattern}) would 
be another signal of enhanced CP-violating $Z^0$ penguins at work. 
\end{itemize}
If future, more accurate $B\to\pi\pi,\pi K$ data will not significantly 
modify the currently observed patterns in these decays 
discussed in Subsections~\ref{ssec:Bpipi-puzzle} and \ref{ssec:BpiK-puzzle},
the scenario of enhanced $Z^0$ penguins with a large CP-violating NP phase 
$\phi$ will remain an attractive possibility for physics beyond the SM. 
It will then be very interesting to confront the corresponding predictions 
for the rare $B$ and $K$ decays discussed above with experiment.

\section{CONCLUSIONS}\label{sec:concl}
\setcounter{equation}{0}
The field of flavour physics and CP violation is very rich and 
represents an exciting topic for theoretical and experimental 
research. In these lectures, we have put our focus on the $B$-meson 
system, which provides a particularly fertile testing ground for 
the SM picture of flavour physics, where CP violation can be 
accommodated by means of the KM mechanism through a single 
phase in the parametrization of the quark-mixing matrix. 
The corresponding UT represents one of the central targets of the
$B$ factories, which govern the current experimental stage of
quark-flavour physics, run II of the Tevatron, and of the LHCb and BTeV 
experiments, which will join these efforts in the not too distant future.  

In 1964, the observation of indirect CP violation, which 
originates from the fact that the mass eigenstates of the neutral 
kaon system are not eigenstates of the CP operator, came as a 
big surprise. After tremendous efforts, also direct CP violation 
could be established in neutral $K$ decays in 1999 by the NA48
and KTeV collaborations. Unfortunately, the calculations of the 
corresponding observable $\mbox{Re}(\varepsilon'/\varepsilon)$,
which is governed by the competition between QCD and EW penguins, 
suffer from large theoretical uncertainties. Consequently, unless
better techniques to deal with the relevant hadronic matrix elements
will be available, $\mbox{Re}(\varepsilon'/\varepsilon)$ does unfortunately
not provide a stringent test of the SM, although the SM analyses 
give results of the same order of magnitude as the experimental 
value. From the theoretical point of view, the rare decays
$K^+\to\pi^+\nu\bar\nu$ and $K_{\rm L}\to\pi^0\nu\bar\nu$ are much more 
promising. On the other hand, these decays exhibit extremely tiny 
branching ratios at the $10^{-10}$ and $10^{-11}$ levels in the SM,
respectively, and are extremely challenging from the experimental point 
of view. Nevertheless, three events for $K^+\to\pi^+\nu\bar\nu$
were already observed at BNL.

Concerning the decays of $B$ mesons, we distinguish between 
leptonic, semileptonic and non-leptonic transitions. The former
exhibit the simplest structure and would be interesting to measure
the non-perturbative decay constants $f_B$, but suffer from 
tiny branching ratios. The semileptonic $B$ decays are more
complicated than the leptonic ones. However, applications of the HQET 
and heavy-quark expansions allow us to determine $|V_{cb}|$ and $|V_{ub}|$, 
which are important ingredients for theoretical predictions and the 
analysis of the UT in the $\bar\rho$--$\bar\eta$ plane. Finally, the 
non-leptonic decays are the most complicated transitions, as far as the
impact of strong interactions is concerned. In order to deal with them
theoretically, low-energy effective Hamiltonians are used, which
consist of perturbatively calculable Wilson coefficients and local
four-quark operators. The former encode the whole short-distance 
dynamics of the decay class at hand, whereas the long-distance 
contributions of a specific channel show up as the corresponding
hadronic matrix elements of the four-quark operators. The same
formalism applies of course also to non-leptonic kaon decays and
is at the basis of the calculations of $\mbox{Re}(\varepsilon'/\varepsilon)$.
The non-leptonic $B$ decays play the key r\^ole for the exploration
of CP violation, since non-vanishing CP asymmetries may be induced by
interference effects in such transitions. In general, the theoretical
interpretation of such CP asymmetries is affected by large hadronic
uncertainties, in analogy to $\mbox{Re}(\varepsilon'/\varepsilon)$. 
However, the $B$-meson system provides tools to deal with these 
uncertainties: there are fortunate cases, where relations between 
various decay amplitudes allow us to {\it eliminate} the -- essentially 
unknown -- hadronic matrix elements, and we may exploit mixing-induced 
CP asymmetries, where the hadronic matrix elements {\it cancel} if the 
decay is governed by a single CKM amplitude. The latter observables
can also be nicely combined with amplitude relations. Following these 
lines, we may also determine -- in addition to the angles of the UT -- 
certain hadronic parameters, which can then be compared with the 
corresponding theoretical calculations, where also a lot of progress 
could be made over the recent years.

Thanks to the efforts of the BaBar and Belle collaborations, CP violation 
could be established in the $B$-meson system in 2001, with the help of 
the ``golden'' mode $B_d\to J/\psi K_{\rm S}$, thereby opening a new era 
in the exploration of this phenomenon. The current experimental status of 
the mixing-induced CP asymmetry of this (and similar) channel(s) implies 
$\sin2\beta=0.736\pm0.049$, in impressive accordance with the
indirect value following from the CKM fits of the UT in the
$\bar\rho$--$\bar\eta$ plane. The physics potential of the $B$ factories 
goes far beyond the famous $B_d\to J/\psi K_{\rm S}$ decay, allowing us 
now to confront many more strategies to explore CP violation with data. 
Here the main goal is to overconstrain the UT as much as possible, 
thereby performing a stringent test of the KM mechanism of CP violation. 
Important $B$-factory benchmark modes to complement the $B\to J/\psi K$ 
system are given by $B\to\pi\pi$ and $B\to\phi K$ decays, and 
exciting data on these channels are already available. The pattern of
the $B\to\pi\pi$ data favours large non-factorizable effects, and
the analyses of CP violation in $B_d\to\pi^+\pi^-$ point towards large 
direct and mixing-induced CP asymmetries, which can be interpreted in 
terms of $\gamma\sim65^\circ$, in accordance with the CKM fits. Although
the BaBar and Belle measurements of these asymmetries are not yet in 
full accordance, they already moved towards each other and it seems 
plausible that they will meet close to the current averages.  On the 
other hand, the Belle measurement of the mixing-induced CP asymmetry 
of $B_d\to \phi K_{\rm S}$ raises the exciting possibility of having 
large NP effects in the $\bar b\to \bar s s \bar s$ quark-level processes. 
However, the corresponding BaBar analysis is consistent with the SM, 
so that we cannot yet draw firm conclusions. Let us hope that this 
unsatisfactory experimental situation will be clarified soon.

As far as the exploration of CP violation with the help of amplitude
relations is concerned, we distinguish between exact and flavour-symmetry 
relations. The prototype of the former is provided by $B^\pm\to K^\pm D$ 
decays, whereas $B_c^\pm\to D_s^\pm D$ transitions offer the ideal 
theoretical realization of the corresponding triangle strategy to 
determine the angle $\gamma$ of the UT. An important example for the
application of flavour-symmetry relations is given by $B\to\pi K$ decays. 
Here the corresponding $B$-factory data point again to a puzzling pattern, 
which may be due to the presence of enhanced EW penguins with a large 
CP-violating NP phase. Although BaBar, Belle and CLEO indicate separately 
the corresponding ``$B\to\pi K$ puzzle'', it is still too early for definite 
conclusions. This kind of NP would yield striking effects in various
rare $B$ and $K$ decays, of which an enhancement of the 
$K_{\rm L}\to\pi^0\nu\bar\nu$ branching ratio by one order of magnitude
and a negative value of $(\sin2\beta)_{\pi\nu\bar\nu}$ would be the most 
spectacular ones. 

Another key element for the testing of the SM description of CP violation 
is the $B_s$-meson system, which is not accessible at the $e^+e^-$ $B$ 
factories operating at the $\Upsilon(4S)$ resonance, BaBar and Belle, 
but can be studied nicely at hadron collider experiments. Interesting 
results on $B_s$ physics are soon expected from run II of the Tevatron, 
where $B^0_s$--$\bar B^0_s$ mixing should be discovered, which is an 
important ingredient for the CKM fits of the UT. The most prominent 
$B_s$ decays include $B_s\to J/\psi \phi$, which is a powerful probe 
for NP contributions to $B^0_s$--$\bar B^0_s$ mixing manifesting themselves 
through a sizeable value of $\phi_s$; $B_s\to K^+K^-$, which can be combined 
with $B_d\to\pi^+\pi^-$ through the $U$-spin flavour symmetry to determine 
$\gamma$; and $B_s\to D^{(\ast)\pm}_s K^\mp$ modes, which allow clean 
determinations of $\phi_s+\gamma$ and can be combined in a variety of
ways with their $B_d\to D^{(\ast)\pm} \pi^\mp$ counterparts, offering 
advantages from the practical point of view. Although the Tevatron 
will provide first insights into these decays, they can only be fully 
exploited at the experiments of the LHC era, in particular LHCb and BTeV. 

Finally, it should be emphasized again that it is crucial to complement 
the studies of CP violation with measurements of rare $B$ 
and $K$ decays, which are sensitive probes for NP. Moreover, it is 
important to keep also an eye on the $D$-meson system, which exhibits 
tiny mixing and CP-violating effects in the SM \cite{D-rev}, as well as 
on various other interesting aspects of flavour physics, such as 
flavour-violating charged-lepton decays (for a very recent study,
see \cite{LFV}), which we could not cover in these lectures. 

In this decade, the successful exploration of flavour physics and CP 
violation will certainly be continued, thereby leading to many further
exciting results and valuable new insights. Let us hope that eventually 
also several ``surprises'' can be established, shedding light on the 
physics beyond the SM!

\vskip1cm
\noindent

\section*{ACKNOWLEDGEMENTS}

I would like to thank the students for their interest in my lectures,
the discussion leaders for their efforts to complement them in the
discussion sessions, and the local organizers for hosting this exciting
school in Armenia. I am also grateful to my collaborators for the fun 
we had working on many of the topics addressed in these lectures.

\end{document}